\newif\ifShowKeys
\ShowKeystrue
\ShowKeysfalse

\documentclass[11pt,a4paper]{article} 
\usepackage[height=8.8in,width=6.45in]{geometry}

\usepackage[hidelinks]{hyperref}

\usepackage{tocloft}

\ifShowKeys \usepackage[notcite]{showkeys} \fi

\usepackage{amsmath, amssymb,amsthm}
\usepackage{amsthm}
\numberwithin{equation}{section}

\usepackage{bm,environ,mathrsfs,array,arydshln}
\usepackage{booktabs,float,slashed,hyperref}
\usepackage[mathcal]{euscript}
\usepackage{tensor} 						
\usepackage{mathabx}
\usepackage{simpler-wick}
\usepackage[vcentermath]{youngtab}

\usepackage{aurical}
\usepackage[T1]{fontenc}
\usepackage[nodayofweek]{date time}
\usepackage{graphicx,epsfig,epic}
\usepackage{tikz} 
\usepackage{tikzfeynman}
\usetikzlibrary{arrows,decorations.pathreplacing,decorations.markings,snakes}
\tikzset{middlearrow/.style={decoration={markings, mark= at position 0.5 with {\arrow{#1}} ,
}, postaction={decorate}}}
\tikzset{decoration={snake,amplitude=.4mm,segment length=2mm,
                       post length=0mm,pre length=0mm}}

\usepackage{framed}						
\definecolor{shadecolor}{rgb}{0.95,0.95,0.97}

\usepackage{comment}

\allowdisplaybreaks

\definecolor{myred}{RGB}{233, 33, 45}

\newcommand{\bs}{\begin{shaded}}
\newcommand{\es}{\end{shaded}}
\def\ba#1\ea{\begin{align}#1\end{align}}		
\newcommand{\be}{\begin{equation}}
\newcommand{\ee}{\end{equation}}
\newcommand{\mc}{\mathcal }

\newcommand{\la}{\label}
\newcommand{\eps}{\varepsilon}

\newcommand{\lp}{\notag \\ & }

\DeclareMathOperator{\tr}{\text{tr}}

\newcommand{\cf}{\textit{cf.} }
\newcommand{\ie}{\textit{i.e.} }
\newcommand{\eg}{\textit{e.g.} }

\newcommand{\N}{\mathcal N}

\newcommand{\fix}[1]{\textcolor{myred}{$\star$ #1 $\star$}}

\makeatletter
\DeclareFontFamily{OMX}{MnSymbolE}{}
\DeclareSymbolFont{MnLargeSymbols}{OMX}{MnSymbolE}{m}{n}
\SetSymbolFont{MnLargeSymbols}{bold}{OMX}{MnSymbolE}{b}{n}
\DeclareFontShape{OMX}{MnSymbolE}{m}{n}{
<-6>  MnSymbolE5
   <6-7>  MnSymbolE6
   <7-8>  MnSymbolE7
   <8-9>  MnSymbolE8
   <9-10> MnSymbolE9
  <10-12> MnSymbolE10
  <12->   MnSymbolE12
}{}
\DeclareFontShape{OMX}{MnSymbolE}{b}{n}{
<-6>  MnSymbolE-Bold5
   <6-7>  MnSymbolE-Bold6
   <7-8>  MnSymbolE-Bold7
   <8-9>  MnSymbolE-Bold8
   <9-10> MnSymbolE-Bold9
  <10-12> MnSymbolE-Bold10
  <12->   MnSymbolE-Bold12
}{}

\let\llangle\@undefined
\let\rrangle\@undefined
\DeclareMathDelimiter{\llangle}{\mathopen}%
 {MnLargeSymbols}{'164}{MnLargeSymbols}{'164}
\DeclareMathDelimiter{\rrangle}{\mathclose}%
 {MnLargeSymbols}{'171}{MnLargeSymbols}{'171}
\makeatother

\catcode`,\active

\catcode`\,12

\def\XXint#1#2#3{{\setbox0=\hbox{$#1{#2#3}{\int}$}
     \vcenter{\hbox{$#2#3$}}\kern-.5\wd0}}

\newcommand{\vev}[1]{\langle  #1 \rangle}

\newcommand{\z}{\zeta}
\renewcommand{\l}{\lambda}
\newcommand{\drg}[1]{\frac{\partial}{\partial #1}}
\newcommand{\W}{\textrm{W}}
\newcommand{\vac}{|0\rangle}
\newcommand{\D}{\mathscr D}
\newcommand{\dtdt}{\widehat{d^{2}\tau}}
\newcommand{\Zzero}{\mathbb{Z}\backslash\{0\}}
\newcommand{\R}{\textrm{R}}
\newcommand{\A}{\textrm{A}}
\renewcommand{\S}{\textrm{S}}
\newcommand{\F}{\textrm{F}}
\newcommand{\ladd}{\textrm{ladder}}

\newcommand{\norm}{\frac{1}{\dim\R}\,}

\def \ov {\over}
\def \ci {\cite}
\def \p {\phi}
\def \WZ {W^{(\z)}}
\def \del{\partial}
\def \te {\textstyle} 
 \def \C  {{\cal C}}
 \def \OO {{\cal O}}
 \def \iffa {\iffalse}

\def \b{\beta}
\def \no {\nonumber}

\newcommand{\foot}{\footnote}
\newcommand{\rf}[1]{(\ref{#1})}
\def \pxi {{\varkappa}}
\def \cW  {{\cal W}}

\def \bigg {\Big}
\def \ed { 
\bibliography{BT-Biblio}
\bibliographystyle{JHEP}
\end{document}}
\def \tr  {\text{Tr}}
\def \gym {g}
\def \ve {\epsilon}
\def \eps {\varepsilon}
\def \pxi {{\varkappa}}
 \def \ha {{1\over 2}}
 \def \ww  {{\rm w}}
\def \ga {{\rm b_1 }}
\def \bxi  {\beta_{\pxi}}
\def \vv {{\rm v}}
\def \bb {{\rm b}}
\def \lu   {u} \def \aa {{\rm a}}  \def \bb {{\rm b}}  \def \cc {{\rm c}}
\def \pz {\psi}
\def \cD  {{\mc D}}
\def \SS {{S^{(4)}}}
\def \ph  {\varphi}
\def \WO {W_{k,0}}
\def \m {\mu}

\def \AA  {{\rm A}} 
\def \WS {{\Gamma}} \def \G  {\Gamma}
\def \xxi {{\bar \xi}}
\def \QR {Q_{\R}}
\def \chn {\eta}
\def \UV {{\rm UV}}  \def \IR {{\rm IR}}  

\def \lze {{{\ell} \neq 0}}   

\begin{document}

\begin{titlepage}



\begin{tabbing}
\hspace*{11.5cm} \=  \kill 
\>  Imperial-TP-AT-2022-01 \\
\> 
\end{tabbing}

\vspace*{15mm}
\begin{center}
{\Large\sc   Wilson loop in general representation  }\vskip 9pt
{\Large\sc    and RG flow  in 1d defect QFT }

\vspace*{10mm}

{\Large M. Beccaria${}^{\,a}$, S. Giombi$^{\,b}$, A.A. Tseytlin$^{\,c, }$\footnote{\ Also at the Institute for Theoretical and Mathematical Physics (ITMP) of Moscow University   and Lebedev Institute.}} 

\vspace*{4mm}
	
${}^a$ Universit\`a del Salento, Dipartimento di Matematica e Fisica \textit{Ennio De Giorgi},\\ 
		and I.N.F.N. - sezione di Lecce, Via Arnesano, I-73100 Lecce, Italy
			\vskip 0.3cm
${}^b$  Department of Physics, Princeton University, Princeton, NJ 08544, USA
			\vskip 0.3cm
${}^c$ Blackett Laboratory, Imperial College London SW7 2AZ, U.K.
			\vskip 0.3cm
			
\vskip 0.2cm
	{\small
		E-mail:
		\texttt{matteo.beccaria@le.infn.it},\ \texttt{sgiombi@princeton.edu}, \  \texttt{tseytlin@imperial.ac.uk}
	}
\vspace*{0.8cm}
\end{center}

\begin{abstract}  
\noindent
The generalized Wilson loop operator interpolating between the supersymmetric and the ordinary Wilson loop
in $\N=4$ SYM theory 
provides an interesting example of renormalization group flow on a line defect: the scalar coupling parameter
$\zeta$ has a non-trivial beta function and may be viewed as a running coupling constant in a 1d defect QFT. 
In this paper we continue 
the study of this operator, generalizing previous results for the beta function and Wilson loop expectation 
value to the case of an arbitrary representation of the gauge group and beyond the planar limit. 
Focusing on the scalar ladder limit where the generalized Wilson loop reduces 
to a purely scalar line operator in a free adjoint theory, and specializing to the case 
of the rank $k$ symmetric representation of $SU(N)$, we also  consider a certain ``semiclassical'' 
limit where $k$ is taken to infinity with the  product $k\,\zeta^2$ fixed. This limit can be conveniently 
studied using a 1d defect QFT  representation 
 in terms of  
 $N$ commuting  
  bosons. Using this representation, we compute the beta function and the circular loop expectation 
value in the large $k$ limit, and use it to derive constraints on the structure of the beta function for general 
representation. We  discuss the corresponding 1d RG flow   and   
  comment on the 
  consistency of the results with 
the 1d defect version of the F-theorem. 
\end{abstract}
\vskip 0.5cm
	{
	}
\end{titlepage}

\tableofcontents
\vspace{1cm}

\newpage
\section{Introduction and summary}

In this paper we  continue our investigation \ci{Beccaria:2017rbe,Beccaria:2018ocq,Beccaria:2019dws,Beccaria:2021rmj}   of  a family  of  operators that  interpolate   between the  supersymmetric Wilson-Maldacena  ($\z=1$) and the 
standard Wilson ($\z=0$)  loop operators \cite{Polchinski:2011im} 
\be
\la{1.1}
W^{(\z)}  (C) = \,\text{Tr}\,{\rm P}\,\exp\oint_{C}d\tau\,\Big[i\,A_{\mu}(x)\,\dot x^{\mu}
+ \z \phi_{m}(x) \, \theta^{m} \,|\dot x|\,\Big], \qquad\qquad  \theta_{m}^{2}=1\ .
\ee
Here $\phi_m$ are the 6 scalars of the  $SU(N)$  $\N=4$ SYM theory. We shall  choose 
  the  unit  vector  $\theta_m$  to be along the 6-th direction, \ie $\phi_m \theta^m =\phi_6\equiv \phi$. The study of \rf{1.1}  is of interest, in particular,  in the context of 1d defect QFT, see \eg \cite{Cooke:2017qgm,Giombi:2017cqn, Liendo:2018ukf,Beccaria:2019dws,Correa:2019rdk,Agmon:2020pde,Cuomo:2021rkm} for related work, and \cite{Giombi:2021uae,Bianchi:2021snj, Cuomo:2021kfm} for other examples of RG flows on line defects.

Let us   first summarize  some previous results. 
In the  simplest case  the trace in \rf{1.1} 
is taken in the fundamental representation; then  the expectation value of 
\rf{1.1} is a function of $\z$, $N$  and 't Hooft coupling   $\l= \gym^{2}\,N$. 
For a smooth contour  $C$,  $\vev{\WZ}$ is logarithmically divergent,  requiring a  renormalization of  the  coupling $\z$. Its renormalized value obeys the renormalization group (RG) equation
 \be 
 \la{1.2}
\vev{\WZ}  \equiv \W\big(\l; \z(\mu), \mu\big)   
 \ , \ \ \ \  \qquad  \Big( \mu {\partial \ov \partial \mu}   + \beta_\z  { \partial \ov \partial \z}\Big) \W =0  \ , \qquad \ \ 
 \beta_\z =\mu  { d \z \ov d  \mu } \ .
\ee
\iffa    The running of the coupling $\zeta$  defines 
a 1d RG flow  between the WL and WML operators.\footnote{The discrete transformation $\p_m \to -\p_m$ is a symmetry of the 
$\N=4$ SYM path integral and implies that   $\W$  is invariant under $\z \to -\z$. In the following, we will assume $\z\geq  0$.} \fi
At weak coupling in the planar limit 
the general structure of $\b_\z$ is expected to be\foot{Note that the one-loop $b_1$  and  two-loop  coefficients $b_{2}, b_{3}$ in   \rf{1.3},(\ref{1.4}),  are scheme independent  as 
 they are invariant  under  redefinitions of $\z$ that do not move the fixed points 
$
\z' = \z+\z\,(1-\z^{2})\,\big[\l\, z_{1}+\l^{2}\,(z_{2}+z_{3}\,\z^{2})+\cdots\big] \ . 
$.}
\ba
\la{1.3}
 \beta_{\zeta} = b_{1}\,\l\,\z(1-\z^{2})+\l^{2}\,\z\,(1-\z^{2})\,(b_{2}+b_{3}\,\z^{2})+\l^{3}\,\z\,(1-\z^{2})\,
(b_{4}+b_{5}\,\z^{2}+b_{6}\,\z^{4})+\mc O(\l^{4})\ .
\ea
The  one-loop term in  the  $\beta_\z$ function   was found in  \ci{Polchinski:2011im}   and the two-loop term in \cite{Beccaria:2021rmj}. Explicitly, 
\be
\la{1.4}
\beta_{\z} = -\frac{\lambda}{8\pi^{2}}\,\z\,(1-\z^{2})+
\frac{\lambda^{2}}{64\pi^{4}}\,\z(1-\z^{4})+\mc O(\lambda^{3})  \ . 
\ee
The WL ($\z=0$) and WML  ($\z=1$) cases   are  the fixed points 
to all orders in $\l$. The running of $\z$ may be considered as an RG flow in an  effective 1d  defect theory coupled  to the bulk SYM theory.  
For a circular contour, F $=\log \W$ has   an  interpretation of  (minus) 
1d  defect theory free energy on  $S^1$, and $\log \W$  obeys \cite{Beccaria:2021rmj} 
 the defect analog of the
 F-theorem  \ci{Klebanov:2011gs,Giombi:2014xxa}  F$^{(\rm UV)} > $F$^{(\rm IR)}$
 (cf. also 
\ci{Kobayashi:2018lil}).
 One may also  define  a defect entropy  function that is monotonically decreasing  along the flow  from UV to IR \ci{Cuomo:2021rkm}.
On general   grounds, consistent with this interpretation,  we should have 
 \be 
 \la{1.5}
 {\partial \ov \partial \z } \log   \W  =   \C \, \beta_\z  \ ,
 \ee
 where    $\C=\C(\l,\z)$   admits the weak  coupling expansion $\C= \frac{\l}{4}+ \OO(\l^2)$ \ci{Beccaria:2017rbe}.

  The  expectation value 
$\W=\vev{\WZ}$   on a circle  has the following structure \ci{Beccaria:2017rbe,Beccaria:2018ocq}
(consistent with  \rf{1.3}  and \rf{1.5}) 
\be
\la{1.6}
\W = \vev{W^{(1)}}\,\Big[1+w_{1}\,\l^{2}\,(1-\z^{2})^{2}+\l^{3}\,(1-\z^{2})^{2}(w_{2}+w_{3}\,\z^{2})+\cdots\Big]\ , \ee
where
 $N^{-1} \vev{W^{(1)}} = \frac{2}{\sqrt\lambda}\,I_{1}(\sqrt\lambda) = 1+\frac{\lambda}{8}+\frac{\lambda^{2}}{192}+\mc O(\lambda^{3})$
\cite{Erickson:2000af}. 
The coefficients $w_1= \frac{1}{128\pi^{2}}$ \ci{Beccaria:2017rbe}  and $w_2$ (which is 
presently unknown) are scheme-independent. The  scheme-dependent coefficient 
 $w_3 $ is  finite after the renormalization of $\z$ \ci{Beccaria:2018ocq}
 \be \la{61}
 w_{3} =- \frac{1}{256\,\pi^4}\big(\log \mu  + \tfrac{5}{6} \big)\ . \ee
 Here $\mu$ is a  renormalization scale (in general multiplied  by the radius of the circle  which is set to 1 here); the coefficient   of $\log \mu$  is  related (via \rf{1.2}) to 
  the coefficient in the  one-loop   beta-function \rf{1.4}  while the constant  $5\ov 6$  is  scheme-dependent.
 
The coefficients of the 
highest $\z$ powers  at  each $\l^n$  order in \rf{1.3}, \ie $b_{1}, b_{3}, b_{6}, \dots$, 
may be computed by restricting to diagrams  with maximal  number of scalar  propagators  attached to the Wilson line. In particular, 
these are diagrams that do not  have  internal vertices, \ie  they are of (scalar)  ladder type. Using the vertex renormalization method
of \ci{Dotsenko:1979wb}, we computed them to five-loop order 
 \cite{Beccaria:2021rmj}
 \ba
& \beta_{\z}^{\rm ladder} =
q_{1}\,\frac{\lambda}{4\pi^2}\,\z^{3}
+q_{2}\,\Big(\frac{\lambda}{4\pi^2}\Big)^{2}\,\z^{5}
+q_{3}\,\Big(\frac{\lambda}{4\pi^2}\Big)^{3}\,\z^{7}
+q_{4}\,\Big(\frac{\lambda}{4\pi^2}\Big)^{4}\,\z^{9}
+q_{5}\,\Big(\frac{\lambda}{4\pi^2}\Big)^{5}\,\z^{11}
+\cdots\ , \la{1.7}   \\
& 
\te q_{1} = \frac{1}{2},\qquad
q_{2} = -\frac{1}{4},\qquad
q_{3} = \frac{1}{4}-\frac{\zeta(2)}{8}, \qquad
q_{4} = -\frac{17}{48}+\frac{\zeta(2)}{3}-\frac{\zeta(3)}{12}, \la{1.8} \\
& \qquad  \qquad 
\te q_{5} = \frac{29}{48}-\frac{37\,\zeta(2)}{48}+\frac{29\,\zeta(3)}{96}+\frac{25\,\zeta(4)}{128}.  \no
\ea
Here  $\zeta(n)$ are the Riemann zeta-function values and  
$q_{3}$  and higher coefficients are scheme dependent.
In this ladder approximation, the  expectation  value of the operator defined 
 on a closed contour parameterized by $\tau\in (0, 2 \pi)$ reduces to 
\be 
\la{1.9}
\vev{\WZ}^{\rm ladder} = \vev{ \text{Tr}\,\text{P}\,\exp\int_0^{2\pi}
d\tau'\, \z\,  \phi(\tau') }= \W(\xi) \ , \qquad \qquad \xi\equiv \l\, \z^2 \ ,
\ee
where we set  $\phi(\tau) \equiv \phi(x(\tau))$  
and  $\vev{...}$  is computed in the free adjoint scalar theory
 \be \langle ... \rangle = \int d \phi \  e^{-S} ...\ , \qquad \qquad 
 \la{1.11}
 S = \frac{1}{\gym^2}\,\int d^4 x\  \text{Tr} (\partial_\alpha \phi \partial^\alpha \phi)  \ .
 \ee
 Redefining the scalar $\phi\to \z^{-1}\phi$  we get the one-coupling theory with $\l= g^2 N $ in $S$ 
 replaced by $\xi$ defined in \rf{1.9}.\foot{Note also that after factoring out one power of $\z$, the expansion in  (\ref{1.8}) 
may be written  in terms of the effective coupling $\xi$.} 
 In the circular or straight line  cases  the associated 1d  propagator $D(\tau-\tau') = \langle  \phi(\tau)\,\phi(\tau')\rangle$  
 has then the  following form\footnote{\la{f2} 
 We recall that the (bulk) $\mc N=4$ SYM action is schematically of the form 
 $S= \frac{1}{\gym^{2}}\,\int d^4 x\  \text{Tr} (F^{2}+D \phi D \phi+\phi^{4}+\dots)$,
and  $\l=\gym^{2}\,N$. Here   we  also took  into account a factor $\ha $ from the relation  $T^{a}T^{a} = \frac{1}{2} N \bm{1}$, 
 for the generators $T^a$ of $SU(N)$ in the fundamental representation.}
\be
\la{1.12}
\text{circle:}\quad D(\tau) =  \frac{\xi}{8\pi^{2}}\frac{1}{4\,\sin^{2}\frac{\tau}{2}}, \qquad\qquad \qquad 
\text{line:}\quad D(\tau) =  \frac{\xi}{8\pi^{2}}\frac{1}{\tau^{2}}.
\ee
Let us note that the study of the partition function $\W$  of  the scalar loop model \rf{1.9}, (\ref{1.11})
is an interesting problem  on its own right, as  this is  an example of  a  particularly simple
 defect QFT. Note that for the case of $SU(2)$, the scalar defect (\ref{1.9}) may also be thought as describing an impurity in the (free) $O(3)$ vector model  
 (see e.g. \cite{Vojta:1999,Liu:2021nck} and references therein, and also \cite{Cuomo:2021kfm} for a related discussion).

The  motivation behind  the present  paper is to try to generalize the expression for 
 the beta function  \rf{1.4} and the Wilson loop expectation value \rf{1.6}  to the case
when the trace in (\ref{1.1}) is taken in a   generic  representation R of $SU(N)$ and   beyond the planar limit.
Let us  consider a generic simple group $G$  with coupling 
$\gym$.  
 Then  for  the circular supersymmetric  WML ($\z=1$) in a general representation R  of a group $G$ 
one finds   \cite{Fiol:2018yuc,Fiol:2013hna} (see also \rf{B.16})  
\be
\la{1.13}
\norm\vev{W^{(1)}} = 1+C_{\R}\,\frac{\gym^2}{4}+\bigg(C_{\R}^{2}-\frac{1}{6}C_{\R}C_{\A}\bigg)\,\frac{\gym^{4}}{32}+\bigg(
C_{\R}^{3}-\frac{1}{2}C_{\R}^{2}C_{\A}+\frac{1}{12}C_{\R}C_{\A}^{2}\bigg)\frac{\gym^{6}}{384}+\cdots.
\ee
Here  $C_\A$   and $C_\R$ are  the quadratic Casimirs  for the adjoint  and R representations 
   ($C_\A= N$ for $G=SU(N)$  and $T^{a}T^{a} = C_{\R}\,\dim \R$,  see Appendix \ref{app:sun} for our conventions). 
   For  any $\z$  we then expect to find for the corresponding generalization of the two-loop part of   \rf{1.6}
\be
\la{1.14}
\norm \vev{W^{(\z)}} = 1+C_{\R}\,\frac{\gym^2}{4}+\bigg[C_{\R}^{2}-\frac{1}{6}C_{\R}C_{\A}+(1-\z^{2})\big(k_{1}+k_{2}\z^{2}\big)\bigg]\,\frac{\gym^4}{32}+\cdots.
\ee
The coefficients $k_{1}$ and $k_{2}$ may  be determined by the methods of \cite{Beccaria:2017rbe} and  we will find that 
\be
\la{1.15}
\norm \vev{W^{(\z)}} = 1+C_{\R}\frac{\gym^2}{4}+\bigg[C_{\R}^{2}
-\frac{1}{6} C_{\R}C_{\A}\,\Big(1- \frac{3}{\pi^{2} }(1-\z^{2})^{2 }\Big)
\bigg]\frac{\gym^4}{32}+\cdots
\ee
Similarly, the beta function for general representation  generalizing the  one-loop 
term in \rf{1.4}  is found to be
\be 
\beta_\z =- C_{\A} \,\z\,(1-\z^{2})\frac{\gym^2}{8\pi^{2}}+\cdots
\la{be-1}
\ee 
Note that (\ref{1.15}) and (\ref{be-1}) are related as expected according to (\ref{1.5}), with ${\cal C}=\ha C_{\R}g^2 +...$. 
In the  formal Abelian limit $C_{\A}=0$ we recover the expected exponentiation of the one-loop term in \rf{1.15}, and the vanishing of the beta function.  

Let us note also that the  coefficients of the higher  powers of $C_{\R}$ in \rf{1.13},\rf{1.15} are  related to the leading one: 
for the general case of  WL in  YM  theory with matter one expects that powers  of $C_{\R}$  exponentiate, 
i.e.  the  non-trivial part of $\log \vev{W}=\sum_{k=1}^\infty  g^{2k} \gamma_k $  
should  start with a  term linear   in $C_{\R}$ ($\gamma_k$ are ``maximally non-abelian''   colour factors; this is a manifestation 
of  the ``non-abelian exponentiation''  \ci{Gatheral:1983cz,Frenkel:1984pz}; see also \ci{Korchemskaya:1992je,Drummond:2007au} in the case of  light-like WL).
 Here $\gamma_1 \sim C_\R$, 
$\gamma_{2,3}$ depend also on $C_\A$, while starting at 4 loops $\gamma_k$  contain  higher Casimir invariants  like $Q_\R$ in 
\rf{1200} \ci{Henn:2016wlm,Henn:2019swt}. This  then suggests (in view of \rf{1.2}  and \rf{1.5} 
 with ${\C}\sim C_\R\, g^2+...$)   that the  one- and two-loop terms  in the corresponding $\beta_\z$ should 
 depend only on $C_\A$  while  the three-loop  term  should  have  dependence on $Q_\R$.\foot{We  are grateful to  G. Korchemsky for this observation and related explanations.}
 We will confirm this below   in the ladder   approximation (see \rf{1.20},
\rf{1.23}).

The ``ladder'' part of \rf{1.15}  (given by highest power of $\z$ at each order in $g$) 
 may be written as 
\be
\la{1.16}
\norm \vev{W^{(\z)}}^{\ladd} = 1+
 C_{\R}C_{\A}\, \frac{\z^4\gym^4}{64\pi^2}+\cdots\  . 
\ee
For the fundamental representation  R=F of $SU(N)$   using that  $C_{\rm F}=\frac{N^{2}-1}{2N}$
and $C_{\A}=N$  we observe  that  (\ref{1.15}) reduces to 
\ba\la{1.17}
\R=\F:\ \ \qquad  
\frac{1}{N}\,\vev{W^{(\z)}} = &1+ \Big[1-\frac{1}{N^{2}}+\mc O\Big(\frac{1}{N^{4}}\Big)\Big]\frac{\l}{8}\lp
+\Big[\frac{1}{192}-\frac{5}{384N^{2}}+\frac{(\z^{2}-1)^{2}}{128\pi^{2}}\,\Big(1-\frac{1}{N^{2}}\Big)
+\mc O\Big(\frac{1}{N^{4}}\Big)\Big]\,\l^{2}+\mc O(\l^{3}).
\ea
This generalizes the  previous planar two-loop  result \rf{1.6} to subleading terms in $1/N$.

We may  parametrize the three-loop term in \rf{1.15} as 
\ba\la{1.18}
\vev{W^{(\z)}} &= \vev{W^{(1)}}\bigg[1+C_{\R}C_{\A} (1 - \z^2)^2\frac{\gym^4}{64\pi^{2}}
+ (1 - \z^2)^2 (\ww_{2} + \ww_{3} \z^2)\, \gym^6+\cdots\bigg], 
\ea
where $\vev{W^{(1)}}$ is given by   (\ref{1.13})  and $\ww_2,\ww_3$   are the analogs 
  of $w_2,w_3$ in 
 \rf{1.6}. In particular, we expect that 
\be
\la{1.19}
\ww_{3} = - \frac{1}{128\, \pi^4} \, C_{\R}C_{\A}^2 (\log \mu +c_3)\ , 
\ee
where  in the $SU(N)$ fundamental representation case and at large $N$  (when $  C_{\R}C_{\A}^2 \to \ha N^3$)  we should  find that  $\ww_3 \to N^3  w_3$ 
 (so that $c_3=\frac{5}{6}$  in the same scheme as \rf{61}). 

For a generic representation $\R$, the  structure of the ladder-limit part of 
three-loop beta function is expected to be the following generalization of (\ref{1.8})\footnote{This follows from inspection of the possible color structures.
We also  impose the condition that the beta-function  has to vanish in the Abelian limit $C_{\A}=0$.}
\ba
\la{1.20}
\beta^{\ladd}_{\z} &= q_{1}'\,C_{\A}\z^{3}\,\frac{\gym^2}{4\pi^{2}}+(q_{2}'\,C_{\A}^{2}+q_{2}''\,C_{\A}C_{\R})\,\z^{5}\,\Big(\frac{\gym^2}{4\pi^{2}}\Big)^{2}\lp\ \ \ 
+\bigg(q_{3}'\,C_{\A}^{3}+q_{3}''\,C_{\A}^{2}C_{\R}+q_{3}'''\,C_{\A}C_{\R}^{2}
+q_{3}'''' \,\QR 
\bigg)\,\z^{7}\,\Big(\frac{\gym^2}{4\pi^{2}}\Big)^{3}+\mc O(g^8),\\
&\qquad \qquad \QR \equiv  \frac{d_{\A}^{abcd}d_{\R}^{abcd}}{C_{\R}\,\dim \R} \ . \la{1200}
\ea
Here in $\QR $  the tensor $d_{\R}^{abcd}$ is the 4-index symmetrized trace $\text{STr}(T^{a}T^{b}T^{c}T^{d})$ (see Appendices \ref{app:sun} and  \ref{app:chern}). 
The $q_n$-coefficients are numerical constants independent of representation. 
We will show that 
\ba
\la{1.21}
\beta^{\ladd}_{\z} &= \frac{1}{2}\,C_{\A}\z^{3}\,\frac{\gym^2}{4\pi^{2}}-\frac{1}{4}\,C_{\A}^{2}\,\z^{5}\,\Big(\frac{\gym^2}{4\pi^{2}}\Big)^{2} 
+\bigg[q_{3}'\,C_{\A}^{3}-3\,\zeta(2)\, \QR 
\bigg]\,\z^{7}\,\Big(\frac{\gym^2}{4\pi^{2}}\Big)^{3}+\mc O(g^8),
\ea
where $q_{3}'$ is a scheme dependent constant  (equal to $\frac{1}{4}$ in the same  regularization scheme  that led to  (\ref{1.8})). Note that \rf{1.18},\rf{1.19}   and \rf{1.21} 
  are consistent with each other via the RG equation \rf{1.2}.

To  justify  (\ref{1.21}) and extract further information about the representation dependence, we will consider the case of R  being  the {\it  $k$-symmetric} representation
$\S_{k}$ of   $SU(N)$.
Using  perturbation theory in large  $k$ at  fixed $ k\,\z^{2}\,\gym^{2}$  and fixed $N$ 
and comparing with (\ref{1.20}) we  will  confirm  (\ref{1.21}).

Our starting point will  be the following 1d path integral 
representation  for  the  Wilson loop  in the $k$-symmetric representation of $SU(N)$ (see,  e.g., \cite{Gomis:2006sb, Gomis:2006im}) 
\foot{This is an example of  representing the trace in some representation  in terms of an integral  over group orbit \ci{Polyakov:2005ss}, cf. also \ci{Hoyos:2018jky} and 
\ci{Affleck:2021jls}   for a  more general  discussion.}
\ba
\la{1.22}  &\W_k = \langle W_k \rangle , \qquad \qquad
W_{k} = \int D\chi D\bar\chi\ \delta(\bar\chi\chi- R^2)\ e^{-S} \  , \qquad 
\qquad   R^2 \equiv k+\tfrac{N}{2},
 \\
 &\qquad  S  = \int ^{2\pi}_0 d \tau\Big[ \,\bar\chi\,\del_\tau\chi+\,\z\,\phi^{a}(\tau)\, 
 \bar\chi\,T^{a}\,\chi\Big] \ , \la{1.23}
\ea
where we specialized to the purely scalar operator (\ref{1.9}), and the averaging $\langle ...\rangle$  is done over the scalar $\p$ as in \rf{1.11}. Here 
 $\phi(\tau) = \phi(x(\tau))$, $\tau\in[0,2\pi ]$  and 
$\chi, \bar\chi$ are periodic bosons transforming in the fundamental representation of $SU(N)$
 ($T^a$  are   generators  in the fundamental representation). 
After the  integration over  the free adjoint scalar field $\p$  we  obtain an effective 
 non-local 1d theory with the action  of the following structure 
\be
\la{1.24}
S= \int d\tau\, \bar\chi\,\del_\tau\chi -   \z^{2}\, g^2\,  \int d\tau d\tau'\, D(\tau- \tau')\, \bar\chi(\tau)T^{a}\chi(\tau) \ \bar\chi(\tau')T^{a}\chi(\tau'),
\ee
where $D(\tau- \tau') = \vev{\phi(\tau)\phi(\tau')} $ (on the line $D \sim   {1\ov (\tau -\tau')^2}$, cf. \rf{1.11},\rf{1.12}).

The rank $k$ of the symmetric representation 
 enters only through $R^2$ in  the delta-function constraint in (\ref{1.22}). 
 Rescaling $\chi$ by $R$  so that now    $\bar \chi \chi =1$  we get  (e.g. on the straight  line) 
 \ba
\la{1244}
S=  R^2 \Big[  \int d\tau\, \bar\chi\,\del_\tau\chi & -   \pxi  \int {d\tau\,  d\tau'\, \ov (\tau- \tau')^2}\, \bar\chi(\tau)T^{a}\chi(\tau) \ \bar\chi(\tau')T^{a}\chi(\tau')\Big]\ , 
\\ 
\la{1.25} &
\pxi \equiv  \frac{\z^{2}\, \gym^{2}\,R^{2}}{8\pi^{2}}  \ . 
\ea
We may then  develop a systematic ``semiclassical'' 
 large $R^2$  or large $k$ perturbation theory  at  fixed $\pxi$  and $N$   for 
 $\W_k$  and the   beta function $\beta_{\pxi}$  for the  coupling  $\pxi$.
 Note that  since in the ladder   approximation 
 the bulk theory is free, the coupling $g$  can take   any value
 (and can actually be absorbed into $\z$   defining $\bar \xi = \z^2 g^2$, cf. \rf{1.9})
 so the large $k$  limit at  fixed $\pxi$  means also   small $\z$  limit. 

Explicitly, we will find that for the $k$-symmetric representation 
\be
\la{1.26}
\bxi = \mu {d\pxi\ov d \mu} = \frac{2N}{R^2}\,\frac{\pxi^{2}}{1+ \pi^2 \pxi^2} -\frac{2N^{2}}{R^{4}}\,\frac{\pxi^{3}\, (1-\ga\, \pi^{2}\pxi^{2})}{(1+\pi^{2}\pxi^{2})^{3}}+\mc O\Big(\frac{1}{R^{6}}\Big), 
\ee
where the  coefficient $\ga$ is scheme  dependent with $\ga=1$ in a particular momentum cutoff scheme (see also discussion below \rf{534}). 

Since $\gym$  and $R^2$ are not running, $\bxi$  is directly related to the  ladder beta function for $\z$ in \rf{1.21}. 
In general, the   large $k$ expansion of $\beta_\pxi$ 
gives  an all order prediction for the small $\z$  expansion of $\beta_{\z}^{\rm ladder}$: 
 it fixes  the coefficient of  the highest power 
of  $k$ at each order in $\z$. 
In particular, expanding the ``one-loop'' term in \rf{1.26} in powers of $\z$ yields 
\begin{equation}
\beta_{\z}^{\rm ladder}=\frac{N g^2}{8\pi^2}\z^3-\frac{N g^6}{512\pi^2}k^2\z^7+\ldots 
\end{equation}
Noting that for the $k$-symmetric representation $Q_\R=k^2\frac{N}{4}+O(k)$, this  allows to  fix the coefficient
of the $Q_\R$ part of the three-loop term in \rf{1.21}. 


Note also that 
in the case   when $k$ is fixed and $N$ is large 
the leading $\pxi^2$ and $\pxi^3$  terms in the small  $\pxi $ expansion in \rf{1.26} are in  agreement with the one-loop and two-loop terms in $\beta_{\z}^{\rm ladder}$ in \rf{1.7}.\foot{
 Let us also note that  the representation \rf{1.22}   applies for any finite $k$, in particular  also to the   $k=1$   case of the fundamental representation. Then naively 
          the  large $R^2$  perturbation  
         could  still be applied   by taking $N$ large   at fixed $\pxi$ in \rf{1.25}
         that  then   becomes        $\pxi \to {1\ov 16 \pi^2} \xi$,      
where $\xi$ was defined in \rf{1.9}. However,  since $N$ here  is as large as $R^2$ 
the $1/R^2$ expansion of the beta-function  in \rf{1.26}  no longer makes sense,\
  i.e. needs to be resummed. One can still unambiguously  extract 
  the lowest order  terms  in the  small $\xi$   expansion  and  match  them with the 
  $\z^3$ and $\z^5$ terms in  $\beta_{\z}^{\rm ladder}$.}

For the  renormalized  value of the  scalar  ladder  Wilson loop expectation value 
on a circle (of unit radius) in \rf{1.22}  defined in the $k$-symmetric representation 
we will find that\foot{In the case of $SU(2)$ group  the 
prefactor $\big(1+\pi^{2}\pxi^{2}\big)^{1/ 2}$  was found earlier in \ci{Komar}.}
\ba \la{444}
 & \W_k  = \dim \S_{k} \,  \big(1+\pi^{2}\pxi^{2}\big)^{N-1\ov 2}\, \Big[ 1 +\frac{\vv_1}{R^{2}}\,\frac{N(N-1)\pxi^{3}}{(1+\pi^{2}\pxi^{2})^{2}}   +  \mc O\big({1\ov R^4}\big)   \Big], \\  &
\qquad \qquad \vv_1= - 2\pi^{2} (\log \mu  + c_3) \ , \la{044}
\ea
where $ \dim \S_{k}= {(N+k-1)!\ov (N-1)!\,  k!}$  is the dimension of the $k$-symmetric representation of $SU(N)$  and $c_3$ is a scheme-dependent  constant as in  \rf{1.19}. Note that the expression 
(\ref{444}) effectively resums an infinite set of terms in the ordinary perturbative expansion in powers of $\zeta$.\footnote{A similar large $k$ limit with $k\gg N$ for the case of the Wilson-Maldacena loop 
was studied in \cite{Correa:2015wma, Correa:2015kfa}, where it was observed that an exponentiation of the one-loop result occurs in this limit.} Expanding (\ref{444}) in powers of $\varkappa$, one finds
\begin{equation}
\frac{1}{\dim \S_{k} }\W_k = 1+  \frac{\pi^2}{2}(N-1)\,  \pxi^2+ \frac{\vv_1}{R^{2}}\,N(N-1)\, \pxi^3+\ldots\ .
\end{equation}
Noting that $C_{\S_k}\sim k^2(N-1)/2N$ at large $k$, one can see that the term quadratic in $\pxi$ matches (\ref{1.16}), 
while the cubic term  
matches  the $\ww_{3}  \z^6 g^6$  term in \rf{1.18},\rf{1.19}.\foot{Indeed,
from \rf{437}  
\ \ 
$\frac{1}{128\pi^4} C_{\R}C_{\A}^2  g^6 \z^6 =  \frac{1}{256 \pi^4}\, N (N-1) k (k+N)  
g^6 \z^6 $
while from \rf{1.25}   we have 
 $\frac{2\pi^2 }{R^{2}}\,N(N-1)\pxi^3= 
  \frac{1}{256 \pi^4}\, N(N-1) (k +  \ha N)^2   g^6 \z^6   $
  so we get agreement at large $k$.} 
  
  The expression \rf{444} satisfies  the  RG equation   as in \rf{1.2}  and also the  analog of the relation \rf{1.5}    with $\bxi$ given by the one-loop term in \rf{1.26}
  \be \la{777}
  \bigg(\mu\frac{\partial}{\partial\mu}+\bxi\frac{\partial}{\partial\pxi}\bigg)\,\W_k=0 \ ,\ \qquad \ \ \ \ 
  \frac{\partial}{\partial\pxi}\,\log \W_k = \bar {\cal C}\,  \bxi \ ,
   \ \ \qquad    \bar {\cal C }= \tfrac{(N-1) \pi^2 R^2 }{ 2 N \pxi}  > 0 \ . \ee
   Let us now  discuss  properties of the  RG flow implied by the $\beta_\pxi$ function in 
      \rf{1.26}.  At the  leading $1/k$ order we find (using that $\pxi \geq 0$)
\ba \la{133} &
{d\pxi\ov d t } =  \frac{2N}{k}
\frac{\pxi^{2}}{1+ \pi^2 \pxi^2}  
 \ ,  \ \  \ \ \ \ \ \ \ \ \  \ t\equiv \log \mu\ , \\
&   \pxi (t) = \gamma  t + \frac{1}{\pi}\sqrt{ 1 + \pi^{2}\gamma^2 t^2}\ , 
 \ \   \ \ \ \ \ \ \   \gamma \equiv  \frac{N}{\pi^2 k}  \ , \la{134} \ea
  so that the IR  ($\mu\to 0$)  and  UV ($\mu\to \infty$)  asymtotics are 
\ba \la{135} 
\ \ {\IR}:  \ \  \pxi(t \to -\infty) =  {1\ov 2 \pi^{2}\gamma\,  |t|} \to    0 , \qquad \qquad 
 \ \ {\UV}: \   \pxi(t\to +\infty)=   2 \gamma\,  t  \to \infty\ .   \ea 
This   asymptotic behaviour is, in fact, exact, i.e. 
 not changed by higher $1/k$ corrections   in $\beta_\pxi$  since     
   the  exact $\beta_\pxi$   satisfies\foot{As 
    is clear from \rf{1.26}, the  $1/R^4$ term vanishes  at large $\pxi$. The same   should be true 
    also at higher orders  as for large $\pxi$  the propagator   goes as $\pxi^{-1}$ 
    while  vertices in the action \rf{1244} are proportional to $\pxi$.} 
\ba 
\beta_\pxi \Big|_{\pxi\to 0} \to  0 \ , \ \ \ \ \ \ \ \ \ \ \beta_\pxi \Big|_{\pxi\to \infty} \to  \frac{2N}{k+ \ha N }={\rm const}  \ .\la{136}\ea 
The corresponding asymptotic behaviour of  the  WL expectation value in  \rf{444} 
\ba\la{138}
   \IR:  \ \  \W_k \Big|_{\pxi \to 0} \to &  \dim \S_{k}  ,  \qquad \qquad 
 \ \ \UV: \  \  \W_k  \Big|_{\pxi \to \infty } \to   \dim \S_{k}  \  \pxi^{N-1} \ ,  \\
& \log  \W_k^{(\UV) } >  \log  \W_k^{(\IR)} \ . \la{139} \ea
 This is 
  consistent   with 1d version of  F-theorem  for $  \W_k $    as   partition function on $S^1$.
  Furthermore, one may consider  the  line defect   entropy  defined in \ci{Cuomo:2021rkm}
  (here $\aa$  is the radius of $S^1$) 
  \be \la{1396}
{\rm s} \equiv  \Big(1 -  \aa {\del \ov \del \aa}\Big) \log  \W_k = \Big (1 -  \mu  {\del \ov \del \mu}\Big) \log  \W_k\ ,  
\ee
which is equal to $\log  \W_k$ at fixed points. 
Using \rf{777}   and $\bar { \cal C} >0 $ 
(which is true at least in perturbation theory) 
we get
\be\la{88}  
   {\rm s} = \log  \W_k  +  \bar {\cal C}\,   \bxi^2    \ \  \geq  \ \
\log  \W_k \ .    \ee
To leading order in the $1/k$ perturbation theory, ${\rm s}=\log \W_k\approx \log \dim \S_k+\ha (N-1) \log (1 + \pi^2 \pxi^2)$, and so both functions monotonically decrease along the RG trajectory. According to the arguments in \ci{Cuomo:2021rkm}, the defect entropy ${\rm s}$ should 
be monotonically decreasing also non-perturbatively. 

Let us mention also  that  if one considers  the  defect line in a  bulk scalar theory  in 
$d=4-\ve$ dimensions  then  the 
coupling   $g^2$  and thus $\pxi\sim g^2 \z^2 k $  will get dimension  $\ve \to 0 $. Then 
the $\bxi$ function  gets an extra term $ -\ve\, \pxi$, and,  in addition to the trivial 
UV fixed point $\pxi=0$,  there are two Wilson-Fisher-type UV and IR  fixed points
\ba
 \la{544} 
&\bxi = -\ve\, \pxi + \frac{2N}{k} \frac{\pxi^2}{1+\pi^2 \pxi^2}+\mc O\big({1\ov k^{2}}\big) \ , \\
\la{5455}
&\bxi=0: \ \ \ \ \  \pxi_{\pm } = \frac{N}{\pi^{2}k\, \ve} \Big( 1 \pm \sqrt{1- \frac{\pi^2 k^2 \, \ve^2}{N^2}} \Big)  + \mc O \big( {1\ov k^2}\big)\,.
\ea 
In order for these fixed points to be real, one should take the small $\ve$ and large $k$ limits in such a way that the condition $\ve k \le \frac{N}{\pi}$ is satisfied (for $\ve k =\frac{N}{\pi}$ the two fixed points 
coincide, and for $\ve k > \frac{N}{\pi}$ they become complex). Taking 
the $\ve \to 0$ limit first, the fixed points reduce to
\ba
&  {\rm  UV}:   \ \ \  \pxi_+ = \frac{2N}{\pi^{2}k}\frac{1}{\ve}+\mc O(\ve^{0}) \to \infty \ ,\qquad \qquad 
  {\rm IR}:   \ \ \ \ \  \pxi_- = \frac{k}{2N}\ve+\mc O(\ve^{2})\to 0 \  .\la{555}
\ea
Like  the asymptotics in \rf{135}
these fixed points are expected to be  stable under higher order $1/R^2$ or $1/k$  corrections to $\bxi$.

\

The structure of the rest of the paper is as follows.
 In section \ref{sec:vertex} we compute the two-loop  $\beta_{\z}$
function in ladder approximation (for any $N$)  by applying the vertex renormalization method  described in \cite{Beccaria:2021rmj}. We also 
discuss the  structure of    $\beta_{\z}$  at three-loop level. 
In section \ref{app:twoW} we 
derive the two-loop expression (\ref{1.15}) for the expectation value $\vev{W^{(\z)}}$ in  any  representation, thus generalizing our previous result
in the  fundamental representation \cite{Beccaria:2017rbe}.

 In section \ref{sec:1d} we introduce  the bosonic 1d path integral expression 
  \rf{1.22},\rf{1.23} for the ladder Wilson loop
in the $k$-symmetric $SU(N)$ representation  and 
 discuss some of its general features. 
 It  is different from  the more standard fermionic 1d path integral (reviewed in Appendix \ref{app:GN}) 
 and  convenient for the  study of  the large $k$ limit  considered  in 
section \ref{sec:pt}.
There we  first discuss the free $\pxi=0$ case  (clarifying   the role of the constant 
 zero modes of $\chi$)  and  then 
 compute the Wilson loop at leading order in  large $k\sim R^2$  for $\pxi\neq 0$.
Finally, we  present  the calculation of  the  $1/R^{2}$ corrections
and, in particular, the logarithmically divergent contributions
 that determine the leading term in the $\beta_{\pxi}$ function. 

In section \ref{sec:beta-line} we show that $\beta_{\pxi}$ may be computed starting from a  
two-point correlator of the adjoint scalars inserted on  the 
 Wilson line. 
We  first reproduce the $1/R^{2}$ term in  $\beta_{\pxi}$  found in 
section \ref{sec:pt} and then    study in detail  the 
order $1/R^{4}$  correction. 

In Appendix \ref{app:sun} we recall our group theoretic conventions. 
Appendix \ref{app:GN} reviews the 1d  fermionic path integral 
representation \cite{Gervais:1979fv} for a Wilson loop in any representation.
Appendix \ref{app:Y3} presents 
 details of the  calculation of the $1/R^{4}$ contribution to the $\beta_{\pxi}$ function in section \ref{sec:beta-line}.  
Appendix \ref{app:chern} is devoted to a general proof
of the universality, in planar limit, of the coefficient of the  three-loop $\z^{7}$ term in $\beta_{\z}^{\ladd}$ in (\ref{1.21}).
 In
Appendix \ref{app:2pointline} we compute the two-loop $\beta_{\z}$ for  generic representation 
using   a  two-point scalar correlator on the line.
In Appendix \ref{app:wound}
 we apply  similar   method  as in   \ref{app:twoW}  
to the closely related case of a multiply wound Wilson loop in the fundamental representation, 
finding  the two-loop term in  the weak gauge coupling  expansion for  generic $\z$.

\

{\bf Note added:}  While  completing this paper,  we learned  about  the partially overlapping work  \ci{Cuomo:2022xgw}, which in particular 
studies the scalar line defect and its large $k$ limit in the $SU(2)$ case (extending some results  announced  in \ci{Komar}). 
We thank the  authors  for sharing  their draft prior to submission.

\section{ $\beta_\z$ function
 in ladder approximation from  vertex renormalization} 
 \la{sec:vertex}

As discussed in detail in \ci{Beccaria:2021rmj}
the beta function for the $\z$ coupling  in  \rf{1.9}
can be obtained from the study of the one point function on a  long  interval  $(-L,L)$\foot{The 
renormalization of $\z$  is universal for any contour   and thus can be determined 
 by considering the simplest straight-line Wilson loop.}   \be
\la{2.1}
\frac{\vev{\tr\Big(\phi(\tau_{0})\,\text{P}\,e^{\int_{-L}^{L}d\tau' \, \z\, \phi(\tau')}\Big) }}{\vev{\tr\Big(\text{P}\,e^{\int_{-L}^{L}d\tau' \, \z\, \phi(\tau')}\Big)}}, 
\ee
where  the 4d   scalar  $\p$  restricted to the line has a free propagator 
$D(\tau) =\langle  \p(\tau) \p(0) \rangle=  \frac{\gym^{2}}{4\pi^{2}}\frac{1}{\tau^{2}}$.
Here  we shall assume $\tr$ to be in generic representation R of a gauge group. 
The averaging is done with respect to the free adjoint scalar field as in \rf{1.11}.

If we denote by $\tau$ the point on the loop  connected by propagator to $ \tau_{0}$, then
$\beta_\z$ function follows from the renormalization of the vertex $V$ in 
\be\la{2.2}
D(\tau_{0}-\tau)\, V(\tau, \z) = 
\begin{tikzpicture}[line width=1 pt, scale=0.5, baseline=0]
\draw (-2,0)--(0,0);
\node[left] at (-2,0) {$\tau_{0}$};
\draw[line  width=2 pt] (0,0) arc(0:45:2);
\draw[line  width=2 pt] (0,0) arc(0:-45:2);
\draw[fill = lightgray] (0,0) circle(0.5);
\node[right] at (0.6,0) {$\tau$};
\end{tikzpicture}, \qquad \qquad 
V(\tau, \zeta) = \zeta + \text{corrections}.
\ee
The point $\tau_0$ is at some far part of the Wilson line. 
We may also choose  point $\tau$ to be at the origin, $\tau=0$.

\subsection{One-loop order}

 In dimensional regularization the propagator is 
\be\la{2.3} 
D(\tau) =\frac{\gym^{2}}{4\pi^{2}}\frac{1}{|\tau|^{2-\ve}}, \qquad \qquad d=4-\ve.
\ee
The one-loop planar diagrams in the numerator of (\ref{2.1}) are
\ba\la{2.4}
\begin{tikzpicture}[line width=1 pt, scale=0.5, baseline=0]
\node[above] at (0,2) {$\tau_{0}$}; 
\node[below] at (0,0) {$0$};
\node[right] at (3,0) {$L$};
\node[left] at (-3,0) {$-L$};
\draw (0,2)--(0,0);
\draw[line width=2 pt] (-3,0)--(3,0);
\draw (0.5,0) arc(180:0:1);
\end{tikzpicture}  &=
\begin{tikzpicture}[line width=1 pt, scale=0.5, baseline=0]
\node[above] at (0,2) {$\tau_{0}$}; 
\node[below] at (0,0) {$0$};
\node[right] at (3,0) {$L$};
\node[left] at (-3,0) {$-L$};
\draw (0,2)--(0,0);
\draw[line width=2 pt] (-3,0)--(3,0);
\draw (-2.5,0) arc(180:0:1);
\end{tikzpicture} =  \z^{3}\, \frac{\gym^2}{4\pi^{2}}C_{\R}\,\frac{L^{\ve}}{\ve\,(\ve-1)},
\ea
where  we used  the group  generators 
satisfy $T^{a}T^{a}=C_{\R}\mathbf{1}$. 
We have also a single non-planar diagram
\ba\la{2.5}
\begin{tikzpicture}[line width=1 pt, scale=0.5, baseline=0]
\node[above] at (0,2) {$\tau_{0}$}; 
\node[below] at (0,0) {$0$};
\node[right] at (3,0) {$L$};
\node[left] at (-3,0) {$-L$};
\draw (0,2)--(0,0);
\draw[line width=2 pt] (-3,0)--(3,0);
\draw (-1,0) arc(180:0:1);
\end{tikzpicture}  = \z^{3}\frac{\gym^2}{4\pi^{2}}(C_{\R}-\frac{1}{2}C_{\A})\,\frac{L^{\ve}(2^{\ve}-2)}{\ve(\ve-1)},
\ea
where   we used  (\ref{A.9})  ($C_{\A}$   corresponds to the adjoint representation, 
i.e. is equal to $N$ for $SU(N)$).  Finally, the denominator of (\ref{2.1})  contributes
\ba\la{2.6}
\begin{tikzpicture}[line width=1 pt, scale=0.5, baseline=0]
\node[right] at (2,0) {$L$};
\node[left] at (-2,0) {$-L$};
\draw[line width=2 pt] (-2,0)--(2,0);
\draw (-1,0) arc(180:0:1);
\end{tikzpicture}  = \z^{2}\,\frac{\gym^2}{\pi^{2}}C_{\R}\,\frac{2^{\ve-2}L^{\ve}}{\ve(\ve-1)}\ .
\ea
The total vertex is then 
\be\la{2.7}
V(\z, L) = \z+C_{\A}\,\z^{3}\frac{\gym^2}{8\pi^{2}}\frac{L^{\ve}(2-2^{\ve})}{\ve(\ve-1)}+\mc O(\l^{2}),
\ee
where the  dependence on $C_{\R}$  canceled out. 
$V$  is  renormalized by  $\z_{\rm bare}=\z  \to \z_{\rm ren} = \z(\mu)$ 
\be\la{2.8}
\z  =  \mu^{\ve/2}\bigg[\z(\mu)+\frac{C_{\A} \gym^2  }{8 \pi^2\ve}\z^3 (\mu)+\mc O(\gym^4)\bigg],
\ee
The renormalized vertex is then 
\be\la{2.9}
V_{\rm ren}(\z(\mu), L) = \z(\mu)+  \frac{C_{\A} \gym^2  }{8 \pi^2}\,\z^3 (\mu)\bigg(-1-\log\frac{L\mu}{2}\bigg)+\mc O(\gym^4),
\ee
and obeys the RG equation 
\be
\la{2.10}
\bigg(\mu\drg{\mu}+\beta^{\ladd}_{\zeta}\drg{\z}\bigg)\, V^{\rm ren}\big(\z(\mu), L\big) = 0,
\ee
with 
\be
\la{2.11}
\beta^{\ladd}_{\z} = C_{\A}\,\z^{3}\,\frac{\gym^2}{8\pi^{2}}+\mc O(\gym^4).
\ee
This  shows  that the  one-loop  beta-function in \rf{1.6} 
is universal, i.e.   does not   depend on a particular representation of the gauge group
used to define the WL.\foot{We   have shown that this applies to the ladder part of 
$\beta_\z$ but   since  it should have $\z^2=1$ as its zero this should be  true also for the full 
one-loop expression.}

\subsection{Two-loop order }

The two-loop diagrams contributing $V(\z, L)$ are much more complicated and we found it convenient to use  the propagator with an  explicit cutoff 
as in \cite{Beccaria:2021rmj} 
\be\la{2.12}
D(\tau) = \frac{\gym^{2}}{4\pi^{2}}\frac{1}{(|\tau|+\eps)^{2}},\qquad \ \ \   \eps\to 0 \ . 
\ee
We shall focus on logarithmic UV divergences $\log^n\ve$. 
To compare with dimensional regularization, let us first  repeat the 
 above one-loop calculation. We find the following analogs of \rf{2.4}, \rf{2.5}  and \rf{2.6}
\ba\la{2.13}
\begin{tikzpicture}[line width=1 pt, scale=0.5, baseline=0]
\node[above] at (0,2) {$\tau_{0}$}; 
\node[below] at (0,0) {$0$};
\node[right] at (3,0) {$L$};
\node[left] at (-3,0) {$-L$};
\draw (0,2)--(0,0);
\draw[line width=2 pt] (-3,0)--(3,0);
\draw (0.5,0) arc(180:0:1);
\end{tikzpicture}  &= 
\begin{tikzpicture}[line width=1 pt, scale=0.5, baseline=0]
\node[above] at (0,2) {$\tau_{0}$}; 
\node[below] at (0,0) {$0$};
\node[right] at (3,0) {$L$};
\node[left] at (-3,0) {$-L$};
\draw (0,2)--(0,0);
\draw[line width=2 pt] (-3,0)--(3,0);
\draw (-2.5,0) arc(180:0:1);
\end{tikzpicture} &=  \z^{3}\,C_{\R}\,\bigg(\frac{L}{\eps}+\log\frac{\eps}{\eps+L}\bigg)\frac{\gym^2}{4\pi^{2}},\ea
\ba\la{2.14}
\begin{tikzpicture}[line width=1 pt, scale=0.5, baseline=0]
\node[above] at (0,2) {$\tau_{0}$}; 
\node[below] at (0,0) {$0$};
\node[right] at (3,0) {$L$};
\node[left] at (-3,0) {$-L$};
\draw (0,2)--(0,0);
\draw[line width=2 pt] (-3,0)--(3,0);
\draw (-1,0) arc(180:0:1);
\end{tikzpicture}  = -\z^{3}(C_{\R}-\tfrac{1}{2}C_{\A})\log\frac{\eps(\eps+2L)}{(\eps+L)^{2}}\,\frac{\gym^2}{4\pi^{2}}.
\ea
\ba\la{2.15}
\begin{tikzpicture}[line width=1 pt, scale=0.5, baseline=0]
\node[right] at (2,0) {$L$};
\node[left] at (-2,0) {$-L$};
\draw[line width=2 pt] (-2,0)--(2,0);
\draw (-1,0) arc(180:0:1);
\end{tikzpicture}  = \z^{2}C_{\R}\,\bigg(\frac{2L}{\eps}+\log\frac{\eps}{\eps+2L}\bigg)\,\frac{\gym^2}{4\pi^{2}}, 
\ea
so that the  total  result  for the vertex reads
\ba
V(\z, L) &= \z+\z^{3}\,C_{\A}\log\frac{\eps(\eps+2L)}{(\eps+L)^{2}}\,\frac{\gym^2}{8\pi^{2}}+\mc O(\gym^{4}).
\ea
Dependence on $C_{\R}$ again drops  out and also the  linear  divergent terms 
 ${L\ov \eps} $ cancel. The vertex 
 is renormalized by
\be
\z \equiv \z_{\rm bare} = \z(\mu)-C_{\A}\,\z^3 (\mu)\,\log(\mu\,  \eps)\frac{\gym^2}{8\pi^{2}}+\mc O(\gym^4)\ .
\ee
Then \be\la{2.18}
V_{\rm ren}\big(\z(\mu), L\big) = \z(\mu)-C_{\A}\,\z^3 (\mu) \log\frac{L\mu}{2}\, \frac{\gym^2}{8\pi^{2}}+\mc O(\l^{2}),
\ee
 obeys the Callan-Symanzik equation (\ref{2.10}) with the same 
  beta function as in (\ref{2.11}). 

The same approach can be extended to the two-loop level.
We find that 
the corresponding  coupling redefinition and renormalized vertex are
\ba\la{2.19}
\z \equiv \z_{\rm bare} = &\z(\mu)-C_{\A}\,\z^3 (\mu)\,\log(\mu\, \eps)\frac{\gym^2}{8\pi^{2}}\lp
+C_{\A}^{2}\,\z^5 (\mu)\bigg[
\frac{1}{4}\log(\mu\, \eps)+\frac{3}{8}\log^{2}(\mu\, \eps)
\bigg]\frac{\gym^{4}}{(4\pi^{2})^{2}}+\mc O(\gym^{6}), \\
V_{\rm ren}\big(\z(\mu), L\big) = &\z(\mu)-C_{\A}\,\z^3 (\mu)\log\frac{L\mu}{2}\, \frac{\gym^2}{8\pi^{2}}\lp
-C_{\A}^{2}\,\z^5 (\mu)\,\bigg[\pi^{2}+12\log^{2}2-3\log\frac{\mu L}{2}\bigg(2+3\log\frac{\mu L}{2}\bigg)\bigg]\,\frac{\gym^4}{384\pi^{4}}
+\mc O(\gym^{6}). \la{2.20}
\ea
The  corresponding two-loop   beta-function is then  given by  (in agreement with the 
 Callan-Symanzik equation \rf{2.10})   
\be\la{2.21}
\beta^{\ladd}_{\z} = C_{\A}\,\z^{3}\,\frac{\gym^2}{8\pi^{2}}-C_{\A}^{2}\,\z^{5}\,\frac{\gym^{4}}{64\pi^{4}}+\mc O(\gym^6).
\ee
Thus the ladder  part  is again universal, \ie does not depend on a particular representation. 
This  independence of representation 
is  an accidental  two-loop property --
we shall see  below that it  does not hold  
at three-loop order.

The  full two-loop  beta-function is then expected to be (cf. \rf{1.3},\rf{1.4}) 
\be
\la{2.22}
\beta_\z =- C_{\A} \,\z\,(1-\z^{2})\frac{\gym^2}{8\pi^{2}}
+\z\,(1-\z^{2})\,(\bb_2 +C_{\A}^{2}\,\z^{2})\,\frac{\gym^4}{64\pi^{4}}\,+ \mc O(\gym^6), 
\ee
where $\bb_2 $  may depend on representation  R. 
\iffa 
could be determined by an independent determination of 
\be
\Delta-1 = \beta'(1) = \frac{C_{\A}\lambda}{4\pi^{2}N}-\frac{(b_{1}(N)+C_{\A}^{2})\,\lambda^{2}}{32\pi^{4}N^{2}}+\cdots.
\ee
\fi
Since the  beta-function  should    vanish  in the abelian limit  $\bb_2$  should  not contain 
$C_{\R}^{2}$ term, i.e.  we  should have 
\be \la{2.23}
\bb_2  = p_1 C_{\A}^{2}+ p_2  C_{\A}C_{\R} \ , \ee
where $p_1,p_2$  are  universal  constants.  Comparing to the case of the  fundamental  representation of $SU(N)$ in the planar limit where the two-loop term is given in \rf{1.4} 
(where $\l= \gym^2 N, \ C_{\A}= N, \ C_{\R}= {N^2-1\ov 2 N}\to \ha N $) 
we get  the constraint 
\be\la{2.24}
p_1 + \ha p_2 = 1\ .
\ee
One natural conjecture  
 is that $p_2=0$ so that $C_\R$ does not appear in \rf{2.22}, i.e. 
that like the one-loop  beta function,  the full two-loop   one 
does not depend  on a choice of a particular representation, namely
\be
\la{2.25}
\beta_\z =- C_{\A} \,\z\,(1-\z^{2})\frac{\gym^2}{8\pi^{2}}
+ C^2_{\A}\z\,(1-\z^{4})\,\frac{\gym^4}{64\pi^{4}}\,+ \mc O(\gym^6) \ . 
\ee


\subsection{Three-loop order} 

As already mentioned in the Introduction (cf. \rf{1.20}),  from the  analysis of possible  contributions to the four-loop WL  expectation value  
the general structure of the 
three-loop beta function  in ladder approximation  is expected to be  
\ba
\la{2.26}
(\beta^{\ladd}_{\z})^{(3)} &= 
\bigg(q_{3}'\,C_{\A}^{3}+q_{3}''\,C_{\A}^{2}C_{\R}+q_{3}'''\,C_{\A}C_{\R}^{2}
+q_{3}'''' \,   \QR   
\bigg)\,\z^{7}\,\Big(\frac{\gym^2}{4\pi^{2}}\Big)^{3} \ ,  
\ea
where $\QR$ was defined in \rf{1200}.
This satisfies the condition of vanishing win the abelian limit when $C_{\A}=0$. 
Here  the tensor $d_{R}^{abcd}$ is the  symmetrized trace of the product of four generators 
\be\la{2.27} 
d_{\R}^{a_{1}\cdots a_{n}} =\text{Str}(T^{a}T^{b}T^{c}T^{d})=
 \frac{1}{n!}\tr\sum_{\sigma\in \mc S_{n}}T^{a_{\sigma(1)}}\cdots T^{a_{\sigma(n)}}.
\ee 
To constrain the numerical  coefficients  $q_3', q_3'', ...$ 
 we  shall  consider the  case of  R  being $k$-symmetric representation of $SU(N)$
in the limit of  $k\gg 1$. Then  (see, e.g.,  \cite{Fiol:2018yuc})
 \ba\la{2.28}
C_{\R} &= \frac{k(N-1)(N+k)}{2N}\to  k^{2}\frac{N-1}{2N}, \\ \la{2.29}
\QR=\frac{d_{\A}^{abcd}d_{\R}^{abcd}}{C_{\R}\,\dim \R} &= \frac{N}{24}\big[N^{2}-6N+6k(k+N)\big] \to  k^{2}\frac{N}{4}.
\ea
As we shall demonstrate below, 
  the ladder beta function is expected to vanish (for generic $N$) 
in the ``classical'' limit  \ci{Komar}
\be\la{2.30}
k\to \infty, \qquad \z \to 0 , \qquad \qquad  k\,\z^{2} = \text{fixed} \ . 
\ee
This implies that  $q'''_{3}=0$. 
Then the  remaining terms give the following large $k$ limit
(at fixed  $\z$) 
\ba
\la{2.31}
(\beta^{\ladd}_{\z})^{(3)} 
 &\stackrel{k\gg 1}{=} \bigg(q_{3}''N(N-1)+q_{3}'''' \frac{N}{4}\bigg)\,k^{2}\,\z^{7}\,\Big(\frac{\gym^2}{4\pi^{2}}\Big)^{3}\ . 
\ea
Below  we will  compute (\ref{2.31}) explicitly   determining the two constants $q_{3}''$ and $q_{3}'''$.
Plugging them into  (\ref{2.26}) will lead to the three-loop  expression quoted in  \rf{1.21}.

\section{Two-loop term in  $\vev{W^{(\z)}}$ for  generic representation}
\la{app:twoW}

Here    we   sketch the computation of the  non-trivial  two-loop 
term in $\vev{W^{(\z)}}$   quoted in  (\ref{1.15}). We  
 generalize the calculation in  \cite{Beccaria:2017rbe} in order to determine
the coefficients $k_{1}$ and $k_{2}$ in (\ref{1.14})  which applies to a
 generic representation R of a simple gauge group $G$.  
 
Decomposing   the two-loop contributions to $\vev{W^{(\z)}}$
into planar ladder diagrams, self-energy diagrams, spider diagrams involving 3-vertices, and 
non-planar ladders, we can write using   \cite{Beccaria:2017rbe} (introducing  explicit color factors) 
\ba
&\norm \vev{W^{(\z)}} = 1+2C_{\R}\,\W_{\rm tree}^{(\z)}\, g^2 
+\bigg[ 4C_{\R}^{2}\,\W_{\rm planar\ ladder}^{(\z)}+
   2C_{\R}\, C_{\A}  \,\W_{\rm self}^{(\z)}\no \\& \qquad \qquad \qquad \qquad +
 2C_{\R}\,    C_{\A}  \,\W_{\rm 3-vertex}^{(\z)}
+4C_{\R}(C_{\R}-\frac{1}{2}C_{\A})\W_{\rm non-planar\ ladder}^{(\z)}
\bigg]\, g^4 +\cdots \la{E.1}
\ea
The expressions of all planar pieces in dimensional regularization are  \cite{Beccaria:2017rbe} (in this Appendix  we follow the notation   of   \cite{Beccaria:2017rbe}   where   $d=2\omega=4-2\ve$)
\ba
\no
& \W_{\rm tree}^{(\z)} = \frac{1}{8}-\frac{1}{8}\z^{2}\ve, \qquad 
\W_{\rm planar \ ladder}^{(\z)} =\frac{1}{192}+(1-\z^{2})\left(\frac{1}{64\pi^{2}\ve}+\frac{1}{128\pi^{2}}(7-3\z^{2})+\frac{\log(\pi e^{\gamma_{E}})}{32\pi^{2}}\right),\\
\la{E.4}
&\W_{\rm self}^{(\z)} = \z^{2}\left(-\frac{1}{64\pi^{2}\ve}-\frac{1}{32\pi^{2}}-\frac{\log(\pi e^{\gamma_{E}})}{32\pi^{2}}\right)
+(1-\z^{2})\left(-\frac{1}{64\pi^{2}\ve}-\frac{1}{16\pi^{2}}-\frac{\log(\pi e^{\gamma_{E}})}{32\pi^{2}}\right), \\
\no 
&\W_{\rm 3-vertex}^{(\z)} = -\W_{\rm self}^{(\z=1)}+(1-\z^{2})\left(-\frac{1}{64\pi^{2}\ve}-\frac{1}{64\pi^{2}}-\frac{\log(\pi e^{\gamma_{E}})}{32\pi^{2}}\right).
\ea
The non-planar ladder contribution  is 
\be
\la{E.6}
\W^{(\z)}_{\rm non-planar \ ladder} = \frac{[\Gamma(1-\ve)]^{2}}{64\pi^{4-2\ve}}\int_{\tau_{1}>\tau_{2}>\tau_{3}>\tau_{4}}d^{4}\tau\ \frac{(\z^{2}-\cos\tau_{13})\ (\z^{2}-\cos\tau_{24})}
{(4\sin^{2}\frac{\tau_{13}}{2}\ 4\sin^{2}\frac{\tau_{24}}{2})^{1-\ve}}.
\ee
Computing it by the method described in \cite{Beccaria:2017rbe}, we find 
\be
\la{E.7}
\W^{(\z)}_{\rm non-planar \ ladder} = \frac{\z^{2}-1}{64\pi^{2}\ve}+\frac{1}{384}
+\frac{(\z^{2}-1)(3\z^{2}-7)}{128\pi^{2}}+(\z^{2}-1)\frac{\log(\pi e^{\gamma_{E}})}{32\pi^{2}}.
\ee
Substituting the expressions in 
 (\ref{E.4}) and (\ref{E.7}) into  (\ref{E.1}) and also expressing the bare   coupling 
  $\z$ by its renormalized  value  using 
the one-loop beta function $\beta_{\z}= -C_{\A}\z(1-\z^{2})\frac{g^{2}}{8\pi^{2}}+\cdots$,\foot{That the 
one-loop term  in  (\ref{1.4})  does not depend on representation R  follows from 
 a direct inspection of the possible color factors, and using also  the condition of 
 the vanishing of beta-function  in the Abelian limit.} 
we finally find the expression in (\ref{1.15}).

\section{
1d path integral  for  ladder Wilson loop  in $k$-symmetric $SU(N)$ representation}
\la{sec:1d}

As was   mentioned in \ci{Beccaria:2021rmj}, one may also 
study  the  (fundamental) WL renormalization  and compute $\beta_\z$  using 
more conventional approach in which path ordering is replaced by 
a functional  integral over the auxiliary 1d fermions $\psi_i$ ($i=1, ..., N$)   as in \cite{Gervais:1979fv,Arefeva:1980zd,Brandt:1981kf}.
We will  review this representation in Appendix \ref{app:GN}. 
Then in the ladder  approximation when  the bulk theory reduces to 
  just a free 4d adjoint scalar  field  integrating  the scalar  out   leads to an 
  effective theory of  $\psi_i(\tau) $  with a  non-local  1d action of the form (cf. \rf{1.24})
 \be\la{3.1}
S=   \int d\tau\,\bar \psi_i \del_\tau \psi^i - { \l \z^2\ov 8 \pi^2}    \int d\tau\,d\tau'\, 
 \bar \psi_j(\tau) \, \psi^i(\tau)\,  \frac{1}{(\tau-\tau')^2}\, \bar  \psi_i(\tau') \, \psi^j(\tau')\ . 
\ee
Below  we will   be interested in the case of  $k$-symmetric  $SU(N)$ representation 
in  which a different 1d effective representation in terms of 1d bosons \cite{Gomis:2006sb, Gomis:2006im} is 
more convenient  (cf.  also \ci{Komar}).\footnote{In  \cite{Gomis:2006sb, Gomis:2006im}, this representation was 
discussed in the context of the half-BPS Wilson loop, but it applies the same way to the generalized Wilson loop \rf{1.1}
or the purely scalar loop \rf{1.9}.}


Let us start with  the following partition function of periodic  bosons
 $\chi^{i}$ in the fundamental representation
of $SU(N)$  
\be
\la{3.2}
Z = \int D\chi D\bar \chi \ e^{i\int^{2\pi}_0  d\tau\mathscr L}, \qquad\qquad \mathscr L = i\,\bar \chi\del_\tau \chi+i\,\bar \chi \p(\tau)\chi, \qquad \qquad \p=\p^{a}T^{a}.
\ee
In the case of our interest $\p(\tau) $ will be the adjoint free scalar $\p$  of the ladder model
restricted to the $\tau$-line \rf{1.9},\rf{1.11} (up to rescaling by $\z$). 

In the operator   quantization 
 (with $[\chi^j, \bar \chi_{i}]
  = \delta^{j}_{i}$)  we have  $Z = {\rm tr}_{\chi}[\textrm{T}\exp{i\int d\tau \mc H(\tau)}]$ where 
  the   time dependent local Hamiltonian is $\mc H(\tau) \equiv \hat \phi =-i\,\bar\chi\phi\chi$. Here, time-ordering is interpreted as  path-ordering and we have  
\be
\la{3.3}
Z = {\rm tr}_{\chi}\Big[\textrm{P}\exp\Big(i \int_{0}^{2\pi }d\tau \, \hat \p(\tau)\Big)\Big]\ ,  
\ee
where the trace is over the Hilbert space of $\chi^i, \bar\chi_i$.
The state space is built starting from $\chi^{i}\vac=0$ and acting with $\bar \chi_{i}$.
$Z$   may be written as a sum of  partition functions 
restricted to the subspace where the particle number operator
$\nu=\bar\chi_{i}\chi^{i}$    has  fixed  value. 
On the many-particle states   with $\nu=k$ 
the action of  $\bar\chi T^{a}\chi$ is the same as that of 
the  generator  $T^{a}$ in the $k$-symmetric 
 representation (that we will denote as  $\S_{k}$).\footnote{For example, on 1-particle state  (corresponding to fundamental representation) 
  we have 
   $(\bar\chi T^{a}\chi)\  \bar\chi_{i}\vac = \bar\chi_{k}(T^{a})^k_j \chi^{j}\bar\chi_{i}\vac = (T^{a})^j_{i}\bar\chi_{j}\vac$.}
Hence, $Z$  computes the sum of all ``Wilson loops'' 
 in the $k$-symmetric representations
\be
\la{3.4}
Z = \sum_{k=0}^{\infty}\, W_{k}\ , \qquad \qquad  W_k = \tr_k\,  \textrm{P}\exp\Big(\int_{0}^{2\pi}d\tau \, \p(\tau)\Big)\ . 
\ee
To select   a particular  $W_k$   contribution we may add the constraint  on $\bar \chi \chi$ 
with a Lagrange multiplier $A=A(\tau) $  as 
\be \la{3.5}
\mathscr L = i\,\bar \chi\del_\tau \chi+i \bar \chi \p(\tau)\chi + A ( \bar \chi \chi - k - {N\ov 2}) 
\ . \ee
The extra  constant shift  by ${N\ov 2}$ is due to   the choice of Weyl ordering.\foot{In the  path integral integral formulation  the number operator $\nu$ corresponds to 
  $\bar\chi\chi-\frac{N}{2}$:   if  $\chi,\bar \chi$ are operators,   using the  
 symmetric (Weyl) ordering prescription we have  
 $\nu = \frac{1}{2}({\bar\chi}_{i}\chi^{i} + \chi^{i}{\bar\chi}_{i}) - \frac{N}{2} $.  
}
In what follows   we shall use the   notation 
\be   R^2 = k + {N\ov 2} \ . \la{3.6} \ee
Note  that  $R^2$  appears in the action 
  as the coefficient of the  1d Chern-Simons term $\int d\tau A$, and  one may argue as usual 
 that it   should not be renormalized since it is quantized. 

We shall  see shortly that this shift by $\frac{N}{2}$  leads indeed to the correct 
result for $W_k$  in the  simplest case of $\phi=0$, namely,  that it is equal to
  the dimension $\tr_k {\rm 1}=  \dim {\rm S} _k= {(N+k-1)!\ov k! \, (N-1)!}$
of the $k$-symmetric representation 
\be
\la{3.7}
{\WO}=\int D\chi D\bar\chi \ e^{- \int d\tau\,\bar\chi\del_\tau\chi} \ 
\delta ( {\bar\chi\chi -  R^{2}} )=  \dim {\rm S} _{k}=\binom{N+k-1}{k} \ .
\ee
This  requires   careful definition of the path integral over the Lagrange multiplier 
$A$, which can be interpreted as a 1d  $U(1)$   gauge field. Indeed, 
the path integral 
\be
\la{3.8}
{\WO}  = \int D\chi D\bar\chi  DA\  \exp\Big( {i\int^{2\pi}_0  d\tau\,\big[i\bar\chi\del_\tau\chi+A(\bar\chi\chi-R^{2})\big]} \Big)
\ee
is invariant  under 
\be \la{3.9}
\chi^i \to e^{i\alpha }\chi^i \ , \qquad \bar\chi_i \to e^{-i\alpha }\bar\chi_i\ , \qquad \qquad 
A\to A + \del_\tau \alpha \ , \qquad \alpha=\alpha(\tau) \ .  \ee
The  function  $\alpha$ compatible with periodic boundary conditions on $\chi$ 
should satisfy $\alpha(2\pi)-\alpha(0) = 2\pi n$ where $n$ is an integer, i.e. 
\be
\la{3.10}
\alpha(\tau) = \alpha_0(\tau) +  n \tau \ , \qquad \qquad 
 \alpha_0(2\pi) =\alpha_0 (0) \ , 
\ee
$\alpha_0$  corresponds to  the ``small'' gauge transformation. 
It  allows  to   gauge fix $A$ to be  a constant 
 \be 
 A=\mu \ , \qquad \qquad 
 \mu = \frac{1}{2\pi} \int_{0}^{2\pi} d\tau\,  A \ .
 \ee
 Under  the ``large'' gauge transformation $ \alpha(\tau) =  n \tau$, 
    $A$  changes by  an integer $n$. Naively one would expect this to be a   symmetry of the path integral \rf{3.8}
    only if $R^2=k + {N\ov 2}$  is an integer, which would require $N$ to be even. However, as we shall see below, $\mu \to \mu + n$ is in fact a symmetry 
for any $N$, due to an ``anomalous''  contribution of 
the  functional determinant  coming from  integration over $\chi$ and $\bar\chi$.
 The redundancy under $\mu\to \mu+n$ can be fixed by restricting the  integration over $\mu$ 
 to the interval $[0,1]$
 \be
\la{3.12}
{\WO}  = \int_{0}^{1} d\mu\int D\chi D\bar\chi\  \exp\Big({i\int^{2\pi}_0  d\tau\,\big[i\bar\chi\del_\tau \chi+\mu(\bar\chi\chi-R^{2})\big]}\Big)  \ . 
\ee
The  functional integral over $\chi$ and $\bar\chi$    gives
$\big[ \det(i\partial_\tau +\mu)]^{-N}$ 
 where   the determinant that can be defined as usual with  the  $\zeta$-function prescription 
  (recall that $\chi(2\pi)=\chi(0)$) 
\ba
\la{3.13}
\det(i\partial_\tau +\mu) &= \prod_{n=-\infty}^{\infty}(n+\mu) = \mu\prod_{n=1}^{\infty}(\mu^{2}-n^{2}) = \mu\prod_{n=1}^{\infty}\frac{n^{2}-\mu^{2}}{n^{2}}\prod_{n=1}^{\infty}(-n^{2}) = \frac{\sin(\pi\mu)}{\pi}\prod_{n=1}^{\infty}(-n^{2})\lp
= \frac{\sin(\pi\mu)}{\pi}\,e^{\log(-1)\zeta(0)-2\zeta'(0)} = -2i\sin(\pi\mu) \ . 
\ea
This leads to  the expected result in  (\ref{3.7})\foot{Here  in computing the integral we use 
analytic continuation in $N$.}
\ba
{\WO}  = & \int_{0}^{1} d\mu \ e^{-2\pi i \mu R^{2}}\,\big[-2i \sin(\pi \mu)\big]^{-N} =
\int_0^1 d\mu \ \frac{e^{-2\pi i k \mu}}{(1-e^{2\pi i \mu})^N}
=   \binom{R^{2}+\frac{N}{2}-1}{R^{2}-\frac{N}{2}} = \binom{N+k-1}{k}\ . \la{3.14}
\ea
Note  that as was claimed  above, the integrand  here is indeed 
invariant under $\mu \to \mu + n$. 

Before proceeding, let us point out as a side remark that a similar 1d action (\ref{3.5}), with $\chi$ taken to be $N$ anticommuting fermions with antiperiodic 
boundary conditions, describes instead the Wilson loop in the rank $k$ antisymmetric representation \cite{Gomis:2006sb} (note that this is different from the 
fermionic representation of \cite{Gervais:1979fv,Arefeva:1980zd,Brandt:1981kf} reviewed in the Appendix \ref{app:GN}). Further generalizations with (bosonic or fermionic) $\chi$ fields carrying 
an additional $U(M)$ index  and a 1d $U(M)$ gauge field on the defect can also be used to describe Wilson loops in representations corresponding to 
a  general Young tableau.

\iffa 
Do not think we need a comment here.
but it could be  like this:
An alternative procedure would be  not to introduce the Lagrange multiplier field $A$
  and to solve the delta-function  constraint  explicitly   by expressing of the components 
 of $\chi$ in terms of the  others. But even reproducing the result \rf{3.14} then becomes a non-trivial
problem.  

\paragraph{Local approach} 
An alternative procedure, which we call the ``local approach'', is to impose(\ref{2.13}) by inserting a delta function directly in (\ref{3.7}), what is 
clearly equivalent to (\ref{3.8}) after integration over all unrestricted gauge fields $A$. However, already in the free limit, proving the relation (\ref{3.7})
in the explicit form 
\be
\la{3.15}
\int D\chi D\bar\chi \ \delta(\bar\chi\chi - R^{2})\ e^{i\int d\tau\,i\bar\chi\del_\tau\chi}=\binom{N+k-1}{k},
\ee
does not seem trivial. \fix{we need some definitive comment here, maybe giving some hope that all is ok in perturbation
theory ?}
\fi

\section{Large $k$ perturbative expansion in scalar ladder model}
\la{sec:pt}


In this section we will work out the large $k$ expansion of the  
Wilson loop in symmetric representation $\S_{k}$ in  the scalar ladder approximation.
We will  begin with the free  theory ($\z=0$)  case  to explain the  strategy of perturbative $1/k$  expansion 
  and then  move on to the  general $\z\neq 0$  case .

\subsection{Free theory }

Since the parameter $k$ appears only in the combination (\ref{3.6}), it will be convenient to work out the large $k$ expansion as an expansion  in inverse powers $1/R^2$.
Thus, our aim  will be  to reproduce  the large $R^2$ expansion of 
\be
\la{4.1}
\WO= \dim S_{k} = \binom{N+k-1}{k} = \binom{R^{2}+\frac{N}{2}-1}{R^{2}-\frac{N}{2}} = \frac{R^{2(N-1)}}{(N-1)!}\,\bigg[1-\frac{N(N-1)(N-2)}{24\,R^{4}}+\cdots\bigg].
\ee
Starting with the exact integral representation (\ref{3.14}) for ${\WO}$ we  may write it in the form 
amenable to $1/R^2$ expansion\foot{We shifted $\mu$ by -1  which is a symmetry of the integral in \rf{3.14}.} 
\ba
\no 
{\WO} &= \int_{-1/2}^{1/2} d\mu\ e^{-2\pi i \mu R^{2}}\,\big[-2i \sin(\pi \mu)\big]^{-N} \\ & = 
\Big(-\frac{2\pi i}{R^{2}}\Big)^{-N}\frac{1}{R^{2}}\int_{-{R^{2}\ov2}}^{{R^{2}\ov 2}} d\mu'\
e^{-2\pi i \mu'}\,\mu'^{-N}\,\Big(1+\frac{\pi^{2}\mu'^{2}N}{6R^{4}}+\cdots\Big).\la{4.2}
\ea
Taking $R$ large and thus setting the  integration limits to $\pm\infty$,\footnote{The 
effect of 
large gauge  transformations is not visible in large $R^2$   perturbation theory   so the  restriction on the range of $\mu$-integration  can be relaxed.
\iffa
{\tiny 
\fix{we need to check what is the error, exponentially small or not. Anyway, we seem to reproduce what was expected so should be ok,
but question is are there corrections at all ? Actually no. Logic could be that analytic continuation in $N$ breaks symmetry under large gauge transformations so the range should be consistently $\mu\in(-\infty,\infty)$. ????   not sure --   needs discussion } }
\fi}
and using
the analytic continuation in the integral 
\be
\la{4.3}
\int_{-\infty}^{\infty}d\mu\, e^{-2\pi i \mu}\mu^{\alpha} = -\frac{1}{(2\pi)^{\alpha}}e^{\frac{i\pi\alpha}{2}}\,\frac{\alpha}{\Gamma(1-\alpha)}, \qquad -1<\text{Re}(\alpha)<0,
\ee
we find
\ba
\la{4.4}
{\WO} &= \frac{R^{2N-2}}{\Gamma(N)}\Big(1-\frac{N(N-1)(N-2)}{24R^{4}}+\cdots\Big),
\ea
in agreement with (\ref{4.1}).

This perturbative procedure has a direct counterpart at the level of the path integral (\ref{3.12}), \ie 
before integrating out 
$\chi, \bar\chi$  in terms of a functional determinant. 
Once again, we expect to find 
\ba
\la{4.5}
{\WO} &=
 \int_{-1/2}^{1/2} 
 d\mu\int D\chi D\bar\chi\  e^{i\int d\tau\,[i\bar\chi\del_\tau\chi+\mu(\bar\chi\chi-R^{2})]} 
= \frac{1}{R^{2}} \int_{-{R^{2}\ov2}}^{{R^{2}\ov 2}} d\mu'\, e^{-2\pi i \mu'} \ \Big[J({\mu'\ov R^{2}})\Big]^{N}, \\
\la{4.6}
J(\m) &\equiv  \int D\chi D\bar\chi\  e^{i\int d\tau\,(i\bar\chi\del_\tau\chi+\m\,\bar\chi\chi)} 
= \big[-2\,i\,\sin(\pi \m)\big]^{-1} 
= \frac{i}{2\pi \m}+\frac{i \pi \m}{12}+\frac{7i \pi^{3}\m^{3}}{720}+\cdots,
\ea
where in \rf{4.6}  $\chi$ is now a singlet field. 
Let us   show how to reproduce   \rf{4.6} in 
small mass expansion. This requires isolating the contribution of the 
 constant zero mode of the $\partial_\tau$ kinetic operator, i.e. 
\ba
\la{4.7}
&\chi=n+\chi', \qquad \ \ \  \bar\chi=\bar n+\bar\chi', \ \ \  \qquad \int d\tau\, \chi' = \int d\tau\,\bar\chi' = 0, \\
\la{4.8}
&S = \int d\tau (i\bar\chi\del_\tau\chi+\m\bar\chi\chi) = \int d\tau (i\bar\chi'\del_\tau\chi'+\m\bar\chi'\bar\chi'+\m\,\bar n\, n).
\ea
The Gaussian integration over the constants $n$ and $\bar n$ gives the $1\ov \mu$ factor 
and the rest of  the small $\m$ expansion is then regular\foot{The factor $\frac{i}{2\pi}$  comes from 
 $\det'(i\partial_{t})^{-1}$:  starting    from (\ref{3.13}),  removing the zero mode and then sending $\mu\to 0$  one finds  $(-2i\frac{\sin(\pi\mu)}{\mu})^{-1}\to \frac{i}{2\pi}$.} 
\ba
\la{4.9}
J(\m) &= \frac{1}{\m}\int D\chi' D\bar\chi'\  e^{-\int d\tau\,\bar\chi'\del_\tau\chi' }\bigg(1+i\m\int d\tau\bar\chi'\chi'-\frac{\m^{2}}{2}\int d\tau\bar\chi'\chi'\int d\tau'\bar\chi'\chi'+\cdots\bigg) \lp
= \frac{i}{2\pi \m}\bigg(1+i\m\,\vev{\int d\tau\bar\chi'\chi'}-\frac{\m^{2}}{2}\vev{\int d\tau\bar\chi'\chi'\int d\tau'\bar\chi'\chi'}+\cdots\bigg).
\ea
The 
expectation values in (\ref{4.9}) are computed using the propagator for the non-constant mode, \ie
\be
\la{4.10}
\D(\tau) = \D(\tau+2\pi) = \vev{\chi'(\tau)\bar\chi'(0)} = \sum_{\ell\neq 0}\frac{1}{2\pi i \ell}e^{i\ell\tau} = \frac{1}{i\pi}\sum_{\ell=1}^{\infty}\frac{\sin(\ell\tau)}{\ell}, \qquad \D(\tau) = -\D(-\tau),
\ee
so that $\D(0)=0$.
The explicit form of $\D$  restricted to  the interval  $\tau\in(0,2\pi)$   is 
\be
\la{4.11}
\D(\tau) = i\,\frac{\tau-\pi}{2\pi},\qquad  \qquad 0<\tau<2\pi.
\ee
Thus $\vev{\int d\tau\bar\chi'\chi'}=0$ and 
\ba
\la{4.12}
&\vev{\int d\tau\bar\chi'\chi'\ \int d\tau'\bar\chi'\chi'} = \int_{0}^{2\pi} d\tau\int_{0}^{2\pi} d\tau'\ \big[\D(\tau-\tau')\big]^{2} =  2\pi \int_{0}^{2\pi} d\tau\ [\D(\tau)]^{2} = -\frac{\pi^{2}}{3},
\ea
where we used (\ref{4.11}). 
As a result, we  reproduce (\ref{4.2}).

\subsection{Interacting case}

Starting with  the ladder scalar model  on the circle (\ref{3.2}), 
let is   make the dependence on the   coupling  $\z$ explicit  by 
$\phi\to \z\phi$ and integrate out the scalar field. 
This gives the  effective 1d action
\be
\la{4.13}
S = \int d\tau\bigg[i\,\bar \chi\del_\tau \chi +\mu(\bar\chi\chi-R^{2})\bigg]
-\frac{i\z^{2}\,g^{2}}{8\pi^{2}} \int \frac{d\tau\,d\tau' }{4\sin^{2}\frac{\tau-\tau'}{2}}\,\bar \chi(\tau) T^{a}\chi(\tau)\,
\bar \chi(\tau') T^{a}\chi(\tau')\ , 
\ee
where we used the explicit form  \rf{1.12} of the scalar propagator restricted to the circle. 
Let us introduce a compact notation  for the integration measure  
\be
\la{4.14}
\dtdt = \frac{d\tau\,d\tau' }{4\sin^{2}\frac{\tau-\tau'}{2}}\ . 
\ee
The effective  coupling that will play a central role  below is 
\be
\la{4.15}
\pxi \equiv \frac{\z^{2}g^{2}R^{2}}{8\pi^{2}}
 \ , \qquad \qquad  R^2= k + {N \ov 2} \ .
\ee
Redefining  $\chi$ and $\bar\chi$ by a factor of $R$   we  may then  write \rf{4.13}   as 
\be
\la{4.16}
S = S_{2}+S_{4} = R^{2}\int d\tau\bigg[i\,\bar \chi\del_\tau \chi +\mu(\bar\chi\chi-1)\bigg]
-i\,\pxi\,R^{2}\int\dtdt\,\bar \chi(\tau) T^{a}\chi(\tau)\,
\bar \chi(\tau') T^{a}\chi(\tau'),
\ee
where $S_4$   stands for the quartic term. 
As in (\ref{4.7}) let us  separate  the  constant part of $\chi$ as 
\be
\la{4.17}
\chi=n+\frac{1}{R}\chi', \qquad \bar\chi=\bar n+\frac{1}{R}\bar\chi', \qquad \int d\tau\, \chi' = \int d\tau\,\bar\chi' = 0\ . 
\ee
Then  
\ba
\la{4.18}
&S_{2} = \int d\tau\bigg[i\,\bar \chi'\del_\tau \chi' +\mu'\,\Big(\bar n n-1 +\frac{1}{R^{2}}\bar\chi'\chi' \Big)\bigg] , \qquad \qquad \mu'= R^2 \mu \ , \\
\la{4.19}
&S_{4} = -i\,\pxi\,R^{2}\int\dtdt\,\bigg(\bar n +\frac{1}{R}\bar\chi'(\tau)\bigg)T^{a}\bigg(n+\frac{1}{R}\chi'(\tau)\bigg)\,
\bigg(\bar n +\frac{1}{R}\bar\chi'(\tau')\bigg)T^{a}\bigg(n+\frac{1}{R}\chi'(\tau')\bigg).
\ea
Note that in addition to $1/R^2$ term in \rf{4.18} the action \rf{4.19} contains 
$1/R$ cubic and $1/R^2$ quartic interaction vertices. 
Integrating  over $\mu'$ we get for the resulting  path  integral  measure 
\be
\la{4.20}
\int  D \chi \, D \bar \chi  \ \to \  \int  D \chi' \, D \bar \chi' \int d n \, d \bar n \  
\delta\Big( \bar n n-1 +\frac{1}{2\pi R^{2}}\int d\tau \, \bar\chi'\chi' \Big)
\ . \ee

\subsubsection{Leading (one-loop)  order at large $R$}

Expanding  \rf{4.18},\rf{4.19}   at large $R^2$ for fixed $\pxi$ 
we  note that  at leading order the delta-function in \rf{4.20} 
imposes that 
\be
\la{4.21}
\bar n_i n_i  = 1\ .  
\ee
Then 
\ba
S_2= & i\int d\tau \,\bar \chi'\del_\tau \chi', \ \ \    S_4= -i\,\pxi\,\int\dtdt\,\bigg[
\bar nT^{a}\chi'(\tau)\ \bar nT^{a}\chi'(\tau')
+
\bar nT^{a}\chi'(\tau)\ \bar\chi'(\tau')T^{a}n
\lp\qquad \qquad \qquad \qquad \qquad 
+
\bar\chi'(\tau)T^{a}n\ \bar nT^{a}\chi'(\tau')
+
\bar\chi'(\tau)T^{a}n\
\bar\chi'(\tau')T^{a}n\bigg]+\mc O(R^{-1}). \la{4.22}
\ea
We used  the following remarkable property of the measure  $\dtdt$ in \rf{4.14} 
 (valid in dimensional regularization, or up to power divergences that we will neglect):  
  for a generic function $f(\tau)$
\be
\la{4.23}
\int \dtdt\  f(\tau) = 0. 
\ee
Using that  for the $T^a$ in the fundamental representation 
\be \la{4.24}
(\bar \alpha\,   T^{a} \beta )\ (\bar \gamma\,  T^{a} \delta ) = \frac{1}{2}(\bar \alpha  \delta )(\bar \gamma \delta  )-\frac{1}{2N}(\bar \alpha \beta )(\bar \gamma \delta )\ , \ee
we then have from \rf{4.19} 
\ba
\la{4.25} &\qquad \qquad S= S^{(2)} + {1\ov R} S^{(3)}  + {1\ov R^2} S^{(4)}      \ ,\\ 
S^{(2)} = &  i\int d\tau \,\bar \chi'\del_\tau \chi' 
 -\frac{i}{2}\pxi\,\int \dtdt \bigg[\Big(1-\frac{1}{N}\Big)\Big(
\chi_{i}'(\tau')\bar n_{i}\bar n_{j}\chi_{j}'(\tau)
+ \bar\chi_{i}'(\tau') n_{i} n_{j}\bar\chi_{j}'(\tau)\Big) \lp\qquad \qquad \qquad\qquad \qquad\ \ \
+2\bar\chi_{i}'(\tau')\Big( \delta_{ij}-\frac{1}{N}n_{i}\bar n_{j}\Big)\chi_{j}'(\tau)\bigg], \la{4.26}
\ea
where we used \rf{4.21}. The explicit form of the 
cubic $S^{(3)} $ and quartic $ S^{(4)} $ terms in the action 
will be discussed  later. 
In momentum space representation
\be
\la{4.27}
\chi'(\tau) = \sum_{\ell\in\mathbb Z\backslash\{0\}} a(\ell)\, e^{i\ell\tau}, \qquad
\bar\chi'(\tau) = \sum_{\ell\in\mathbb Z\backslash\{0\}} \bar{a}(\ell)\,  e^{i\ell\tau}, 
\ee
$S_2$   in \rf{4.23}  becomes 
\ba
\la{4.28}
 i \int d\tau\, \bar\chi'\del_\tau\chi' = 2\pi\sum_{\ell\in\mathbb Z\backslash\{0\}}\ell\,\bar a(\ell)\,  a(-\ell).
\ea
Using  that\footnote{This follows, for instance,  from $\frac{1}{2}\log(1+b^{2}-2b \cos\theta) = -\sum^\infty_{n=1}\frac{b^{n}}{n}\cos(n\theta)$, after applying 
$(b\partial_{b})^2$ and setting $b=1$.} 
\be \sum^\infty_{\ell=1}(-\ell)\cos(\ell\tau) = \frac{1}{4\sin^{2}\frac{\tau}{2}}\ , \la{4.29}
\ee 
 we   have 
\ba
\la{4.30}
& \int \dtdt \ \bar\chi_{i}'(\tau')\chi_{j}'(\tau) = -2\pi^{2}\,\sum_{\ell=1}^{\infty}\ell\,\big[\bar a_{i}(\ell)\, a_{j}(-\ell)+\bar a_{i}(-\ell)\, a_{j}(\ell)\big ] = 
-2\pi^{2}\sum_{\ell\in\mathbb Z\backslash\{0\}}|\ell|\,\bar a_{i}(\ell)a_{j}(-\ell),
\ea
and  a similar expression for the integral of  two $\chi'$'s or two $\bar\chi'$'s.
The resulting quadratic part  \rf{4.26}
of the total action that determines the leading contribution at large $R$ is
\ba
\la{4.31}
S^{(2)} &= 2\pi\,\sum_{\ell\in\mathbb Z\backslash\{0\}}\bigg\{
\ell\,\bar a_{i}(\ell) a_{j}(-\ell) +\frac{i\pi\pxi}{2}\,|\ell|\,\bigg[
\bigg(1-\frac{1}{N}\bigg)\Big( \bar n_{i}\bar n_{j} a_{i}(\ell)a_{j}(-\ell)
+n_{i} n_{j} \bar a_{i}(\ell)\bar a_{j}(-\ell) \Big)\lp\qquad\qquad \ \ \ \ 
+2\bigg(\delta_{ij}-\frac{1}{N}n_{i}\bar n_{j}\bigg)\,\bar a_{i}(\ell) a_{j}(-\ell)
\bigg]
\bigg\} = 2\pi \sum_{\ell\in\mathbb Z\backslash\{0\}} \AA_{u}\, (\ell)\, Q_{uv}(\ell)\, \AA_{v}(-\ell),
\ea
where $\AA_{u} = (a_{1}, \dots, a_{N}, \bar a_{1}, \dots, \bar a_{N})$ and $Q_{uv}$ is the $2N\times 2N$ matrix
\be
\la{4.32}
Q(\ell) = \frac{1}{2}\ell\ 
\begin{pmatrix}
0 & -1 \\ 1 & 0
\end{pmatrix}
+\frac{i\pi\pxi}{2}\,|\ell|\,
\begin{pmatrix}
(1-\frac{1}{N}) n\otimes n & 1-\frac{1}{N}  n\otimes  \bar n  \\
1-\frac{1}{N} \bar n\otimes  n  & (1-\frac{1}{N}) \bar n\otimes \bar n 
\end{pmatrix} \ .
\ee
Using that  $\bar n n =1$ its determinant  evaluates to  
\be
\la{4.33}
\det Q(\ell) = \frac{1}{4^{N}}\,\ell^{2}\, \big(\ell^{2}+\pi^{2}\pxi^{2}|\ell|^{2}\big)^{N-1} = \frac{1}{4^{N}}\ell^{2N}(1+\pi^{2}\pxi^{2})^{N-1}.
\ee
A short-cut  way to  this  result  is to use the rotational symmetry of the problem 
implying that  determinant can only depend on length on $n_i$   which is 1 
and then  to choose this constant vector 
$n_i=(1, 0, ..., 0)$.  

Thus the integral over $\AA_u=(a_i, \bar a_i)$ gives 
\be
\la{4.34}
\prod_{\lze}
[\det Q(\ell)]^{-1/2} \ \propto\  (1+\pi^{2}\pxi^{2})^{\frac{N-1}{2}} \ , 
\ee
where we used that in the  $\zeta$ function regularization\foot{ 
This derivation of the 
  $(1+\pi^{2}\pxi^{2})^{\frac{N-1}{2}}$ prefactor  in $\W_k$  is formally  very similar to 
  the one of the Born-Infeld  factor in the disc   partition function of an open string in external 
   abelian gauge field \ci{Fradkin:1985qd}.}
\be \la{4.35} \prod_{\lze}
c = \prod_{\ell=1}^{\infty}c^{2} = \exp\big(\zeta(0)\log c^{2}\big) = c^{-1} \ . \ee
The 
$\pxi$-independent proportionality constant in \rf{4.34}   and the  normalization of the 
path integral measure can be accounted for  at the end 
  by  observing that 
for $\pxi=0$  the action \rf{4.16}  becomes free  and thus  the partition function should be given by 
  \rf{3.14}  (or its large $R$ expansion in \rf{4.1})  as discussed above. 

We thus  find  for the  ladder  Wilson loop expectation value 
\ba   & \W_k  = \dim \S_{k} \,  \big(1+\pi^{2}\pxi^{2}\big)^{N-1\ov 2}\, \big( 1 +
 \WS\big), \qquad
 \WS= \G_2 + \G_4 + ...\ , \ \ \   \G_{2n} = \mc O( R^{-2n})  \ , \la{4.36}\ea
or,  equivalently,\foot{In the case of $SU(2)$ equivalent result was announced earlier 
in \ci{Komar}.}   
\be \la{4.37}
\log \W_k  
{=} \log \dim \S_{k} + \frac{N-1}{2}\log\Big(1+\frac{\z^{4}\,g^4\, R^{4}}{64\pi^{2}}\Big) +  \G_2 + 
\mc O( R^{-4})  \ , 
\ee
where $ \WS$ stands for 
 subleading corrections at large $R$ and fixed $\pxi= \frac{\z^{2}\,g^2\, R^{2}}{8\pi^{2}}$.


Using that $R^2= k + {N \ov 2}$  and expanding in powers of $\z^2 g^2$  we     may 
compare  \rf{4.37}  with the   ladder part of the two-loop expression for the WL expectation value in \rf{1.16}. Since for $k$-symmetric representation of $SU(N)$  one has 
\be \la{437}
C_{\A}=N\ , \qquad \qquad C_{\R} = \frac{k(k+N)(N-1)}{2N}\ee
 so that 
$C_{\A}\, C_{\R}= \ha (N-1) ( k^2 + {N k})  $  and thus 
we find  the  agreement  with the  leading $\z^4 g^4$ term in the expansion of \rf{4.37} 
in  both  leading and subleading orders in large $k$  expansion.

\iffa
In terms of $k$, at the precision we are working, this reads
\be
\la{4.33}
\log \W_{k} \stackrel{k\gg 1}{=} \log \dim \S_{k} + \frac{N-1}{2}\log\Big(1+\frac{\z^{4}\,k^{2}\,(1+\frac{N}{k})\,g^{4}}{64\pi^{2}}+\mc O(k^{0})\Big).
\ee
and we can  compare this expression with  (\ref{1.15}). For the scalar loop we have only the $\z^{4}$ term of that expression, and using $C_{\A}=N$ and $C_{\R} = \frac{k(k+N)(N-1)}{2N}$ we fully agree
with the first term of the expansion of (\ref{4.33}) -- and for this term the $\mc O(k^{0})$ corrections are absent.
\fi

\subsubsection{Propagators for the $\chi', \bar\chi'$ fluctuation}
\la{sec:prop}

To develop  perturbation theory in $1/R^2$  starting with \rf{4.18},\rf{4.19}, i.e. 
to compute the effect of interaction terms  that  complement the quadratic part of the action \rf{4.31} we need to find the  propagators for  the corresponding fluctuation fields. 
\iffa To go beyond the leading order, we will need to compute the effect of the quadratic $1/R^{2}$ term  in (\ref{4.18}) and the additional $\mc O(R^{-2})$ interaction terms
from the expansion of the quartic action (\ref{4.19}). These corrections will be treated perturbatively starting from the quadratic action $S_{2}^{(0)}$ at order $\mc O(R^{0})$ in 
(\ref{4.26}), (\ref{4.31}). Here, we start discussing the associated propagators.
\fi
 Using the covariance with respect to the  rotation of the  constant 
 vector $n_i$ 
 in (\ref{4.17}),(\ref{4.21})  and of the fluctuation fields 
we  may  write the  quadratic action  (\ref{4.26}) 
in the special frame where  \be\la{4.38}
 n=\bar n = ( 0, \dots, 0,1) \ . \ee
Let us     label the components of  non-constant  fluctuation $\chi_{i}' $   as 
 \be
\la{4.39}
\chi_{i}' = (\eta_{1}, \dots, \eta_{N-1},\ph), \qquad\qquad
\bar\chi_{i}' = ( \bar\eta_{1}, \dots, \bar\eta_{N-1},\bar\ph),
\ee
 Then the  quadratic action (\ref{4.26}) reads  ($r=1, ..., N-1$)
\ba
\la{4.40}
S^{(2)} =&i \int d\tau\,\big(\bar \ph\del_\tau \ph+\,\bar \eta_r\del_\tau \eta_r\big) \lp\ 
 - i \, \pxi\,\int \dtdt \bigg[\Big(1-\frac{1}{N}\Big)\Big(
\ha \ph(\tau')\, \ph(\tau)+\ha \bar\ph(\tau') \, \bar\ph(\tau)+ \bar\ph(\tau')\, \ph(\tau)\Big)
+ \bar\eta_r(\tau')\ \eta_r(\tau)\bigg]. 
\ea
Going to momentum space,  inverting the $2\times 2$ matrix in the $\ph, \bar\ph$ sector
 and using \rf{4.28}, we find for the corresponding   propagators 
\ba
\la{4.41}
\D_{\ph\ph}(\tau-\tau') &= \vev{\ph(\tau)\ph(\tau')} = \vev{\bar\ph(\tau)\bar\ph(\tau')} 
= -\frac{N-1}{2N}\pxi \sum_{\lze}\frac{1}{|\ell|}\,e^{i\ell(\tau-\tau')}, \nonumber\\
\D_{\bar\ph \ph}(\tau-\tau') &= \vev{\bar\ph(\tau)\ph(\tau')} = \sum_{\lze}\bigg(\frac{i}{2\pi\ell}+\frac{N-1}{2N}\,\pxi\,\frac{1}{|\ell|}\bigg)\,e^{i\ell(\tau-\tau')},\nonumber\\
\D_{\eta\eta}(\tau-\tau') &= \frac{1}{2\pi}\sum_{\lze}\frac{i}{\ell+i\pi\pxi\,|\ell|}\,e^{i\ell(\tau-\tau')},\qquad \qquad 
 \vev{\bar\eta_{r}(\tau)\eta_{s}(\tau')}= \delta_{rs}\D_{\eta\eta}(\tau-\tau') .
\ea
Computing the sums  and restricting to the interval $0<\tau<2\pi$    the  propagators may be written explicitly as 
\ba
\la{4.42}
\D_{\ph\ph}(\tau) &= \vev{\ph(\tau)\ph(0)} = \vev{\bar\ph(\tau)\bar\ph(0)} = \pxi\, \frac{N-1}{2N}\log\Big(4\sin^{2}\frac{\tau}{2}\Big),\no \\
\D_{\bar\ph\ph}(\tau) &= \vev{\bar\ph(\tau)\ph(0)} = \frac{1}{2\pi}(\tau-\pi) -\pxi\, \frac{N-1}{2N}\, \log\Big(4\sin^{2}\frac{\tau}{2}\Big), \nonumber \\
\D_{\eta\eta}(\tau) &= \frac{1}{2\pi}\,\frac{1}{1+\pi^{2}\pxi^{2}}\left[\tau-\pi-\pi \pxi \log\Big(4\sin^{2}\frac{\tau}{2}\Big)\right] \ . 
\ea
Then they can be  extended to all $\tau$  by periodicity.  Note that 
 the linear in $\tau$ part is not
continuous at  $\tau=0$ where it has a jump.
 For the  corresponding propagators in momentum space   we then  have\footnote{
In all cases  $\vev{A(\tau)B(\tau')} = \sum_{\ell\neq 0}K_{\ell}\ e^{i\ell(\tau-\tau')}$. Hence
$
\vev{A_{p}B_{q}} = \frac{1}{(2\pi)^{2}}\int d\tau d\tau'
\sum_{\ell\neq 0}K_{\ell}\, e^{i\ell(\tau-\tau')}\, e^{-ip\tau-iq\tau'} $

\quad $
  = K_{p}\,\delta_{p+q,0}.
$
}
\ba
\la{4.43}
\vev{\ph_{p}\ph_{q}} &= \vev{\bar\ph_{p}\bar\ph_{q}} = -\pxi\frac{N-1}{2N}\,\frac{1}{|p|}\,\delta_{p+q,0}, \qquad \qquad 
\vev{\bar\ph_{p}\ph_{q}} = \bigg(\frac{i}{2\pi\,p}+\pxi\,\frac{N-1}{2N}\frac{1}{|p|}\bigg)\,\delta_{p+q,0}, \nonumber \\
\vev{\bar\eta_{r, p}\eta_{s, q}} &=  \delta_{rs}\,\frac{1}{2\pi}\frac{i}{p+i\pi\,\pxi\,|p|}\,\delta_{p+q,0}, \qquad r,s=1, \dots, N-1.
\ea
In the following, it will be convenient to use these propagators in the  more 
general case when $\bar n n =u$ (where $u$ is a positive constant)   and thus 
when  $n = \bar n = \sqrt{u}\,( 0, \dots, 0,1)$.
  Then (\ref{4.43})--(\ref{4.52}) generalize simply  by 
   the replacement $\pxi\to u\,\pxi$, \cf \rf{4.23},(\ref{4.25}).

\subsubsection{$1/R^{2}$ order:  logarithmic divergence and one-loop beta function}

The next  step is to compute the leading $1/R^2 $ term $\G_2$  in $\G$ in 
(\ref{4.37}). It is given by the sum of the three contributions
(which are effectively  two-loop ones from the path integral point of view)
\be
\la{4.44}
\WS_2 = \mathsf{D}+\Sigma_{4}+\Sigma_{3}  \ .
\ee
Here $\mathsf{D}$ is the contribution of the $1/R^2$ term in $S_2$ in \rf{4.18} or in the delta-function in \rf{4.20}. 
$\Sigma_4$  is the  contribution of the quartic interaction terms in \rf{4.19} or $S^{(4)}$ in \rf{4.25} 
and $\Sigma_{3}$   comes from the  contraction of two cubic $1/R$ vertices in $S^{(3)}$  in \rf{4.25}. 
We will  focus on the logarithmic UV  divergent part of \rf{4.44}. 
Its renormalization 
  will determine  the leading  one-loop $1/R^{2}$  term 
in the 
beta function for $\pxi$.   

\paragraph{D-term}

The $\mathsf D$-contribution in (\ref{4.44})    comes from the delta-function constraint in \rf{4.20}.
 Expanding this delta-function  in $1/R^2$ gives 
\ba
& \delta\big(\bar n n-1+ M \big) = \delta(\bar n n-1)+\delta'(\bar n n-1) M +...
= \delta(\bar n n-1)-\frac{\partial}{\partial u}\delta(\bar n n-u)\, M\bigg|_{u=1}+..., \la{4.45} \\
&M\equiv  \frac{1}{2\pi R^{2}}\int d\tau\, \bar\chi'\chi'
= \frac{1}{R^{2}}\sum_{i=1}^{N}\sum_{\lze}\bar \chi_{i}'(\ell)\chi_{i}'(-\ell)\ , \la{4.46}
\ea
where we introduced an auxiliary parameter $u$. 
Then in the subleading term  the integration in \rf{4.20}  is done with the constraint $\bar n n =u$ 
with $u$ set to 1  at the end. We get using \rf{4.39}--\rf{4.41} 
\ba
& \sum_{i=1}^{N}\sum_{\lze}\vev{\bar \chi_{i}'(\ell)\chi_{i}'(-\ell)}_{_{\bar n n=u}} = \sum_{\lze}\Big[ \vev{\bar\ph_{\ell}\ph_{-\ell}}_{_{\bar n n=u}} 
+\sum_{r=1}^{N-1}\vev{\bar\eta_{r, \ell}\eta_{r,\ell}}_{_{\bar n n=u}} \Big]  \la{4.47} \\ &
\qquad = \sum_{\lze}\bigg[\frac{i}{2\pi\,\ell}+u\,\pxi\,\frac{N-1}{2N\, |\ell|}+\frac{i(N-1)}{2 \pi(\ell+i\pi\,u\,\pxi\,|\ell|)}\bigg]
= \frac{u \,  \pxi (N-1)  (N+1+\pi ^2 u^2 \pxi ^2)}{N (1+\pi ^2 u^2 
\pxi ^2) }\, I_0 \ , \no \\
&\qquad I_0 \equiv  \sum_{\ell=1}^{\infty}\frac{e^{-\eps\,\ell}}{\ell} = -\log\eps+\mc O(\eps), \la{4.48}
\ea
where we introduced an exponential cut-off in the sum over $\ell$.
Then  the contribution of the correction  term in (\ref{4.45})  is found to be 
\ba
\la{4.49}
\mathsf D &= -\frac{1}{R^{2}} I_0\, \frac{1}{(1+\pi^{2}\pxi^{2})^{\frac{N-1}{2}}}\frac{\partial}{\partial u}\bigg[u^{N-1}(1+u^{2}\pi^{2}\pxi^{2})^{\frac{N-1}{2}}\,\frac{u \,  \pxi  \, (N-1) 
(N+1+\pi ^2 u^2 \pxi ^2)}{N (1+\pi ^2 u^2 
\pxi ^2) }\bigg]\Big|_{u=1} \,.
\ea
Here in the square bracket  we included the factor of $u^{N-1}$ from the integral over $n$, i.e. 
$\int dn d \bar n\,  \delta( \bar n n - u)$, 
 and the leading  $\pxi$ dependent factor  in  (\ref{4.34}),\rf{4.36} 
  generalized to the present case of  $u\neq 1$.
  Computing the derivative over $u$ at $u=1$   we finish with
\be
\la{4.50}
\mathsf D  = -\frac{1}{R^{2}}I_0\,  \frac{(N-1)\, \pxi}{N (1+\pi ^2 \pxi ^2)^2}\, 
\Big[N(N+1)+\pi^{2}\pxi^{2}(2N^{2}-1)+\pi^{4}\pxi^{4}(2N-1)\Big]
\ . 
\ee

\paragraph{Quartic terms}

The contribution $\Sigma_{4}$ in (\ref{4.44}) comes from the quartic terms in (\ref{4.19})
 after expanding   $e^{i S}$ 
\be
\la{4.51}
\Sigma_{4} =    i \langle \SS \rangle\ , \qquad  \qquad 
\SS= -i \frac{\pxi}{R^{2}}\int\dtdt\,[\bar\chi'(\tau) T^{a} \chi'(\tau)]\  [\bar\chi'(\tau') T^{a} \chi'(\tau')].
\ee 
Using again the $SU(N)$ fusion relation \rf{4.24}, i.e. 
\be
\la{4.52}
\bar \chi'(\tau) T^{a}\chi'(\tau)\,
\bar \chi'(\tau') T^{a}\chi'(\tau') = \frac{1}{2}[\bar \chi'(\tau')\chi'(\tau)][\bar \chi'(\tau)\chi'(\tau')]-\frac{1}{2N}[\bar \chi'(\tau)\chi'(\tau)][\bar \chi'(\tau')\chi'(\tau')],
\ee
we obtain 
\ba
\la{4.53}
\Sigma_{4} = &\frac{1}{R^{2}}\frac{N-1}{2N}\,\pxi\,\int \dtdt \bigg[
[\D_{\bar\ph\ph}(0)]^2+\D_{\bar\ph\ph}(\tau -\tau ') \D_{\bar\ph\ph}(-\tau +\tau ')-2 \D_{\bar\ph\ph}(0) \D_{\eta\eta}(0)+[\D_{\eta\eta}(0)]^2\lp
+N \D_{\bar\ph\ph}(-\tau +\tau ') \D_{\eta\eta}(\tau -\tau ')+N 
\D_{\bar\ph\ph}(\tau -\tau ') \D_{\eta\eta}(-\tau +\tau ')-\D_{\eta\eta}(\tau -\tau ') \D_{\eta\eta}(-\tau +\tau ')\lp
-N \D_{\eta\eta}(\tau -\tau ') 
\D_{\eta\eta}(-\tau +\tau ')+N^2 \D_{\eta\eta}(\tau -\tau ') \D_{\eta\eta}(-\tau +\tau ')+[\D_{\ph\ph}(\tau -\tau ')]^2
\bigg].
\ea
Ignoring  constant divergent terms (that drop out after  integrating with measure $\dtdt$, \cf (\ref{4.23})) and using the symmetry under $\tau\leftrightarrow \tau'$, we get 
\ba
\Sigma_{4} = &\frac{1}{R^{2}}\frac{N-1}{2N}\,\pxi\,\int \dtdt \bigg[
\D_{\bar\ph\ph}(\tau -\tau ') \D_{\bar\ph\ph}(-\tau +\tau ')
+2N \D_{\bar\ph\ph}(-\tau +\tau ') \D_{\eta\eta}(\tau -\tau ')\no \\ &\qquad \qquad +(N^{2}-N-1)\D_{\eta\eta}(\tau -\tau ') \D_{\eta\eta}(-\tau +\tau ')
+[\D_{\ph\ph}(\tau -\tau ')]^2
\bigg]. \la{4.54}
\ea
Using the translation invariance gives
\ba
\la{4.55}
\Sigma_{4}=& \frac{2\pi}{R^{2}}\frac{N-1}{2N}\,\pxi\,\int^{2\pi}_0\frac{d\tau}{4\sin^{2}\frac{\tau}{2}}  \bigg[
\D_{\bar\ph\ph}(\tau) \D_{\bar\ph\ph}(2\pi-\tau)+2N \D_{\bar\ph\ph}(\tau) \D_{\eta\eta}(2\pi-\tau)\lp
\qquad \qquad +(N^{2}-N-1)\D_{\eta\eta}(\tau) \D_{\eta\eta}(2\pi-\tau)+[\D_{\ph\ph}(\tau)]^2
\bigg].
\ea
The propagators $\D$ in \rf{4.42}  have a linear part $\sim \tau-\pi$ and a log part $\sim \log(4\sin^{2}\frac{\tau}{2})$. Due to parity around $\tau=\pi$
there cannot be crossed contributions. The logarithmic divergences may come only 
 the linear in $\tau$ terms.\foot{The contributions of   purely logarithmic terms in $\D$ 
 are finite. This may be  easily  shown  in dimensional regularization. 
In mode regularization, where we add  a factor $\exp(-\eps\,\ell )$ to the $\ell$-th Fourier mode \cite{Beccaria:2017rbe},  this is also true up to a power-like 
divergence $\frac{\log\eps}{\eps}$. 
Indeed, using $\log(4\sin^{2}\frac{\tau}{2}) \to  -2\sum_{n=1}^{\infty}\frac{1}{n}e^{-\eps\, n}
\, \cos(n\tau)$, we have
\ba
\int_{0}^{2\pi} & \frac{d\tau}{4\sin^{2}\frac{\tau}{2}}[\log4(\sin^{2}\frac{\tau}{2})]^{2} = 4\sum_{n,p,q=1}^{\infty}e^{-(n+p+q)\eps}\frac{-n}{pq}
\int_{0}^{2\pi}d\tau\cos(n\tau)\cos(p\tau)\cos(q\tau)
=4\bigg[-\frac{\pi}{2}\sum_{p,q=1}^{\infty}\frac{p+q}{pq}e^{-2(p+q)\eps}\lp -2\times \frac{\pi}{2}\sum_{1\le q<p<\infty}e^{-2p\eps}\frac{p-q}{pq}
\bigg] 
= 4\pi\frac{1+(1+e^{2\eps})\log(1-e^{-2\eps})}{-1+e^{2\eps}} = \frac{2\pi}{\eps}(1+2\log 2+2\log \eps)-6\pi+\mc O(\eps).\notag
\ea
}
The linear in $\tau$ parts are 
\ba
\la{4.56}
\D^{\rm lin}_{\eta\eta}(\tau) = \frac{1}{2\pi}\frac{1}{1+\pi^{2}\pxi^{2}}(\tau-\pi),\qquad
\D^{\rm lin}_{\ph\ph}(\tau) = 0,\qquad
\D^{\rm lin}_{\bar\ph\ph}(\tau) = \frac{1}{2\pi}(\tau-\pi).
\ea
Using \rf{4.29}   with  mode regularization, i.e. 
$\frac{1}{4\sin^{2}\frac{\tau}{2}} \to \sum^\infty_{\ell=1}e^{-\eps\,\ell} (-\ell)\cos(\ell\tau) $
  we have (cf. \rf{4.48}) 
\ba
\la{4.57}
\int_{0}^{2\pi}\frac{d\tau}{4\sin^{2}\frac{\tau}{2}} = 0, \qquad 
\int_{0}^{2\pi}\frac{d\tau}{4\sin^{2}\frac{\tau}{2}}\, \tau = 0, \qquad 
\int_{0}^{2\pi}\frac{d\tau}{4\sin^{2}\frac{\tau}{2}}\,\tau^{2} = -4\pi\,\sum_{\ell=1}^{\infty}\frac{e^{-\eps\,\ell}}{\ell }=- 4 \pi I_0, 
\ea
and then finally the  logarithmically divergent part of $\Sigma_4$ 
is found to be
\be
\la{4.58}
\Sigma_{4}^{\rm UV} = \frac{1}{R^{2}} I_0 \,\frac{(N-1)\, \pxi}{N\,  (1+\pi^{2}\pxi^{2})^{2}}\,\big[
N(N+1)+2\pi^{2}\pxi^{2}(N+1)+\pi^{4}\pxi^{4}\big]\ . 
\ee
Combined with \rf{4.50} this  gives 
\ba
\la{4.59}
\mathsf{D}+\Sigma_{4}^{\rm UV} = -\frac{\pi^{2}}{R^{2}} I_0 \,\frac{(N-1)\, \pxi^{3}}{N\, (1+\pi^{2}\pxi^{2})^{2}}\big[
-3-2N+2N^{2}+2\pi^{2}\pxi^{2}(N-1)\big]\ . 
\ea

\paragraph{Cubic terms}

The term $\Sigma_{3}$ in (\ref{4.44}) is coming from contraction of two   cubic  vertices 
vertices $S^{(3)}$ in the action \rf{4.19},\rf{4.25} (i.e.  from the quadratic term 
in the expansion of $e^{i S_4}$). 
Explicitly, expanding $S_4$  in \rf{4.19} near $n=\bar n=(1,0,...,0)$  using   \rf{4.17},\rf{4.39} 
gives 
\ba
\la{4.60}
 S_4= &-i\pxi R^{2}\int \dtdt\bigg(\frac{1}{2}[\bar \chi(\tau')\chi(\tau)][\bar \chi(\tau)\chi(\tau')]-\frac{1}{2N}[\bar \chi(\tau)\chi(\tau)][\bar \chi(\tau')\chi(\tau')]\bigg)\longrightarrow \lp
-\frac{i\pxi}{R}\int \dtdt\bigg[
\frac{1}{2}(\bar \ph(\tau')+\ph(\tau))[\bar \chi'(\tau)\chi'(\tau')]
+\frac{1}{2}[\bar \chi'(\tau')\chi'(\tau)](\bar \ph(\tau)+\ph(\tau'))\lp\qquad \qquad 
-\frac{1}{2N}(\bar \ph(\tau)+\ph(\tau))[\bar \chi'(\tau')\chi'(\tau')]
-\frac{1}{2N}[\bar \chi'(\tau)\chi'(\tau)](\bar \ph(\tau')+\ph(\tau'))\bigg] \lp
= -\frac{i\pxi}{R}\int \dtdt\bigg[
(\bar \ph(\tau')+\ph(\tau))[\bar \chi'(\tau)\chi'(\tau')]
 -\frac{1}{N}(\ph(\tau)+\bar \ph(\tau))[\bar \chi'(\tau')\chi'(\tau')]\bigg] .\ea
 Thus (using  that  $\bar\chi' n = \bar\ph$, {\em etc.})
 \ba S^{(3)} 
 =  -\frac{i\pxi}{R}\int \dtdt\bigg[ &
(\bar \ph(\tau')+\ph(\tau))[\bar \ph(\tau)\ph(\tau')+\bar \eta(\tau)\eta(\tau')]\no \\ &
 -\frac{1}{N}(\ph(\tau)+\bar \ph(\tau))[\bar \ph(\tau')\ph(\tau')+\bar \eta(\tau')\eta(\tau')]
\bigg] \no\\
 = -\frac{i\pxi}{R}\int \dtdt\bigg[ &\Big(1-\frac{1}{N}\Big)\ph(\tau)\bar\ph(\tau)[\ph(\tau')+\bar\ph(\tau')]
+[\bar \ph(\tau')+\ph(\tau)] \, \bar \eta(\tau)\eta(\tau')\no \\ &
 -\frac{1}{N}[\ph(\tau)+\bar \ph(\tau)]\,\bar \eta(\tau')\eta(\tau')
\bigg].\la{4.61}
\ea
For
three generic  non-constant functions of $\tau$ we have the following expression in terms of their Fourier modes 
\be
\la{4.62}
\int \dtdt A(\tau) B(\tau) C(\tau') = 
 -2\pi^{2}\mathop{\sum_{\ell,p\in \Zzero}}_{
 \ell\neq p}|n|\,A_{p}B_{\ell-p}C_{-\ell}.
\ee
Hence, introducing the mode regularization  factor we get 
\ba
\la{4.63}
S^{(3)} &= -\frac{2\pi^{2}i\pxi}{R}\mathop{\sum_{\ell,p\in \Zzero}}_{\ell\neq p}|\ell|\,e^{-\eps\,|\ell|}\,
\bigg[
\Big(1-\frac{1}{N}\Big)\ph_{p}\bar\ph_{\ell-p}(\ph_{-\ell}+\bar\ph_{-\ell})
+\bar\ph_{p}\eta_{\ell-p}\bar\eta_{-\ell}\lp\qquad \qquad \qquad \qquad
+\ph_{p}\bar\eta_{\ell-p}\eta_{-\ell}-\frac{1}{N}\bar\eta_{p}\eta_{\ell-p}(\ph_{-\ell}+\bar\ph_{-\ell})\bigg].
\ea
Taking  the expectation value of $[S^{(3)}]^2$ 
using the momentum space propagators in  \cf (\ref{4.43}) 
we obtain a sum of triple products of propagators,
that after the integration can be  
 reduced  to a set of double sums. Regulating all infinite sums with an exponential mode cutoff  and 
dropping  power-like  singular terms of the form $1\ov \eps$ or $\log\eps\ov \eps$, we find for the UV logarithmically  divergent  part
\be
\la{4.64}
\Sigma_{3}^{\rm UV} = \frac{1}{2}\vev{\big[i\, S^{(3)}\big]^{2}} = \frac{\pi^{2}}{R^{2}} \log\eps\, \frac{(N-1)\, \pxi^{3}}{N\, (1+\pi^{2}\pxi^{2})^{2}}\big[2N+3- 2\pi^{2}\pxi^{2}(N-1)\big]\ .
\ee
Summing this up with \rf{4.59}  gives the 
total log divergence at order $1/R^{2}$  
\be
\la{4.65}
\G_2=\mathsf{D}+\Sigma^{\rm UV}_{4}+\Sigma_{3}^{\rm UV} = \frac{2\pi^{2}}{R^{2}}\,N(N-1)\,\frac{\pxi^{3}}{(1+\pi^{2}\pxi^{2})^{2}}\,\log\eps+\cdots.
\ee

\paragraph{Log divergence and beta-function}

Using  \rf{4.37} the  ladder Wilson loop expectation  value   is thus
\ba
\la{4.66}
\log \W_{k} &= \log \dim \S_{k}+\frac{N-1}{2}\log(1+\pi^{2}\pxi^{2})+\G_2 + \mc O(R^{-4}) \lp
= \log \dim \S_{k}+\frac{N-1}{2}\log(1+\pi^{2}\pxi^{2})+ \frac{2\pi^{2}}{R^{2}}\,N(N-1)\,\frac{\pxi^{3}}{(1+\pi^{2}\pxi^{2})^{2}}\,\log\eps+ \cdots 
\ea
where dots stand for finite parts and higher $R^{-4}$  corrections. 

The divergence in \rf{4.66} can be absorbed into renormalization of $\pxi$ 
(which is equivalent to renormalization of $\z$ as this is the only running coupling, cf. \rf{2.19})
\ba\la{4.67}
\pxi \equiv \pxi_{\rm bare} \to \pxi(\mu) - \frac{2N}{R^{2}}\frac{\pxi^{2}(\mu) }{ 1+\pi^{2}\pxi^{2}(\mu)}\log (\mu \eps) 
+\mc O(R^{-4}), 
\ea
so that  the renormalized $\W_k$   expressed in terms of renormalized $\pxi(\m) $ (cf. \rf{444}) satisfies 
\ba
\la{4.68} &
\bigg(\mu\frac{\partial}{\partial\mu}+\bxi \frac{\partial}{\partial\pxi}\bigg)\, \W_k = 0,\\
\la{4.69}
&\bxi =\mu  {d\pxi \ov d  \mu } 
 = \frac{2N}{R^{2}}\frac{\pxi^{2}}{ 1+\pi^{2}\pxi^{2} }+\mc O(R^{-4})\ . 
\ea
The  corresponding 
$\pxi^3 \log \mu \sim  \z^6 g^6 \log\mu$ term in renormalized $\log \W_k$ 
  is in agreement with  the   $\z^6 g^6 \log \mu $ term in \rf{1.18},\rf{1.19}: 
\be
\la{4.70}
-\frac{1}{128\pi^{4}}\, C_{\R}C_{\A}^{2}\,\z^{6}g^{6} = -\frac{2\pi^{2}}{k}N(N-1)\,\pxi^{3}+\cdots.
\ee
Here  we used \rf{437}, \rf{2.28} and expanded at large $k$
with fixed $\pxi= {1\ov 8 \pi^2} \z^2 g^2  R^2, \  R^2=k + \ha N $. 
 The beta-function \rf{4.69}  written in terms of $\z$   gives 
 \be \la{4.71}
 \bxi\ \to \ \beta^{\ladd}_{\z}= \mu  {d\z \ov d  \mu }  
 = \frac{ \z^3 N g^2      }{8\pi^2  \big(1+ {1\ov 64 \pi^2}   \z^4 g^4 R^4 \big)}
 +\mc O(R^{-4})\ . 
\ee
Expanding in small $\z$ 
 the  leading $\z^3$ term here is in agreement  with the one-loop beta-function in \rf{1.4}
 (in the  large $N$  limit  
  $\l= N g^2 $).
 The first correction from the  denominator in \rf{4.71}  comes only at order $\z^7 \l^3$. 

We also  get the following analog of the relation \rf{1.5}
\be 
 \la{4.72}
 {\partial \ov \partial \pxi } \log   \W_k  =   \bar \C \, \bxi  \ ,\qquad \qquad 
 \bar \C= {(N-1)\pi^2 R^2\ov 2 N \pxi}+ ... \ , 
 \ee
where  the leading one-loop term \rf{4.69} in $\bxi $  comes directly from 
the leading  finite one-loop term  $\frac{N-1}{2}\log(1+\pi^{2}\pxi^{2})$ in \rf{4.66}.\foot{An apparent singularity of $\bar \C$ in $\pxi$ is just an artifact of the expression of \rf{4.72} in terms of $\pxi \sim \z^2$ 
rather than $\sqrt {\pxi} \sim \z$.}

\section{$\beta_\pxi$-function   from  two-point  correlator  on  
Wilson line}
\la{sec:beta-line}

The Wilson line   may be viewed   as  defining a 
defect  1d CFT  with basic correlation functions of local operators inserted on the line 
 defined as (here $\cW=\exp [\z \int d \tau \phi] $ is the  scalar Wilson factor)
\be
\la{5.1}
\llangle \mc O_{1}(\tau_{1})\cdots \mc O_{n}(\tau_{n})\rrangle \equiv
\frac{\vev{\tr\big[\text{P}\, \mc O_{1}(\tau_{1})\cdots \mc O_{n}(\tau_{n})\,  \cW
 \big]}}{\vev{\tr\big[\text{P}\,   \cW \big]}} \ . 
\ee
 If we consider the scalar ladder model 
as a subsector of 
 $\N=4$ SYM  then for the  two-point function of a ``transverse'' scalar $\phi_{\perp}$ 
not coupled to the loop (i.e. not appearing in the  Wilson factor  $\cW$)
there is no genuine anomalous dimension,   \ie  all divergences in 
\be
\la{5.2}
G_{\perp}(\tau_{12}) = \pxi\, \llangle \phi_{\perp}(\tau_1)\ \phi_{\perp}(\tau_2)\rrangle, 
\ee
can be absorbed into $\z$ or   $\pxi$ only (i.e. no extra $Z$ factor is needed). Thus, 
the renormalized  two-point function should  satisfy
\be
\la{5.3}
\Big(\mu\frac{\partial}{\partial\mu}  + \bxi \frac{\partial}{\partial\pxi}\Big) G_{\perp}^{\rm ren} =0\ . 
\ee
In this section we discuss how we can use this relation to extract 
the beta-function $\bxi$ from the two-point function $G_{\perp}^{\rm ren}$.

This  way of deriving  $\bxi$    has several advantages.
First, the propagator on the line
is simpler than on the circle, \cf (\ref{1.12}). Second, there are
 no  constant zero modes on the line and
 thus  it will be possible to treat the delta-function  constraint  $\bar \chi \chi=R^2$ 
 (\cf (\ref{3.7}) in the free case)
by  solving it directly. 

Our starting point   will be 
  the bosonic 1d action on the line   (\cf (\ref{4.13}); we rescaled $\chi$ by $R$)
\be
\la{5.4}
S = i R^2 \,\int d\tau\,\bar \chi\del_\tau \chi
- i \pxi  R^{2}  \int \frac{d\tau\,d\tau' }{(\tau-\tau')^{2}}\,\bar \chi(\tau) T^{a}\chi(\tau)\,
\bar \chi(\tau') T^{a}\chi(\tau')\ , \qquad \qquad \bar\chi\chi=1 \ . 
\ee
This  action has the same  local $U(1)$  invariance as in \rf{3.9}\foot{Due to the constraint, the kinetic term is invariant up to an 
irrelevant total derivative.}  
\be
\la{5.5}
\chi_{i}\to e^{i\alpha(\tau)}\chi_{i}, \qquad\qquad 
\bar\chi_{i}\to e^{-i\alpha(\tau)}\bar\chi_{i}\ . 
\ee
We can use  this symmetry to gauge fix $\chi_{N}$  to be real;  
then solving the constraint we get 
\be
\la{5.6}
\chi_{N}=\bar\chi_{N}=\big(1 - \bar \chn_r \chn_r 
\big)^{1/2}, \ \ \  \qquad \   \chn_r \equiv  (\chi_1, ..., \chi_{N-1} )\ . 
\ee
In the following, we shall  use the notation $\chn$ for the  $N-1$  independent 
components $\chi_r$. 
 The kinetic term in \rf{5.4}   becomes  simply $\bar \chn_r\del_\tau \chn_r$
(since $\chi_N$ is real, it  contributes only a total derivative). 

The two-point function (\ref{5.2}) may be written as 
\be
\la{5.7}
G_{\perp}(\tau_{12}) = \pxi\ \llangle [\bar\chi \phi_{\perp} \chi (\tau_1)]\  [\bar\chi \phi_{\perp} \chi (\tau_2)] \rrangle,
\ee
where the indices of  the adjoint scalar $\phi_\perp  = \phi^{a}_\perp T^{a}$  
are contracted with the 1d bosons.
The average is done 
with the effective action (\ref{5.4}) (already incorporating 
the effect of the integral  over free coupled scalar) and with  the free 
scalar bulk action \rf{1.11} 
 for $\phi_{\perp}$.\foot{
One may view \rf{5.7}  as originating from  the generating functional 
with the coupling $\z \phi +  h(\tau) \phi_\perp$  and then differentiating twice over the source function 
$h(\tau)$.}
Computing first the expectation value with respect to the bulk field $\phi_{\perp}$  one gets 
($\tau_{12}=\tau_1 - \tau_2$)
\be
\la{5.8}
G_\perp (\tau_{12}) 
= \frac{\pxi\,g^{2}}{8\pi^{2}\tau_{12}^{2}}  \langle\Big( [\bar\chi(\tau_{2})\chi(\tau_{1})][ \bar\chi(\tau_{1})\chi(\tau_{2})] - {1\ov N}\Big) \rangle,
\ee
where $\langle...\rangle$   is the remaining averaging over 1d bosons $\chi$. 

\subsection{One-loop $1/R^2$   contribution}

Then writing this in terms of independent $N-1$ components 
$\chn_r= (\chi_1, ..., \chi_{N-1})$ in \rf{5.6}
 we get 
\ba
\la{5.9}
G_{\perp}(\tau_{12})&=\frac{\pxi \, g^{2}}{8\pi^{2}\tau_{12}^{2}}\,\vev{\Big[1-\frac{1}{N}+\bar\chn (\tau_{2})\chn (\tau_{1}) + \bar\chn (\tau_{1})\chn (\tau_{2}) -\bar\chn (\tau_{1})\chn (\tau_{1})
-\bar\chn (\tau_{2})\chn (\tau_{2})  +\mc O(\chn^{4}) \Big]}\lp
=\Big(1-\frac{1}{N}\Big)\frac{\pxi\, g^{2}}{8\pi^{2}\tau_{12}^{2}}\bigg[1+ {N\ov R^2} \,\big [\D(\tau_{12})+\D(-\tau_{12})-2\D(0)\big]+\mc O(\vev{\chn^{4}})\bigg] \ . 
\ea
Here $\D$ is the  infinite line  analog of the exact propagator \rf{4.41} on the circle
that is found from  the action \rf{5.4}   after using \rf{5.6}\foot{We explicitly extracted  the  
$1\ov R^2$ prefactor  which  is due to  the normalization of $\chi_r\equiv \chn_r $ in \rf{5.4}; 
$\langle \bar \chn_r (\tau)\chn_s(0) \rangle=  \delta_{rs} \D(\tau)$.}
\ba
\la{5.10}
\D(\tau) = \int_{-\infty}^{\infty} \frac{dp}{2\pi} \frac{i}{p+i\,\pi\,\pxi\,|p|}\,e^{ip\tau} 
= \frac{1}{\pi(1+\pi^{2}\pxi^{2})}\, \int_{0}^{\infty}\frac{dp}{p}\, \big[\pi\,\pxi\,\cos(p\tau)-\sin(p\tau)\big]\ . 
\ea
While $\D(\tau)$  is singular in the IR  (at $p=0$) 
the  combination   appearing   in (\ref{5.9}) 
\be
\la{5.11}
\D(\tau)+\D(-\tau)-2\D(0) = \frac{2\pxi}{1+\pi^{2}\pxi^{2}}\, \int_{0}^{\infty}dp\ \frac{\cos(p\tau)-1}{p}
\ee
 is regular at $p=0$. Its UV divergence 
   at $p\to \infty$  can be regularized   with a  hard cutoff $|p|<\Lambda$:
\be
\la{5.12}
\int_{0}^{\Lambda}dp\ \frac{\cos(p\tau)-1}{p} = -\log(\bar\Lambda\tau) +\mc O(\bar\Lambda^{-1}),\qquad \bar\Lambda=\Lambda\,e^{\gamma_{\rm E}}.
\ee
This is equivalent to mode regularization $e^{-\eps \, \ell}$ if used in \rf{4.41} 
after identifying $\eps = \bar\Lambda^{-1}$ (cf. also \rf{4.48}), i.e. 
\be
\la{5.13}
\int_{0}^{\infty}dp \frac{\cos(p\tau)-1}{p}e^{-p/\bar\Lambda} = -\frac{1}{2}\log\Big(1+\bar\Lambda^{2}\tau^{2}\Big) = -\log(\bar\Lambda\,\tau) +\mc O(\bar\Lambda^{-1})\ . 
\ee
Thus we find for the log divergent part of  \rf{5.9} (we assume $\tau >0$)\foot{ 
Note that  here 
 the UV scale $\bar\Lambda$ enters only  together with  $\tau$ 
   so that there are no IR divergences.  
   Thus we  can safely take the limit of the  infinite length of the  line
    as in the  similar computations in \cite{Beccaria:2021rmj}.}

\be
\la{5.14}
G_{\perp}(\tau)= \Big(1-\frac{1}{N}\Big)\,\frac{\pxi\, g^{2}}{8\pi^{2}\tau_{12}^{2}}\left[ 1  - \frac{1}{R^{2}} 
f^{(1)}_{1}(\pxi) \log(\bar\Lambda\, \tau) + \cdots\right], \qquad 
f^{(1)}_{1}(\pxi) =\frac{2N \pxi}{1+ \pi^2 \pxi^2} \ . 
\ee
Then  renormalizing $\pxi$ as in \rf{4.67}   we find that the renormalized $G_{\perp}(\tau)$ satisfies 
the CS equation \rf{5.3} with \be
\la{5.15}
\bxi =  \frac{1}{R^2}\,\pxi\, f^{(1)}_{1}(\pxi)  + \mc O\big(\frac{1}{R^4}\big) =   \frac{2N}{R^2}\frac{\pxi^2}{1+ \pi^2 \pxi^2}+   \mc O\big(\frac{1}{R^4}\big)  \ , 
\ee
which is the same as in \rf{4.69}.

\subsection{Subleading $1/R^{4}$ contribution}

At the next order   we expect 
to find   the following $1/R^{4}$  corrections  in   (\ref{5.14})  
\ba
\la{5.17}
G_{\perp}(\tau)=& \Big(1-\frac{1}{N}\Big)\,\frac{\pxi\, g^{2}}{8\pi^{2}\tau_{12}^{2}}\,\bigg[ 1  - \frac{1}{R^{2}}\,f^{(1)}_{1}(\pxi)\,\log(\bar\Lambda\tau)\lp\qquad \qquad\qquad \qquad
  -   \frac{1}{R^4} \bigg( f^{(2)}_{0}(\pxi)+f^{(2)}_{1}(\pxi)\,\log(\bar\Lambda\tau)+f^{(2)}_{2}(\pxi)\,\log^{2}(\bar\Lambda\tau)\bigg)+\cdots\bigg]\ . 
\ea
Assuming  renormalizability  or using the  CS equation (\ref{5.3})   we  have (prime is derivative over $\pxi$)
\ba
\la{5.18}
&f^{(2)}_{2} = -\frac{1}{2}f^{(1)}_{1}\,\big[f^{(1)}_{1}+\pxi (f^{(1)}_{1})'\big] = - 4 N^2\,\frac{\pxi^2}{(1+\pi^2 \pxi^2)^3} \ , \\
&\la{5.19}
\bxi = \frac{1}{R^2}\,  \pxi\,f^{(1)}_{1} + \frac{1}{R^4}\, \pxi\,f^{(2)}_{1} + \mc 
O\big( {1\ov R^6}\big) \ , 
\ea
where in \rf{5.18} we used the one-loop expression in  (\ref{5.14}).

In \rf{5.17}  we assumed that all IR divergences cancel, i.e. the UV cutoff enters together with $\tau $.  Thus to check \rf{5.18}   and to 
find the two-loop coefficient  $f^{(2)}_{1}$  we may   concentrate on extracting  the 
${1\ov R^4} \log \tau$ terms.

To compute corrections to \rf{5.9}   we note that in general they   come from the following 
expectation value   computed with the effective propagator $\D$ in \rf{5.10} 
 (we again use $\chn \equiv (\chi_r)$, \, $r=1, ..., N-1$)
\ba
\la{5.28}
X = \langle \bigg[(1-\bar\chn (\tau_{2})\chn (\tau_{2}))^{1/2}(1-\bar\chn (\tau_{1})\chn (\tau_{1}))^{1/2}+\bar\chn (\tau_{2})\chn (\tau_{1})\bigg]\bigg[\tau_{1}\leftrightarrow\tau_{2}\bigg]\, e^{i S_{\rm int}}\rangle , 
\ea
where $S_{\rm int}$   contains  interacting (higher than quadratic in $\chn $)  parts 
of the quartic part of the action in \rf{5.4}   after one eliminates $\chi_N$ using \rf{5.6}. 
The relevant  quartic interaction term in $S_{\rm int}$  is given by 
\ba
\la{5.29}
{S}^{(4)} _{\rm int} = - \ha i \pxi R^2   \int \frac{d\tau d\tau'}{(\tau-\tau')^{2}}
\bigg\{
& [\bar\chn (\tau)\chn (\tau')][\bar\chn (\tau')\chn (\tau)]
-[\bar\chn (\tau)\chi(\tau')][\bar\chn (\tau)\chn (\tau)]\lp
-[\bar\chn (\tau')\chi(\tau)][\bar\chn (\tau)\chn (\tau)]
+[\bar\chn (\tau')\chi(\tau')][\bar\chn (\tau)\chn (\tau)]
\bigg\}.
\ea
In general 
\be \la{5.30}
X= X_1 + X_2 + X_3 + ...\ , \ee
  where $X_n$ is given by  sums of   products of $n$ propagators 
$\D$. 
The $1/R^4$  correction  will come from  either 
 doing   contractions of four  $\chn $  in the prefactor in 
\rf{5.28} between themselves ($X_2$ term) 
or  from contractions of  two   $\chn $ 
  with one power of ${S}^{(4)} _{\rm int} $   from  the expansion of
 $e^{i S_{\rm int}}$ ($X_3$ term).  We do not need to include disconnected  contractions 
 as they cancel against the  contributions of the normalization factor  in \rf{5.1}.
 
 We thus  find 
 \ba
\la{5.26}
G_{\perp}(\tau_{12})&
=\Big(1-\frac{1}{N}\Big)\frac{\pxi\, g^{2}}{8\pi^{2}\tau_{12}^{2}}\bigg[1
+\frac{N}{N-1} (X_1+ X_2 + X_3 ) +   \mc O \big({1\ov R^6}\big)  
\bigg],\\ \la{526}
X_1 &=  \frac{1}{R^{2}}(N-1) \,\big[\mc D(\tau_{12})+\mc D(-\tau_{12})\big]  \ ,    \qquad \qquad 
\mathcal D(\tau) \equiv \D(\tau)-\D(0)\ ,   \\ \la{527}
X_2 & = \frac{1}{R^{4}}N(N-1)\,{\mc D(\tau_{12})\ \mc D(-\tau_{12})}\ ,\\
X_4& =  \pxi\frac{N(N-1)}{4R^{4}}\int \frac{d\tau d\tau'}{(\tau-\tau')^{2}}  \ Y_3 (\tau, \tau', \tau_{12}) \ , 
\la{528}
\ea
where $Y_3$  is the  relevant connected  part  given by  the sum of products of three  propagators 
\iffa 
\ba
\la{5.33}
& Y_{1} = \mathcal{D}(-\tau ) \mathcal{D}(\tau ) \mathcal{D}(\tau -\tau ')
-2 \mathcal{D}(-\tau ) \mathcal{D}(\tau -\tau ') \mathcal{D}(\tau ')
+\mathcal{D}(\tau -\tau ') \mathcal{D}(-\tau ') \mathcal{D}(\tau ')\lp
+\mathcal{D}(-\tau ) \mathcal{D}(\tau ) \mathcal{D}(-\tau +\tau')
-2 \mathcal{D}(\tau ) \mathcal{D}(-\tau ') \mathcal{D}(-\tau +\tau ')
+\mathcal{D}(-\tau ') \mathcal{D}(\tau ') \mathcal{D}(-\tau +\tau '), \\
\la{5.34}
& Y_{2} = 2(N-1) \mathcal{D}(\tau -\tau ') \mathcal{D}(-\tau +\tau '), \\
\fi
\ba\la{5.35}
 Y_{3} & = -2 {\mathcal{D}(\tau -\tau _{12})} \big[ \mathcal{D}(-\tau ) \,\mathcal{D}(\tau -\tau ')\,
 +\mathcal{D}(-\tau ) \,\mathcal{D}(-\tau +\tau ')-2 \mathcal{D}(-\tau ') \,\mathcal{D}(-\tau +\tau ')\big]
 \\
  &\ \ \ 
-2 {\mathcal{D}(-\tau +\tau _{12})} \big[\mathcal{D}(\tau )\,  \mathcal{D}(\tau -\tau ')-
2 \mathcal{D}(\tau -\tau ') \, 
\mathcal{D}(\tau ')+\mathcal{D}(\tau )\,  \mathcal{D}(-\tau +\tau ')\big]\lp
+2 {\mathcal{D}(\tau -\tau _{12})\, \mathcal{D}(-\tau +\tau _{12}) } \big[\mathcal{D}(\tau -\tau ')+ \mathcal{D}(-\tau +\tau ')\big]
-4 {\mathcal{D}(-\tau +\tau _{12})\,  \mathcal{D}(-\tau _{12}+\tau ')} \,  \mathcal{D}(\tau -\tau '). \no 
\ea
All terms in \rf{5.26} are expressed in terms of the shifted propagator $\mathcal{D}(\tau)$
  that is regular in the IR  (cf. \rf{5.10},\rf{5.11}) 
  \ba
 \mathcal{D}(\tau)=R^{2}\big[\D(\tau)-\D(0)\big ]=&
\int_{-\infty}^{\infty}\frac{dp}{2\pi}\frac{i}{p+i\pi\pxi |p|}(e^{ip\tau}-1) \no \\ = &%
\frac{1}{\pi(1+\pi^{2}\pxi^{2})}\, \int_{0}^{\infty}\frac{dp}{p}\, \Big(\pi\,\pxi\,\big[\cos(p\tau)-1 \big]-\sin(p\tau)\Big) \ . \la{5100}  
\ea
The terms in the third  line of \rf{5.35}
become independent of  $\tau_{12}$ after shifting   of $\tau$ and $\tau'$ by $\tau_{12}$
under  the integral in \rf{528}. 
The remaining terms (in the first and the second line) can be written, 
using also the symmetry  $(\tau, \tau')\leftrightarrow (-\tau, -\tau')$  of \rf{528}  as 
\ba\no
Y_{3} = &-4\,\big[\mc D(-\tau)\,\mc D(\tau-\tau')-\mc D(-\tau')\,\mc D(-\tau+\tau')\big]\,
 \big[\mc D(\tau-\tau_{12})+\mc D(\tau+\tau_{12})\big] \\
\la{5.37}
 =& -4\,\mc D(-\tau)\mc D(\tau-\tau')\,\big[\mc D(\tau-\tau_{12})+\mc D(\tau+\tau_{12})-\mc D(\tau'-\tau_{12})-\mc D(\tau'+\tau_{12})\big].
\ea
Thus \rf{5.26}  is given by 
\ba
\la{5.38}
G_{\perp}(\tau)=& \Big(1-\frac{1}{N}\Big)\frac{\pxi\, g^{2}}{8\pi^{2}\tau_{12}^{2}}\bigg[1
-\frac{2N}{R^{2}}\,\frac{\pxi}{1+\pi^{2}\pxi^{2}}\,\log(\bar\Lambda\tau)\lp
+\frac{N^{2}}{R^{4}}\,\mc D(\tau_{12})\,\mc D(-\tau_{12})
+\frac{N^{2}}{4R^{4}}\, {\pxi}\, \int\frac{d\tau d\tau'}{(\tau-\tau')^{2}}\,Y_{3}(\tau, \tau', \tau_{12})
+  \mc O \big({1\ov R^6}\big)  \bigg].
\ea
The  computation  of the logarithmically divergent part of the $1/R^4$ correction in the second line of \rf{5.38}   is quite non-trivial 
and is presented in Appendix \ref{app:Y3}. Here we just quote the result
\ba
\la{5.39}
G_{\perp}(\tau) &= \Big(1-\frac{1}{N}\Big)\frac{\pxi\, g^{2}}{8\pi^{2}\tau_{12}^{2}}\bigg[1
-\frac{2N}{R^{2}}\,\frac{\pxi}{1+\pi^{2}\pxi^{2}}\,\log(\bar\Lambda\tau)\lp\qquad \qquad 
+\frac{N^{2}}{R^{4}}\frac{4\pxi^{2}}{(1+\pi^{2}\pxi^{2})^{3}}\,\log^{2}(\bar\Lambda\tau)
+\frac{N^{2}}{R^{4}}\ \frac{2\,\pxi^{2}\,(1- \ga\,  \pi^{2}\pxi^{2})}{(1+\pi^{2}\pxi^{2})^{3}}\,\log(\bar\Lambda\tau)+\cdots
\bigg]\ .
\ea
The coefficient $\ga$ in general is scheme dependent;  in the 
 momentum cutoff scheme  we  found that  (see Appendix \ref{app:Y3})
 \be  \ga=1  \ . \la{531} \ee
The $\log^2$ term  obeys  the RG  condition  (\ref{5.18})  while the $\log $ term  
  leads to the two-loop term in the beta-function  \rf{5.19}
\be
\la{126}
\bxi = \frac{2N}{R^2}\,\frac{\pxi^{2}}{1+ \pi^2 \pxi^2} -\frac{2N^{2}}{R^{4}}\,\frac{\pxi^{3}\, (1- \ga\, \pi^{2}\pxi^{2})}{(1+\pi^{2}\pxi^{2})^{3}}+\mc O\Big(\frac{1}{R^{6}}\Big)\ . 
\ee
As already discussed   below \rf{1.26}  the 
 lowest  $ \pxi^3$  term in  the $1/R^4$    correction  corresponds precisely to the $\z^5$ term 
 in the two-loop ladder beta function for $\z$ in \rf{1.4},\rf{1.21}.

\subsection{Comments on scheme dependence  
and  
 three-loop $\beta^{\ladd}_\z$  in general representation}

Let us comment on the scheme dependence of of the beta-function \rf{126}.  In general,
in this one-coupling theory (with only $\z$ or $\pxi$ running  and expansion  going in powers of 
$\hbar= {1\ov R^2}$)  the scheme freedom should  correspond to coupling $\pxi$  redefinitions 
\ba &\pxi \to  \pxi +   { 1 \ov R^2} q_1(\pxi) + ...\ , \la{533} \\
\bxi =   \mu {d\pxi \ov d \mu}  &= {1\ov R^2}   b_1 (\pxi ) + {1\ov R^4}   b_2 (\pxi ) +  ...
   \to \beta_{\pxi}   +   {1\ov R^4}  \Big[ q_1(\pxi) \,  b'_1 (\pxi )  -  b_1(\pxi)\,   q_1'(\pxi) \Big] + ...  \la{534}
   \ea
Thus unless $q(\xi)$ is exactly proportional to the one-loop  beta function term   $b_1(\xi)$
(as it happens  in simplest  cases of one-coupling theories) the two-loop $1/R^4$ term 
 is not, in general,  invariant. 
For example, considering small $\pxi$ expansion, 
 with $q_1= c_1 \pxi^2 + c_2 \pxi^4 + ...$  and using that  the one-loop term in \rf{126} is  
$b_1= 2N(\pxi^2 - \pi^2  \pxi^4 + ...)$
we find that 
$q_1 b'_1 - b_1 q_1'= - 4 N ( c_2 + \pi^2 c_1) \pxi^5 + ....$. 

Thus   while the  coefficient of the leading $\pxi^3$ term in the two-loop  correction in \rf{126}
is invariant, the  coefficient $\ga$ of the  first  subleading $\pxi^5$ term is,  in  general, 
scheme 
dependent. 
At the same time the denominator $(1 + \pi^2 \pxi)^{-3}$  structure 
originating from $(1 + \pi^2 \pxi^2)^{-1}$  factors in the  exact  propagator \rf{5100}
appears to be universal (at least in a  natural class  of regularization
schemes that  do not substantially modify the structure 
of \rf{5.11},\rf{5100}).

Next, let us elaborate  on the implications of the structure of $\bxi$ in \rf{126}
(see   comments below \rf{4.69}). 
Using the definition of $\pxi$   we  may 
turn  (\ref{126}) into a perturbative large $R^{2}\sim k$ 
expansion  of $\beta^{\ladd}_{\z}$
\be
\la{5.40}
\beta^{\ladd}_{\z} = \frac{N}{2}\z^{3}\frac{g^{2}}{4\pi^{2}}
-\frac{N^{2}}{4}\z^{5}\Big(\frac{g^{2}}{4\pi^{2}}\Big)^{2}-\frac{\pi^{2}\,N\,k^{2}}{8}\,\z^{7}\Big(\frac{g^{2}}{4\pi^{2}}\Big)^{3}+...
\ee
Note that  three loop $g^6 \z^7$  term in (\ref{5.40}) comes
 entirely from the expansion of the denominator in the first term in (\ref{126})
 or from \rf{4.71}. Indeed,    
  the $1/R^{4}$  term  in \rf{126}   produces only $\z^5 g^4 + \z^9 g^8 +...$ terms. 
   Comparing with (\ref{2.31}) for general $N$   fixes the coefficients there as 
\be
q_{3}''=0, \qquad \qquad q_{3}'''' = -3\,\zeta(2)= - \ha \pi^2\ .\la{5.31}
\ee
We thus obtain the following 3-loop ladder beta function for general representation (cf. \rf{1.20},\rf{1200}) 
\ba\la{5.32}
\beta^{\ladd}_{\z} &= \frac{1}{2}\,C_{\A}\z^{3}\,\frac{g^{2}}{4\pi^{2}}-\frac{1}{4}\,C_{\A}^{2}\,\z^{5}\,\Big(\frac{g^{2}}{4\pi^{2}}\Big)^{2}
+\bigg(q_{3}'\,C_{\A}^{3}-3\,\zeta(2)\, \QR 
\bigg)\,\z^{7}\,\Big(\frac{g^{2}}{4\pi^{2}}\Big)^{3}+\mc O(g^{8}).
\ea
  We will  prove in Appendix \ref{app:chern} that in  the planar limit,
for  any  irreducible representation $\R$ of $SU(N)$  the coefficient $\QR$ 
of the $\zeta(2) \z^7$ 
term in \rf{5.32} is universal, i.e.   one has ($\l= g^2 N$) 
\ba
\la{5.43}
\beta^{\ladd}_{\z} = \frac{1}{2}\,\z^{3}\,\frac{\l}{4\pi^{2}}-\frac{1}{4}\,\z^{5}\,\Big(\frac{\l}{4\pi^{2}}\Big)^{2}
+\bigg(q_{3}'-\frac{\zeta(2)}{8}\bigg)\, \z^{7}\,\Big(\frac{\l}{4\pi^{2}}\Big)^{3}+\mc O(\l^{4}),
\ea
Comparing with (\ref{1.7}), we see that the  $\QR$  
term in \rf{5.32}  corresponds to 
 the $\zeta(2)$  transcendental part of the
 coefficient  $q_{3}=\frac{1}{4}-\frac{\zeta(2)}{8}$ in  \rf{1.7}.
 This agreement is remarkable given that the three loop beta function is, in general,  expected to be scheme 
dependent. Indeed, the expansion (\ref{1.7}) has been derived in dimensional regularization while (\ref{126}, \ref{5.40})
have been obtained in a  mode regularization. This suggests that only $q_{3}'$ term in \rf{5.32} 
 is actually scheme dependent  while $q_{3}''''$ in \rf{5.31} 
  is scheme independent. 
 An explanation of this scheme independence is that this coefficient 
 comes from  the $\kappa^4$ term  in the expansion of the one-loop term in $\beta_\pxi$ in \rf{1.26}, \ie from the first  scheme-independent 
 term in the perturbative $1/k$ expansion.

\section*{Acknowledgements}
We are grateful to S. Komatsu  
and G. Korchemsky  for useful  discussions. 
MB was supported by the INFN grant GSS (Gauge Theories, Strings and Supergravity). 
The work of SG is 
supported in part by the US NSF under Grant No. PHY-1914860. 
AAT was  supported by the STFC grant ST/T000791/1. 
Part of this   work was done while AAT was a participant of the program
``Confinement, Flux Tubes, and Large N'' at the KITP in Santa Barbara   where his work was supported 
in part by the NSF under Grant No.  PHY-1748958.

\appendix
\section{$SU(N)$ conventions}
\la{app:sun}

For the $SU(N)$ generators in the fundamental representation we have ($a=1, ..., N^2-1; \ i=1, ..., N$) 
\be\la{A.1}
[T^{a}, T^{b}] = i\,f^{abc}\,T^{c}, \quad \tr\,  T^{a}=0, \quad \tr\, (T^{a}T^{b} )
= \frac{1}{2}\delta^{ab},\quad (T^{a}T^{a})_{ij} =\frac{N^{2}-1}{2N}\,\delta_{ij}, 
\ee
\be
\la{A.2}
T^{a}_{ij}T^{a}_{kl} = \frac{1}{2}\Big(\delta_{il}\delta_{jk}-\frac{1}{N}\delta_{ij}\delta_{kl}\Big), \qquad 
f^{acd}f^{bcd} = N\delta^{ab}.
\ee
Then also  
\ba
\tr(T^{a}T^{a}T^{b}T^{b}) &= \frac{1}{2}\tr(1)\ \tr(T^{b}T^{b})-\frac{1}{2N}\tr(T^{b}T^{b}) = 
\big(\frac{N}{2}-\frac{1}{2N}\big)\frac{N^{2}-1}{2} = 
\frac{(N^{2}-1)^{2}}{4N}, \\
\tr(T^{a}T^{b}T^{a}T^{b}) &= \frac{1}{2}\tr(T^{b})\ \tr(T^{b})-\frac{1}{2N}\tr(T^{b}T^{b}) = -\frac{1}{2N}\frac{N^{2}-1}{2} = -\frac{N^{2}-1}{4N}, \\
\tr(T^{a}T^{b}T^{b}T^{a}) &= \tr(T^{a}T^{a}T^{b}T^{b}) = \frac{(N^{2}-1)^{2}}{4N}.
\ea
For a  
{generic representation R} we define the index  $C_\R$ by  
\be
\la{A.6}
T^{a}T^{a} = C_{\R}\, \mathbf{1}\ , \qquad \qquad \tr (T^{a}T^{a}) = C_{\R}\,\dim \R \ .
\ee
In the special case of the  fundamental representation 
\be\la{A.7} 
\tr (T^{a}T^{a}) = \frac{1}{2}(N^{2}-1)\quad\to\quad C_{\F}=\frac{N^{2}-1}{2N}.
\ee
For  the adjoint representation 
 $(T^{a}_{\rm adj})_{bc} = -i f_{abc}$ so from (\ref{A.2}) we have $C_{\A}=N$. 

 For the $k$-symmetric representation $\textrm{S}_{k}$
\be\la{A.8} 
\dim \textrm{S}_{k} = \binom{N+k-1}{k},\qquad  \qquad C_{\textrm{S}_{k}} = \frac{k(N-1)(N+k)}{2N}.
\ee
Let us note also the following  relation\foot{For other similar relations  
see,  for instance,  sec. 3.1 of \cite{vanRitbergen:1998pn}.}
\be
\la{A.9}
T^{a}T^{b}T^{a} = (C_{\R}-\frac{1}{2} C_{\A})\,T^{b}.
\ee
Also, if  $X$ is some matrix (e.g. a  product of some generators) 
 then  
\be
\la{A.10}
T^{a}T^{b} X T^{a}T^{b} = T^{a}T^{b} X T^{b}T^{a}-\frac{1}{2}C_{\A}\,T^{a}XT^{a}.
\ee
Useful examples are 
\ba
\tr(T^{a}T^{a}T^{b}T^{b}) &= C_{\R}^{2}\dim \R, \\
\tr(T^{a}T^{b}T^{a}T^{b}) &= (C_{\R}-\frac{1}{2}C_{\A})\tr(T^{b}T^{b}) = (C_{\R}-\frac{1}{2}C_{\A})C_{\R}\dim \R.
\ea

\section{1d Fermionic representation for  the Wilson loop
\la{app:GN}}

The Wilson loop  admits a 1d fermionic representation \cite{Gervais:1979fv,Arefeva:1980zd,Brandt:1981kf}
that we we will  review  here for the case of a general representation of gauge group. 
We start with  the path-ordered exponential
\be
\la{B.1}
U_{\aa\bb} = \left[\textrm{P}\,\exp \int_{\tau_{1}}^{\tau_{2}}F(\tau) \right]_{\aa\bb},
\ee
where $F(\tau)=F^{a}(\tau)T^{a}$ is a Lie algebra valued function
 in the representation R  (with  the corresponding indices  being $\aa,\bb$). 
We can write
\ba
\la{B.2}
U_{\aa\bb} &= e^{\Omega}\left[\frac{\delta^{2}}{\delta\bar\lu_{\aa}(\tau_{2})\,\delta\lu_{\bb}(\tau_{1})}
\exp\Big( \int_{\tau_{1}}^{\tau_{2}}d\tau\ \frac{\delta}{\delta\lu_{\cc}(\tau)}F_{\cc\cc'}
(\tau)\frac{\delta}{\delta\bar\lu_{\cc'}(\tau)}\Big) \right]_{\lu=
\bar\lu=0}\ e^{-\Omega},\notag \\
\Omega &= \int_{\tau_{1}}^{\tau_{2}}d\tau d\tau'\ \bar\lu_{\cc}(\tau)\lu_{\cc}(\tau')\,\theta(\tau-\tau'),
\ea
where $\lu$ and $\bar\lu$ are (Grassmann) vectors of  R   and $\theta$ is  the step function. 
$\Omega$ admits  the following  representation  in terms of path integral 
over anticommuting   fields $\pz_\aa$ and $\bar \pz_\aa$ 
(which are vectors in the representation  R)
with the antiperiodic  boundary condition
  $\pz(\tau_{2}) = -\pz(\tau_{1})$ \footnote{This follows from  
$\theta(\tau)$ being the propagator associated with  the first order
 kinetic term $\partial_{\tau}\theta(\tau) = \delta(\tau)$.}
\be
\la{B.3}
e^{-\Omega} = \int D\pz D\bar \pz\ \exp\int d\tau\, \big [\bar \pz\partial_{\tau}\pz+i(\bar\lu \pz+\bar \pz\lu)\big].
\ee
Then for a closed loop the trace of $U$ in  (\ref{B.2}) may be written  as 
\be
\la{B.4}
\tr\,  U = \frac{\delta^{2}}{\delta\bar\lu_{a}(0)\delta\lu_{a}(0)}\left[\log
\int D\pz D\bar \pz\, \exp\int d\tau\bigg(\bar \pz\partial_{\tau}\pz-\bar \pz F \pz+i(\bar \lu \pz+\bar \pz \lu)\bigg)
\right]_{\lu=\bar\lu=0}.
\ee
Eq. \rf{B.4} represents  the  2-point function $\vev{\bar \pz(2\pi)\, \pz(0)}$
whose   perturbative expansion is expressed  in terms of factors of  $F(\tau)$ 
 connected by $z$-propagators, i.e. by theta functions that implement path-ordering. 

As an example,  let us  consider the 
  scalar   ladder  model. Integrating the free scalar field we get
the  corresponding 1d  effective  action  
\be 
S=\int d\tau\, \bar \pz\partial_{\tau}\pz
+\frac{1}{2}\z^2 \int d\tau d\tau' \ D(\tau-\tau') \bar \pz(\tau)  T^{a} \pz(\tau)\, \bar \pz(\tau') 
 T^{a} \pz(\tau') \ . \la{B.5}\ee
Here $D(\tau-\tau') =\vev{\phi(\tau) \phi(\tau') }$ is the scalar propagator restricted to the line. 
Introducing an auxiliary 1d  field $\sigma_{a}(\tau)$  we may write the  
 corresponding ladder WL expectation  value  as  
\be
\la{B.6}
\W = \Big\langle\tr
\Big[\frac{1}{\partial_{\tau}-\sigma_{a}(\tau)\,T^{a}}\Big]\Big\rangle \ , 
\ee
where $\langle ... \rangle$  amounts to Wick contractions of the free fields $\sigma_{a}(\tau)$ with the  propagator $\vev{\sigma_{a}(\tau)\, \sigma_{b}(\tau')} = \delta_{ab}D(\tau-\tau')$.
This  reconstructs  the standard perturbative evaluation of the Wilson loop like 
\rf{1.1} or \rf{1.9}.

The same steps  may  be repeated in the case  of the circular
 $\frac{1}{2}$-BPS loop in $\N=4$ SYM where 
the function $F$  can be  read off  from (\ref{1.1}) with $\zeta=1$.  
Assuming  interaction terms in the SYM action  do not contribute (as turns out to be true) 
 and  integrating out the  free scalar and vector fields we obtain
\be
\la{B.7}
\vev{W^{(1)}} = \frac{\int D\pz D\bar \pz\ \bar \pz(2\pi)\, \pz(0)\, \exp\bigg[\int d\tau\, 
\bar \pz\partial_{\tau}\pz+ \frac{g^{2}}{16\pi^{2}}\big(\int d\tau\,  \bar \pz  T^{a} \pz \big)^2
\bigg]
}
{\int D\pz D\bar \pz\  \exp\Big[ \int d\tau\, \bar \pz\partial_{\tau}\pz+
\frac{g^{2}}{16\pi^{2}}\big(\int d\tau\,  \bar \pz  T^{a} \pz \big)^2 
\bigg]
} \ . 
\ee
We used that   here the  effective propagator  corresponding to the
 combination $(AA + \p\p)^{ab}$ is constant  \ci{Erickson:2000af}, i.e. 
 $D(\tau)=D_0=\frac{g^{2}\delta^{ab}}{8\pi^{2}}$.
Introducing an auxiliary constant field $\sigma_{a}$, we may write the quartic action
 in a  local  form 
\ba
\la{B.8}
&
\int d\tau\,  \bar \pz\partial_\tau \pz+\frac{g^{2}}{16\pi^{2}}\Big( \int d\tau \,  \bar \pz(\tau)
 T^{a}\pz(\tau)\Big)^2  \  \to \  \int d\tau\,  \bar \pz\partial_\tau \pz
-\frac{g}{2\pi}\sigma_{a}\,\int d\tau\bar \pz(\tau)T^{a}\pz(\tau)-\sigma_{a}^{2}.
\ea
Integrating out the fermions $\pz, \bar \pz$, we then  obtain another equivalent representation 
\be
\la{B.9}
\vev{W^{(1)}}  = \Big\langle\tr
\Big[\frac{1}{\partial_{\tau}-\frac{g}{2\pi}\sigma_{a}\,T^{a}}\Big]\Big\rangle, \qquad \qquad 
\langle ...\rangle = \int \prod_{a}d\sigma_{a} \ e^{-\sigma_{a}^{2}}\ ...
\ee
Let us show how  (\ref{B.9})  can be  used to reproduce
 the perturbative expansion  in  (\ref{1.13}). 
We  expand the trace  using $\langle \tau_2 | (\partial_{\tau})^{-1} 
 | \tau_1\rangle= \theta(\tau_{2}-\tau_{1})$. For instance, 
\be
\la{B.10}
\langle 2\pi |( \partial_{\tau})^{-3}|0\rangle = \int_{0}^{2\pi}d\tau d\tau'\,  \theta(\tau-0)\theta(\tau'-\tau)\theta(2\pi-\tau') = \int_{\tau<\tau'}d^{2}\tau.
\ee
We then obtain 
\ba
\no 
\tr\Big[\frac{1}{\partial_\tau-\frac{g}{2\pi}\sigma_{a} T^{a}}\Big] =& \dim \R+\tr(T^{a}T^{b})\sigma_{a}\sigma_{b}\frac{g^{2}}{4\pi^{2}}\int_{\tau_{1}<\tau_{2}} d^{2}\tau\\  &\qquad\quad+
\tr(T^{a}T^{b}T^{c}T^{d})\sigma_{a}\sigma_{b}\sigma_{c}\sigma_{d}\Big(\frac{g^{2}}{4\pi^{2}}\Big)^{2}\int_{\tau_{1}<\cdots<\tau_{4}} d^{4}\tau+\cdots.\la{B.11}
\ea
Taking the average using that 
\ba
\la{B.12}
&\vev{\sigma_{a}\sigma_{b}} = \frac{1}{2} \delta_{ab}, \qquad \vev{\sigma_{a}\sigma_{b}\sigma_{c}\sigma_{d}} = \frac{1}{4}(\delta_{ab}\delta_{cd}+ \delta_{ac}\delta_{bd}+\delta_{ad}\delta_{bc}),\\
\la{B.13}
& \tr(T^{a}T^{b})\delta_{ab} = \tr(T^{a}T^{a}) = \dim \R\, C_{\R}, \ \\
& \tr(T^{a}T^{b}T^{c}T^{d})(\delta_{ab}\delta_{cd}+ \delta_{ac}\delta_{bd}+\delta_{ad}\delta_{bc}) = 2\tr(T^{a}T^{a}T^{b}T^{b})+\tr(T^{a}T^{b}T^{a}T^{b})\lp\qquad \qquad 
 = \dim \R\,\big[ 2C_{\R}^{2}+C_{\R}(C_{\R}-\frac{1}{2}C_{\A})\big],\\
\la{B.15}
&\int_{\tau_{1}<\tau_{2}} d^{2}\tau = 2\pi^{2}, \qquad
\int_{\tau_{1}<\cdots<\tau_{4}} d^{4}\tau = \frac{2}{3}\pi^{4}, 
\ea
we find that  
\be
\la{B.16}
\frac{1}{\dim \R}\vev{W^{(1)}} = 1+\frac{1}{4}C_{\R}\, \gym^2 
+\frac{1}{192}C_{\R}(6C_{\R}-C_{\A})\,\gym^4 +\mc O (\gym^6),
\ee
which is in agreement with (\ref{1.13}).

\section{Computation of divergent part of $1/R^4$ term   in  scalar two-point function}    
\la{app:Y3}

To find   the  divergent part of  the $\int d\tau d\tau'\ (\tau-\tau')^{-2}\,Y_{3}  $  term 
in (\ref{5.38})  we shall   use somewhat eclectic   direct  cutoff  method. 
First, let us   introduce a   UV cutoff $a\to 0$ in the $(\tau-\tau')^{-2}$  kernel
(which originated from  the 4d  scalar propagator restricted to the line) as
\ba
\la{C.2}
\frac{1}{(\tau-\tau')^{2}} &\to \frac{1}{(\tau-\tau')^{2}+a^{2}}
 = \int_{-\infty}^{\infty}dp\ \frac{e^{-a\, |p|}}{2a}\,e^{ip(\tau-\tau')} \ , \qquad a\to 0 \ . 
\ea
Using the expression for the  propagator $\cD$  in \rf{5100}   in  $Y_3$ in \rf{5.37} 
and integrating over  $\tau, \tau'$ we then obtain\foot{Note that  singular terms like 
 $\frac{1}{p+i\pi\pxi |p|}$ at $p\to 0$ that appear  at intermediate steps cancel.}
\ba
\la{C.3}
&\bar Y_3\equiv \int\frac{d\tau d\tau'}{(\tau-\tau')^{2} + a^2 }\,Y_{3} 
= \int_{0}^{\infty}dp_{1}\int_{0}^{\infty}dp_{2}\, f(p_{1},p_{2}),\\
&f(p_{1},p_{2}) \Big|_{p_1 < p_2}= \frac{4 e^{-a (p_1+p_2)} (1-e^{a p_1}) \pxi\,  \cos (p_1 \tau _{12})}{a (1+\pi ^2 \pxi ^2)^3 \, p_1^2 p_2}  \big[1+\pi ^2 \pxi ^2+e^{a 
p_1} (3-\pi ^2 \pxi ^2)+2 e^{a p_2} (1-\pi ^2 \pxi ^2)\big]   \no
\ea 
\iffa
\begin{cases}
\frac{4 e^{-a (p_1+p_2)} (-1+e^{a p_1}) \pxi  (-1-\pi ^2 \pxi ^2+e^{a 
p_1} (-3+\pi ^2 \pxi ^2)+2 e^{a p_2} (-1+\pi ^2 \pxi ^2)) \cos (p_1 \tau _{12})}{a (1+\pi ^2 \pxi ^2)^3 p_1^2 p_2}, & p_{1}<p_{2}, \\
\frac{4 e^{-a (p_1+p_2)} (-1+e^{a p_2}) \pxi  (-1-\pi ^2 \pxi ^2+e^{a 
p_2} (-3+\pi ^2 \pxi ^2)+2 e^{a p_1} (-1+\pi ^2 \pxi ^2)) \cos (p_1 \tau _{12})}{a (1+\pi ^2 \pxi ^2)^3 p_1^2 p_2}, & p_{2}<p_{1}.
\end{cases}
\ee
\fi 
Introducing an extra  hard  momentum cutoff $p_{1},p_{2}<\Lambda$ 
(which we will   later relate to $1/a$) and integrating over $p_{2}$ 
we get 
\ba
 & 
\bar Y_3 = \int_{0}^{\Lambda}dp_{1}\ \frac{4 \pxi  \cos (p_1 \tau _{12})}{a (1+\pi ^2 \pxi ^2)^3 p_1^2}
\bigg[
e^{-a p_1} (1-e^{a p_1}) \big[1+\pi 
^2 \pxi ^2+e^{a p_1} (3-\pi ^2 \pxi ^2)\big]  \text{Ei}(-a \Lambda ) \la{C.5} \\
\no  &+e^{a p_1} (3-\pi ^2 \pxi ^2) \text{Ei}(-a p_1)
+e^{-a p_1} (-3+\pi ^2 \pxi 
^2) \text{Ei}(a p_1)+2 e^{-a p_1} (-1+e^{a p_1}) (-1+\pi ^2 \pxi ^2) 
\log (a \bar{\Lambda })\bigg],
\ea
where 
 $\text{Ei}(z) = -\int_{-z}^{\infty}dt\, \frac{e^{-t}}{t}$, 
and $\bar\Lambda = \Lambda\, e^{\gamma_{\rm E}}$. 
To perform the last integration over $p_{1}$ we consider the integrand in the limit $\Lambda\to\infty$, $a\to 0$. 
Dropping power divergent terms $\sim 1/a$
and integrating over $p_1< \Lambda $ 
we find that 
 the terms depending on $\tau_{12}$ are 
\be
\la{C.6}
\frac{\pxi}{4} \bar Y_2=
-\frac{\pxi^{2}  (-3+\pi ^2 \pxi ^2)}{(1+\pi ^2 \pxi ^2)^3}\  \log ^2(\bar\Lambda\tau_{12})
-\frac{2\pxi^{2}\,(-5+2\,\log(a\bar\Lambda)+\pi^{2}\pxi^{2})}{(1+\pi^{2}\pxi^{2})^{3}}\,\log(\bar\Lambda\tau_{12})+\cdots \ . 
\ee
To this we need to add the contribution of the $X_2$ term in \rf{527}   or 
$\mc D(\tau_{12})\ \mc D(-\tau_{12}) $ in \rf{5.38}. 
Introducing the same momentum cutoff $\Lambda$  in the propagators $\cD$ \rf{5100} 
($\int^\infty_0 dp \to \int^\Lambda_0 dp$)   and integrating over $p$ we get  for $\tau>0$
\iffa \ba
\la{C.7}
\mc D(\tau) &= \frac{1}{\pi(1+\pi^{2}\pxi^{2})}\, \int_{0}^{\Lambda}\frac{dp}{p}[\pi\,\pxi\,(\cos(p\tau)-1)-\sin(p\tau)].
\ea
Integrating and expanding at large $\Lambda$, we have
\fi
\ba
\la{C.8}
&\mc D(\pm \tau) = -\frac{\pxi}{1+\pi^{2}\pxi^{2}}\,\log(\bar\Lambda|\tau|)\mp \frac{1}{2(1+\pi^{2}\pxi^{2})}+\mc O(\Lambda^{-1}) \ , \\
\la{C.9}
&\mc D(\tau_{12})\mc D(-\tau_{12}) = \frac{\pxi^{2}}{(1+\pi^{2}\pxi^{2})^{2}} \ \log^{2}(\bar\Lambda\tau_{12})+ {\rm finite}\ .
\ea
Combining the  contributions of \rf{C.6}   and \rf{C.9}  we get for the relevant  divergent 
terms in  \rf{5.38}
\ba
\la{C.10}
G_{\perp}(\tau) &= \Big(1-\frac{1}{N}\Big)\frac{\pxi\, g^{2}}{8\pi^{2}\tau_{12}^{2}}\bigg[1
-\frac{2N}{R^{2}}\,\frac{\pxi}{1+\pi^{2}\pxi^{2}}\,\log(\bar\Lambda\tau)\lp
+\frac{N^{2}}{R^{4}}\frac{4\pxi^{2}}{(1+\pi^{2}\pxi^{2})^{3}}\,\log^{2}(\bar\Lambda\tau)
+\frac{N^{2}}{R^{4}}\ \frac{2\,\pxi^{2}\,\big[5-2\log(a\bar\Lambda)-\pi^{2}\pxi^{2}\big]}{(1+\pi^{2}\pxi^{2})^{3}}\,\log(\bar\Lambda\tau)+\cdots
\bigg].
\ea
The coefficient  of the $\log^{2}$ term here  
agrees with the  one following from the  RG constraint (\ref{5.18}).

The resulting coefficient  of the  leading ${1\ov R^4} \pxi^3$ term in the beta function 
will then be $5-2\log(a\bar\Lambda)$  (cf. \rf{5.39},\rf{126}). 
To match  the known two-loop  coefficient of $\z^5$ term   in $\beta^\ladd_\z$   in \rf{1.3},\rf{1.7}
we  need to require  that $ a$  and $\bar \Lambda$ are related so that 
$\log(a\bar\Lambda)=1$. 
It is  clearly desirable  to find a  more systematic regularization approach in which this value  will appear automatically.
In principle, it should be sufficient to introduce a UV cutoff 
only in the bulk propagator  kernel $ 1\ov (\tau-\tau')^2$  appearing in the  1d effective action 
\rf{5.4}. Then this cutoff will appear also in the exact $\pxi$-dependent propagator \rf{5.10}.
However, our attempts to use some  natural choices like dimensional regularization led 
to complicated integrals  that we did not manage to evaluate. 

\section{Universal form of planar limit of  three-loop term in $\beta^{\ladd}_\z$}
\la{app:chern}

\newcommand{\ch}{\textrm{Ch}}

The three-loop $\z^7$ term in $\beta^{\ladd}_\z$  in \rf{1.21} 
 contains the 
 group-theoretic coefficient
 \be \la{ff1} \QR= \frac{d_{\A}^{abcd}d_{\R}^{abcd}}{C_{\R}\,\dim \R} \ . \ee
Here  we shall  prove that for any irreducible representation $\R$ of $SU(N)$ one has 
\be
\la{D.1}
\lim_{N\to\infty }\frac{\QR}{N^{3}}\,  
 = \frac{1}{24},
\ee
independently on $\R$, leading to the universal
coefficient of the $\zeta(2) \z^7$  term in   (\ref{5.43}).

To prove (\ref{D.1}) we need  first to recall some definitions.
The Chern character in representation $\R$ 
with generators $T^{a}_{\R}$    is a  function of   $X^{a}$  ($a=1, ..., N^2-1$) 
 defined by 
\ba
\la{D.2}
&\ch({\R})= \tr \big[ e^{X^{a}T^{a}_{\R}}\big]  = \sum_{n=0}^{\infty}\frac{1}{n!}d_{\R}^{a_{1}\dots a_{n}}X^{a_{1}}\cdots X^{a_{n}} \ , \\
\la{D.3}
&\ch(\R_{1}\times \R_{2}) = \ch(\R_{1})\,\ch(\R_{2}), \qquad \qquad \ch(\R_{1}+ \R_{2}) = \ch(\R_{1})+\ch(\R_{2}).
\ea
For symmetric  $\S_{k}$ and antisymmetric $\A_{k}$ representations  it is known that 
\be
\la{D.4}
\ch(\S_{k}) = \sum_{k=\sum_{i}n_{i}m_{i}}\prod_{i}\frac{1}{m_{i}!}\Big[\frac{\ch(n_{i}\F)}{n_{i}}\Big]^{m_{i}}, \qquad 
\ch(\A_{k}) = (-1)^{k}\,\sum_{k=\sum_{i}n_{i}m_{i}}\prod_{i}\frac{1}{m_{i}!}\Big[-\frac{\ch(n_{i}\F)}{n_{i}}\Big]^{m_{i}},  
\ee
where the sums are over all integer partitions of $k$ ($n_{i}$ appears in the partition with multiplicity $m_{i}$). Tensoring representations one can 
obtain expressions for characters in terms of fundamental characters, see 
 examples  below. 
 For a generic irreducible 
representation with $n_{\R}$ blocks in the Young tableau, the leading large $N$ power comes from the term with a maximal power of $\ch(\F)$ 
\be
\la{D.5}
\ch(\R) = \frac{1}{h_{\R}}[\ch(\F)]^{n_{\R}}+\cdots.
\ee
In (\ref{D.5}) $h_{\R}$ is obtained as the product over  all blocks $B$ in the Young tableau of their hook length,
 defined as one plus the number of blocks below and to the right to $B$. 
The relevant terms in (\ref{D.2}) are then
\ba
\ch(\R) &= \frac{1}{h_{\R}}\bigg[N^{n_{\R}}+n_{\R}\,N^{n_{\R}-1}\bigg(\frac{1}{2}d_{\F}^{ab}X^{a}X^{b}+\frac{1}{4!}d_{\F}^{abcd}X^{a}X^{b}X^{c}X^{d}+\cdots\bigg)\lp
+\frac{n_{\R}(n_{\R}-1)}{2!}N^{n_{\R}-2}\bigg(\frac{1}{2}d_{\F}^{ab}X^{a}X^{b}+\frac{1}{4!}d_{\F}^{abcd}X^{a}X^{b}X^{c}X^{d}+\cdots\bigg)^{2}\bigg], 
\ea
where subscript $\F$ refers to fundamental representation. 
Picking the terms with 0,2,4  factors of $X^a$  gives the leading power of $N$ in 
\ba
\la{D.7}
\dim \R = \frac{N^{n_{\R}}}{h_{\R}}+\cdots, \quad 
C_{\R} = \frac{N^{2}-1}{\dim \R}\frac{n_{\R}N^{n_{\R}-1}}{2h_{\R}}+\cdots = \frac{n_{\R}N}{2}+\cdots,
 \quad 
d_{\R}^{abcd} = \frac{n_{\R}N^{n_{\R}-1}}{h_{\R}}\,d_{\F}^{abcd}+\cdots.
\ea
Using now that 
\be
d_{\F}^{abcd}d_{\A}^{abcd} = \frac{N(N^{2}-1)(N^{2}+6)}{48} = \frac{N^{5}}{48}+\cdots, 
\ee
we obtain for \rf{ff1} 
\be
\QR 
= \frac{N^{3}}{24}+\cdots,
\ee
implying (\ref{D.1}). 

Let us now present some explicit examples  of particular representations that  
check  the expansions (\ref{D.7}). 
Let us begin with the case of the representation ${\tiny\yng(2,1)}$ which is the minimal one 
not included in  $\S_{k}$ and $\A_{k}$ series. We start with 
\be
\la{D.10}
{\tiny\yng(1)}\times {\tiny\yng(1)}\times {\tiny\yng(1)} = {\tiny\yng(3)}+2\,{\tiny\yng(2,1)}+{\tiny\yng(1,1,1)}\ .
\ee
Using (\ref{D.3}), (\ref{D.4}) one obtains \cite{Fiol:2018yuc}
\be
\la{D.11}
\ch({\tiny\yng(2,1)}) = \frac{1}{3}\big[\ch(\F)\big]^{3}-\frac{1}{3}\ch(3\F).
\ee
Using  (\ref{D.2}) and expanding to  4th order gives then
\ba
\la{D.12}
\dim_{\tiny\yng(2,1)} = &\frac{N(N^{2}-1)}{3}, \qquad\qquad  C_{\tiny\yng(2,1)} = \frac{3(N^{2}-3)}{2N},
\\ \la{D.13}
 d_{\tiny\yng(2,1)}^{abcd} = &(N^{2}-27)\,d_{\F}^{abcd}+\frac{N}{2}(\delta^{ab}\delta^{cd}+\delta^{ac}\delta^{bd}+\delta^{ad}\delta^{bc}).
\ea
Contracting with $d_{\A}^{abcd}$  for adjoint representation  and using  $d_{\A}^{aabc} = \frac{5}{6}N^{2}\,\delta^{bc}$
\be
\la{D.15}
d_{\tiny\yng(2,1)}^{abcd}d_{\A}^{abcd} = \frac{1}{48}N(N^{2}-1)(N^{4}+39N^{2}-162)
\ , \qquad 
\lim_{N\to\infty}\frac{\QR}{N^{3}}\bigg|_{\R = {\tiny\yng(2,1)}} = \frac{1}{24}.
\ee
As a  next  example  we consider is $\tiny\yng(3,1)$. From 
\be
\la{D.17}
{\tiny\yng(1)}\times {\tiny\yng(3)} = {\tiny\yng(4)}+{\tiny\yng(3,1)},
\ee
we obtain 
\be
\la{D.18}
\ch({\tiny\yng(3,1)}) = \frac{1}{8}\big[\ch(\F)\big]^{4}+\frac{1}{4}\big[\ch(\F)\big]^{2}\ch(2\F)-\frac{1}{8}\big[\ch(2\F)\big]^{2}-\frac{1}{4}\ch(4\F).
\ee
Expanding as in (\ref{D.2}) gives
\ba
\la{D.19}
&\dim_{\tiny\yng(3,1)} = \frac{N(N^{2}-1)(N+2)}{8}, \qquad C_{\tiny\yng(3,1)} = \frac{2(N^{2}+N-4)}{N},
\\
\la{D.20}
&d_{\tiny\yng(3,1)}^{abcd}d_{\A}^{abcd} = \frac{1}{96} {N(N^{2}-1)(N+2)(N^{4}+7N^{3}+74N^{2}+48N-384)},
\ea
and thus again
\be
\lim_{N\to\infty}\frac{\QR }{N^{3}}\bigg|_{\R = {\tiny\yng(3,1)}} = \frac{1}{24}.
\ee
Our  final example is ${\tiny\yng(2,2)}$. Using 
\ba
{\tiny\yng(1)}\times {\tiny\yng(2,1)} = {\tiny\yng(3,1)}+{\tiny\yng(2,2)}+{\tiny\yng(2,1,1)}, \qquad\qquad
{\tiny\yng(1)}\times {\tiny\yng(1,1,1)} = {\tiny\yng(2,1,1)}+{\tiny\yng(1,1,1,1)} 
\ea
and (\ref{D.11}) and (\ref{D.18})  we get
\ba
&\qquad \qquad \ch({\tiny\yng(2,2)}) = \frac{1}{12}\big[\ch(\F)\big]^{4}+\frac{1}{4}[\ch(2\F)\big]^{2}-\frac{1}{3}\ch(\F)\, \ch(3\F), \no \\
& \dim_{\tiny\yng(2,2)} = \frac{N^{2}(N^{2}-1)}{12}, \qquad C_{\tiny\yng(2,2)} = \frac{2(N^{2}-4)}{N},
\qquad 
d_{\tiny\yng(2,2)}^{abcd}d_{\A}^{abcd} = \frac{N^{2}(N^{2}-1)(N^{2}-4)(N^{2}+42)}{144},\no
\\
&\qquad\qquad \qquad \qquad   \lim_{N\to\infty}\frac{\QR }{N^{3}}\bigg|_{\R = {\tiny\yng(2,2)}} = \frac{1}{24}.
\ea

\section{Two-loop ladder beta function 
from two-point correlators on the line }
\la{app:2pointline}

In \cite{Beccaria:2021rmj} we showed how to compute the   two-loop ladder beta function $\beta_{\z}^{\ladd}$ in the planar limit  by considering the 
defect  two-point function of the scalar fields (either coupled to the loop or  ``transverse''  to it) in the case  when the scalar  Wilson loop  in the fundamental representation 
 is defined on a  straight  line of length $2L$.
Here  we will extend that calculation to the case of a 
generic representation of $SU(N)$  at finite $N$. 

\subsection{Transverse scalar}

For  one  ``transverse''  scalar  
 denoted  by   $\phi_{\perp}$
which  does not appear in the  Wilson line  exponent  we want to compute
\be
\la{F.1}
G_{\perp}(\tau) = \llangle \phi_{\perp}(0)\,\phi_{\perp}(\tau) \rrangle = 
\frac{\vev{\tr\big[\text{P}\, \phi_{\perp}(0)\,\phi_{\perp}(\tau)\,\exp\int_{-L}^{L}d\tau'\, \phi(\tau')\big]}}{\vev{\tr\big[\text{P}\,\exp\int_{-L}^{L}d\tau'\, \phi(\tau')\big]}} \ . 
\ee
Here $\phi$ is rescaled  by $\z$   so that 
the relevant coupling   that  appears in  the propagator is 
$\bar \xi= \z^2 g^2=  N^{-1} \xi $ (where $\xi$ was defined in (\ref{1.9})). 
The propagator on the line (cf. \rf{1.11},\rf{1.12})
 in dimensional regularization is given  by (cf. \rf{2.3})
\be\la{F2}
D(\tau) = \frac{N \xxi }{8\pi^{2}}\frac{1}{|\tau|^{2-\ve}}, \qquad \qquad d=4-\ve, \qquad 
 \ \ \ \   \xxi = \z^2 \, g^2  \ . 
\ee
\paragraph{One loop}

At the tree and one loop level  we    have  from \rf{F.1} 
\ba
\la{F.3}
& \vev{\tr\big[\text{P}\,\exp\int_{-L}^{L}d\tau'\, \phi(\tau')\big]} = 1+
\begin{tikzpicture}[line width=1 pt, scale=0.5, baseline=0]
\node[left] at (0,0) {$-L$};
\node[right] at (6,0) {$L$};
\draw[line width=2 pt] (0,0)--(6,0);
\draw (1,0) arc(180:0:2);
\node[below] at (1,0) {$\tau_{1}$};
\node[below] at (5,0) {$\tau_{2}$};
\end{tikzpicture}+\cdots\ ,\\
\la{F.4}
&  \vev{\tr\big[\text{P}\, \phi_{\perp}(0)\,\phi_{\perp}(\tau)\,\exp\int_{-L}^{L}d\tau'\, \phi(\tau')\big]} = 
\begin{tikzpicture}[line width=1 pt, scale=0.5, baseline=0]
\node[left] at (0,0) {$-L$};
\node[right] at (5,0) {$L$};
\draw[line width=2 pt] (0,0)--(5,0);
\draw[densely dashed] (1,0) arc(180:0:1.5);
\node[below] at (1,0) {$0$};
\node[below] at (5,0) {$\tau$};
\end{tikzpicture}\lp\qquad
+\begin{tikzpicture}[line width=1 pt, scale=0.5, baseline=0]
\node[left] at (0,0) {$-L$};
\node[right] at (7,0) {$L$};
\draw[line width=2 pt] (0,0)--(7,0);
\draw[densely dashed] (1,0) arc(180:0:1);
\node[below] at (1,0) {$0$};
\node[below] at (3,0) {$\tau$};
\draw (4,0) arc(180:0:1);
\node[below] at (4,0) {$\tau_{1}$};
\node[below] at (6,0) {$\tau_{2}$};
\end{tikzpicture}
+\begin{tikzpicture}[line width=1 pt, scale=0.5, baseline=0]
\node[left] at (0,0) {$-L$};
\node[right] at (7,0) {$L$};
\draw[line width=2 pt] (0,0)--(7,0);
\draw (1,0) arc(180:0:1);
\node[below] at (1,0) {$\tau_{1}$};
\node[below] at (3,0) {$\tau_{2}$};
\draw[densely dashed] (4,0) arc(180:0:1);
\node[below] at (4,0) {$0$};
\node[below] at (6,0) {$\tau$};
\end{tikzpicture} \lp\qquad
+\begin{tikzpicture}[line width=1 pt, scale=0.5, baseline=0]
\node[left] at (0,0) {$-L$};
\node[right] at (6,0) {$L$};
\draw[line width=2 pt] (0,0)--(6,0);
\draw[densely dashed] (1,0) arc(180:0:2);
\node[below] at (1,0) {$0$};
\node[below] at (5,0) {$\tau$};
\draw (2,0) arc(180:0:1);
\node[below] at (2,0) {$\tau_{1}$};
\node[below] at (4,0) {$\tau_{2}$};
\end{tikzpicture}+
\begin{tikzpicture}[line width=1 pt, scale=0.5, baseline=0]
\node[left] at (0,0) {$-L$};
\node[right] at (6,0) {$L$};
\draw[line width=2 pt] (0,0)--(6,0);
\draw (1,0) arc(180:0:2);
\node[below] at (1,0) {$\tau_{1}$};
\node[below] at (5,0) {$\tau_{2}$};
\draw[densely dashed] (2,0) arc(180:0:1);
\node[below] at (2,0) {$0$};
\node[below] at (4,0) {$\tau$};
\end{tikzpicture}\lp\qquad
+\begin{tikzpicture}[line width=1 pt, scale=0.5, baseline=0]
\node[left] at (0,0) {$-L$};
\node[right] at (5,0) {$L$};
\draw[line width=2 pt] (0,0)--(5,0);
\draw[densely dashed] (1,0) arc(180:0:1);
\node[below] at (1,0) {$0$};
\node[below] at (3,0) {$\tau$};
\draw (2,0) arc(180:0:1);
\node[below] at (2,0) {$\tau_{1}$};
\node[below] at (4,0) {$\tau_{2}$};
\end{tikzpicture}
+\begin{tikzpicture}[line width=1 pt, scale=0.5, baseline=0]
\node[left] at (0,0) {$-L$};
\node[right] at (5,0) {$L$};
\draw[line width=2 pt] (0,0)--(5,0);
\draw (1,0) arc(180:0:1);
\node[below] at (1,0) {$\tau_{1}$};
\node[below] at (3,0) {$\tau_{2}$};
\draw[densely dashed] (2,0) arc(180:0:1);
\node[below] at (2,0) {$0$};
\node[below] at (4,0) {$\tau$};
\end{tikzpicture}+\cdots\ . 
\ea
A given  diagram contributes with factor $[\frac{\xxi}{4\pi^{2}}]^{\nu}, \ \nu=$ number of loops. The planar diagrams have color factor $[C_{\R}]^\nu$, while the last two  two-loop
 non-planar 
diagrams have  a factor of $C_{\R}(C_{\R}-\frac{1}{2}C_{\A})$. 
As a  result,  we find for \rf{F.1}  
\be
G_{\perp}(\tau) = \frac{\tau ^{-2+\ve } C_{\R} \xxi }{4 \pi ^2}+\frac{\tau ^{-2+\ve } 
\big[(L-\tau )^{\ve }+2 \tau ^{\ve }-(L+\tau )^{\ve }\big] C_{\A} C_{\R} \xxi 
^2}{32 \pi ^4 (-1+\ve ) \ve }+\mc O(\xxi ^3),
\ee
where  the dependence on $C_{\R}$ is just  by an overall factor.
This is renormalized by setting (cf. \rf{2.8}) 
\be
\xxi = \mu^{\ve}\bigg[\xxi(\mu)+\frac{p_{1}}{\ve}\, \xxi\,^{2}(\mu)+\cdots\bigg],\qquad \ \ \ p_{1} = \frac{C_{\A}}{4\pi^{2}}\ . 
\ee
One can then take $L\to \infty$ and  finally  
\be
\la{F.8}
G_{\perp}^{\rm ren}(\tau, \mu) = C_{\R}\,\frac{\xxi}{4\pi^{2}}\frac{1}{\tau^{2}}\bigg[1-C_{\A}\frac{\xxi}{4\pi^{2}}\big(\log(\mu\tau) + 1 \big)+\mc O(\xxi^{2})\bigg].
\ee
Note that here  
there is no need for an  additional $Z$-factor  so  that   we have 
\be\la{F8}
\bigg(\mu\drg{\mu}+\beta_\xxi\,\drg{\xxi}\bigg)\,G_{\perp}^{\rm ren}(\tau; \mu) =0.
\ee

\paragraph{Two loops}

At two loops, the most convenient scheme is the regularization discussed in \cite{Beccaria:2021rmj} where the propagator is  as in \rf{2.12}
\be
\la{F.10}
D(\tau) = \frac{N \xxi}{8\pi^{2}}\frac{1}{(|\tau|+\eps)^{2}}, \qquad \eps \to 0  \ . 
\ee
 We find that the renormalized two-point function  is then\foot{Compared to \rf{F.8} found in dimensional regularization  in this regularization 
 scheme  the  one-loop  correction contains just $\log (\mu \tau)$ term, i.e. the two $\mu$ parameters in \rf{F.8}  and in \rf{E.10} are  related  by a  factor of $e$.} 
\be\la{E.10}
G_{\perp}^{\rm ren}(\tau; \mu) = C_{\R}\,\frac{\xxi}{4\pi^{2}}\frac{1}{\tau^{2}}\bigg[1-C_{\A}\frac{\xxi}{4\pi^{2}}\log(\mu\tau)
+C_{\A}^{2}\Big(\frac{\xxi}{4\pi^{2}}\Big)^{2}\bigg(
\frac{\pi^{2}}{24}+\frac{1}{2}\log(\mu\tau)+\log^{2}(\mu\tau)
\bigg)
+\mc O(\xxi^{3})\bigg] \ , 
\ee
so that \rf{F8}   is satisfied with 
\be
\la{F.12}
\beta_{\xxi} =C_{\A}\,  \frac{\xxi^2}{4\pi^{2}}-\frac{1}{2}C^2_{\A}\,\frac{\xxi^3}{(4\pi^2)^2}+\mc O(\xxi^{4})\  . 
\ee


\subsection{Coupled scalar}

The same analysis for the  scalar field $\phi$ coupled to the Wilson line requires the evaluation of  around $ 200$ different diagrams. 
All of them can be treated with the regularization (\ref{F.10}) with  the final result 
\ba
G^{\rm ren}(\tau; \mu) =& C_{\R}\,\frac{\xxi}{4\pi^{2}}\frac{1}{\tau^{2}}\bigg[1+C_{\A}\frac{\xxi}{4\pi^{2}}\big(1-3\log(\mu\tau)\big)\lp
+C_{\A}^{2}\Big(\frac{\xxi}{4\pi^{2}}\Big)^{2}\bigg(-2+
\frac{5\pi^{2}}{24}-\frac{3}{2}\log(\mu\tau)+6\log^{2}(\mu\tau)
\bigg)
+\mc O(\xxi^{3})\bigg], 
\ea
where  in addition  to renormalization of $\xxi$  one needs to introduce a $Z$-factor, i.e.
$G^{\rm ren} = Z G$   with 
\be\la{f13}
Z = 1-\frac{C_{\A}\,\xxi}{2\pi^{2}}\log(\eps\, L)+\frac{C_{\A}^{2}\xxi^{2}}{16\pi^{4}}\big[2\log(\eps\, L)+\log^{2}(\eps\, L)\big]+\mc O(\xxi^{3}).
\ee
As a result, $G^{\rm ren}$   satisfies  the 
 Callan-Symanzik equation with an anomalous dimension $\Delta$  
 (see a discussion in \cite{Beccaria:2021rmj})
\be\la{f14}
\bigg[\mu\drg{\mu}+\beta_\xxi\drg{\xxi}+2(\Delta-1)\bigg]\,\big( \xxi^{-1}\, G^{\rm ren}(\tau; \mu)\big) =0,
\ee
where $\beta_{\xxi}$ is as in (\ref{F.12}) and
\be
\Delta=1+\frac{3\,C_{\A}\,\xxi}{8\pi^{2}}-\frac{5\,C_{\A}^{2}\,\xxi^{2}}{64\pi^{4}}
+\mc O(\xxi^{3}).
\ee

\section{Multiply wound Wilson loop}
\la{app:wound}

Our results have a simple application to the case of $k$-wound Wilson loop in the fundamental representation F. This generalization amounts to 
the replacement $\tr_{\F} U\to \tr_{\F}(U^{k})$ in the definition of the Wilson loop (\ref{1.1}). To start, let us 
write 
\be
\la{G1}   
\tr_{\F} U^{k} = \sum_{i}c^{k}_{\ i}\tr_{\R_{i}}U,
\ee
where the sum is over all irreducible representations appearing in $\F^{\otimes k}$. Then, (\ref{G1}) implies
the following relation for the associated Wilson loops 
\be
W_{k\rm{-wound}} =\sum_{i}c^{k}_{\ i}W_{\R_{i}}.
\ee
The coefficients $\{c^{k}_{\ i}\}$ in (\ref{G1}) appear in the inversion of the Frobenius formula \cite{Fulton}
\be
\la{G.3}
\tr_{\R}U = \frac{1}{|\R|!}\sum_{\sigma\in S_{|\R|}}\chi_{\R}(\sigma)\prod_{i=1,2,\dots}\tr_{\F} U^{k_{i}(\sigma)}.
\ee
Here  $K=|R|$ is the number of blocks in the Young tableau of $\R$, $k_{i}(\sigma)$ is the length of the $i$-th cycle of the permutation $\sigma$. 
The symmetric group characters $\chi_{\R}(\sigma)$ are obtained as 
\be
\chi_{\R}(\sigma) = \text{coeff. of } x_{1}^{\ell_{1}}\cdots x_{K}^{\ell_{K}}\ \text{in }\ \Delta(x)\prod_{j\ge 1}^{n}P_{j}(x)^{\nu_{j}(\sigma)}\ , 
\ee
where $\lambda_{i}$ are the rows of the Young tableau of R, padded with zero to have $K$ entries, $\nu_{j}(\sigma)$ is the number of cycles of length $j$ in $\sigma$, and 
$\Delta(x) = \prod_{1\le i< j \le K}(x_{i}-x_{j})$. For $K=2$ this gives the well known relations\foot{We use $(s_1, s_2, ....)$ to label  traces in representations R  by the corresponding Young tableau.}
\ba
\tr_{(1,1)}U &= \frac{1}{2}(\tr_{\F} U)^{2}-\frac{1}{2}\tr_{\F} U^{2}, \qquad 
\tr_{(2)}U = \frac{1}{2}(\tr_{\F} U)^{2}+\frac{1}{2}\tr_{\F} U^{2}, 
\ea
so that the inversion of (\ref{G.3}) reads
\be
\la{G.6}
\tr_{\F} U^{2} = \tr_{(2)}U-\tr_{(1,1)}U.
\ee
Repeating the same procedure for higher values of $k$, (\ref{G.6}) generalizes to 
\ba
\tr_{\F} U^{3} &= \tr_{(3)}U-\tr_{(2,1)}U+\tr_{(1,1,1)}U, \notag \\
\tr_{\F} U^{4} &= \tr_{(4)}U-\tr_{(3,1)}U+\tr_{(2,1,1)}U-\tr_{(1,1,1,1)}U, \notag \\
\tr_{\F} U^{5} &= \tr_{(5)}U-\tr_{(4,1)}U+\tr_{(3,1,1)}U-\tr_{(3,1,1,1)}U+\tr_{(1,1,1,1,1)}U,
\ea
and so on.  To evaluate (\ref{1.15}), we need the sum of $C_{\R}$ and $C_{\R}^{2}$ based on the decomposition (\ref{G1}), \ie the effective $k$-dependent coefficients
\be
\la{G.8}
C_{\R}\to \sum C_{\R} = k^{2}\,\frac{N^{2}-1}{2N}, \qquad  C_{\R}^{2}\to \sum C_{\R}^{2} = k^{2}\frac{(N^{2}-1)[N^{2}+k^{2}(-3+2N^{2})]}{12N^{2}},
\ee
leading to 
\ba
\frac{1}{N}\,\vev{W}_{k-\rm wound} = 1+&k^{2}\,\frac{N^{2}-1}{8N}\,\gym^2 
+k^{2}\,(N^{2}-1)\Big[\frac{k^{2}}{192}\Big(1-\frac{3}{2N^{2}}\Big)
+\frac{1}{128\pi^{2}}(1-\z^{2})^{2}\Big]\,g^4+\mc O(g^6).  \la{G9}
\ea
We remark that, for $\z=1$, the  winding is implemented by the simple substitution rule $g^2 \to k^{2}g^2 $, which is clear in the matrix model  model representation of the BPS WML 
(at any finite $N$). For generic $\z$, we notice that the coefficient of the 
$(1-\z^{2})$ term is instead  $\sim k^{2}g^4$, \ie has a different scaling with $k$.

The same analysis applies to the two point functions or more general correlators. Once we write 
$\llangle \tr_{\F}[\mc O(\tau_{1})\cdots \mc O(\tau_{n})\,U^{k}]\rrangle$ as derivatives of $\llangle \tr_{\F} U(\eta)^{k}\rrangle$
where $\eta(s)$ is a local coupling to $\mc O$, and we can treat $\tr_{\F} \big[U(\eta)\big]^{k}$ as above. 

\newpage

\bibliography{BT-Biblio}
\bibliographystyle{JHEP}
\end{document}